\documentclass[aps, pra, reprint, longbibliography, superscriptaddress]{revtex4-1}
\usepackage{amsmath, amssymb, graphicx, xcolor, mathrsfs, hyperref, array, bbm}

\usepackage{gensymb}

\newcommand{\beq}{\begin{equation}}
\newcommand{\eeq}{\end{equation}}

\newcommand{\eq}[1]{Eq.~(\ref{#1})}

\newcommand{\trenv}[1]{\text{tr} \left\{ #1 \right\} }

\newcommand{\expn}[1]{\text{exp} \left\{ #1 \right\}}

\newcommand{\abs}[1]{\left| #1 \right|}

\newcommand{\wt}[1]{\widetilde{#1}}

\newcommand{\intg}{\mathbb{Z}}

\newcommand{\id}{\mathbbm{1}}

\newcommand{\curve}{\mathcal{C}}

\newcommand{\hlt}{\mathcal{H}}

\newcommand{\ve}{\varepsilon}

\newcommand{\ket}[1]{| #1 \rangle}

\newcommand{\vR}{\mathbf{R}}

\newcommand{\va}{\mathbf{a}}

\newcommand{\vh}{\mathbf{h}}

\newcommand{\vk}{\mathbf{k}}

\newcommand{\vq}{\mathbf{q}}

\renewcommand{\vr}{\mathbf{r}}

\newcommand{\bal}{{\boldsymbol{\alpha}}}
\newcommand{\bbe}{{\boldsymbol{\beta}}}

\newcommand{\bsg}{\boldsymbol{\sigma}}

\newcommand{\viz}{\emph{viz}}

\newcommand{\nullv}{\mathbf{0}}

\newcommand{\screwF}{\mathcal{U}_{S4}}

\newcommand{\screwT}{\mathcal{U}_{S2}}

\newcommand{\glide}{\mathcal{U}_G}

\newcommand{\latname}{$8^2.10$-$a$}
\newcommand{\latnameb}{$8^2.10$-$b$}

\newcommand{\lsym}{\boldsymbol{\gamma}}

\begin{document}
\title{Crystalline Kitaev spin liquids}

\author{Masahiko G. Yamada}
\email[E-mail: ]{m.yamada@issp.u-tokyo.ac.jp}
\affiliation{Institute for Solid State Physics, University of Tokyo, Kashiwa 277-8581, Japan.}

\author{Vatsal Dwivedi}
\email[E-mail: ]{vdwivedi@thp.uni-koeln.de}
\affiliation{Institut f\"ur theoretische Physik, Universit\"at zu K\"oln, Z\"ulpicher Stra\ss e 77a, D-50937 K\"oln}

\author{Maria Hermanns}
\email[E-mail: ]{hermanns@thp.uni-koeln.de}
\affiliation{Institut f\"ur theoretische Physik, Universit\"at zu K\"oln, Z\"ulpicher Stra\ss e 77a, D-50937 K\"oln}

\begin{abstract}
Frustrated magnetic systems exhibit many fascinating phases. Prime among them are quantum spin liquids, where the magnetic moments do not order even at zero temperature. A subclass of quantum spin liquids called Kitaev spin liquids are particularly interesting, because they are exactly solvable,  can be realized in certain materials, and show a large variety of gapless and gapped phases.
Here, we show that non-symmorphic symmetries can enrich spin liquid phases, such that the low-energy spinon degrees of freedom form  three-dimensional Dirac cones or nodal chains.
In addition, we suggest a realization of such Kitaev spin liquids in metal-organic-frameworks. 
\end{abstract}

\maketitle

%
\section{Introduction}     \label{sec:intro} 
As a paradigmatic example of nontrivial collective physics associated with interacting quantum many body systems, quantum spin liquids (QSLs) have been a topic of much interest in recent years~\cite{Balents2010spin, WitczakKrempa2014correlatedAnnualReview,Savary2017quantum}.
QSLs are strongly correlated magnetic systems, ordinarily characterized by the following three features: the absence of a spontaneous symmetry breaking, the absence of a long-range magnetic order, and the existence of fractionalized excitations and/or emergent gauge fields~\cite{Misguich2010qsl}. 
They are also long-range entangled~\cite{KitaevPreskill,LevinWen} --- a property that is often used to identify spin liquids theoretically~\cite{Jiang2012identifying}. 
 
Owing to the strong frustration, such systems are usually not analytically tractable.  However, one significant class of exceptions has been \emph{Kitaev spin liquids}(KSLs)~\cite{Hermanns2017physics,TrebstReview,RoserReview}. These are generalizations of Kitaev's exactly solvable honeycomb model~\cite{Kitaev2006anyons} --- a spin-1/2 quantum magnet on the honeycomb lattice described by the Hamiltonian
\beq 
  \hlt_{\text{Kitaev}} = -\sum_{\gamma-\text{bonds}} J_\gamma \sigma_j^\gamma \sigma_k^\gamma,    \label{eq:kitaev}
\eeq 
where $\gamma \in \{x,y,z\}$ labels the $j$--$k$ bond. These models exhibits strong exchange frustration arising from bond directional interactions, which leads to a spin liquid phase. 

The fundamental idea of Kitaev's solution is to represent the spin in terms of Majorana fermion operators, so that the Kitaev Hamiltonian reduces to a model of noninteracting Majorana fermions coupled to a \emph{static} $\intg_2$ gauge field. As the excitations of the gauge field are usually gapped,  one is left with a free hopping Hamiltonian for  Majorana fermions. 
If this band structure associated with this Hamiltonian is gapped/gapless, then so is the resulting quantum spin liquid. Furthermore, the associated bulk bands may carry topological invariants, analogous to the case of topological insulators.  

The Kitaev model is, in fact, exactly solvable on any tricoordinated lattice, and has been studied in a wide variety of two-dimensinal (2D)~\cite{Yao2007exact,Yang2007mosaic,Kamfor2010kitaev,Rachel2016landau} and three-dimensional (3D) systems~\cite{Si2008anyonic, Mandal2009exactly,  Hermanns2014quantum, Hermanns2015weyl, Hermanns2015spin-peierls, Obrien2016classification}. 
A comprehensive classification of 3D KSLs --- based on time-reversal and inversion symmetries --- has been arrived at in Ref.~\cite{Obrien2016classification} recently. This classification is very similar to that of complex fermions (see e.g.~\cite{Chiu2016classification} and references therein), except in some important aspects due to the possibility of a projective representation of the time-reversal and/or lattice symmetries. The 3D KSLs have been shown to harbor many interesting 3D gapless phases of Majorana fermions, some notable examples being Majorana Fermi surfaces~\cite{Hermanns2014quantum,Hermanns2015spin-peierls}, nodal lines~\cite{Mandal2009exactly,Lee2014heisenberg,Kimchi2014three} and Weyl points~\cite{Hermanns2015weyl,Obrien2016classification}. 

Lattice symmetries have been known to enrich the physics of topological insulators and superconductors~\cite{Slager2013,PhysRevB.88.125129,PhysRevB.88.085110,Shiozaki2014topology}, beyond the conventional tenfold classification of topological insulators and superconductors~\cite{Altland1997classification,Schnyder2008classification,Kitaev2009periodic}. A prominent set of examples of such phases are the so-called \emph{topological crystalline insulators}~\cite{Fu2011tci, Hsieh2012tci}, which possess a bulk topological invariant protected by lattice rotation~\cite{PhysRevB.86.115112,PhysRevLett.111.047006} and/or inversion symmetries~\cite{FuKane2007tci,Turner2010entanglement,Hughes2011inversion}. \footnote{Even though lattice symmetries are necessarily broken in real materials due to imperfections, it is sufficient that the symmetry is realized `on average' to see experimental signatures of topological crystalline insulators~\cite{Hsieh2012tci,Tanaka2012SnTe,Yang2012SnTe,Sziawa2012SnTe}. }
Semi-metallic phases, harboring e.g. Dirac nodes or nodal lines, can be classified using similar methods as for the gapped phases~\cite{Volovik2003,Horava2005stability,Matsuura2013protected,Zhao2013topological,Chiu2014classification,Yang2014dirac}. 
In certain cases, the presence of non-symmorphic symmetries can ensure the existence of topologically protected gapless points~\cite{Parameswaran2013nonsymmorphic,Fag2015new,Zhao2016nonsymmorphic,Wang2016topological,Zhijun2016,Takahashi2017topological} in fermionic systems at half filling, which would conventionally be expected to be gapped.

\begin{figure*}[bt]
	\centering
	\includegraphics[ width=2\columnwidth]{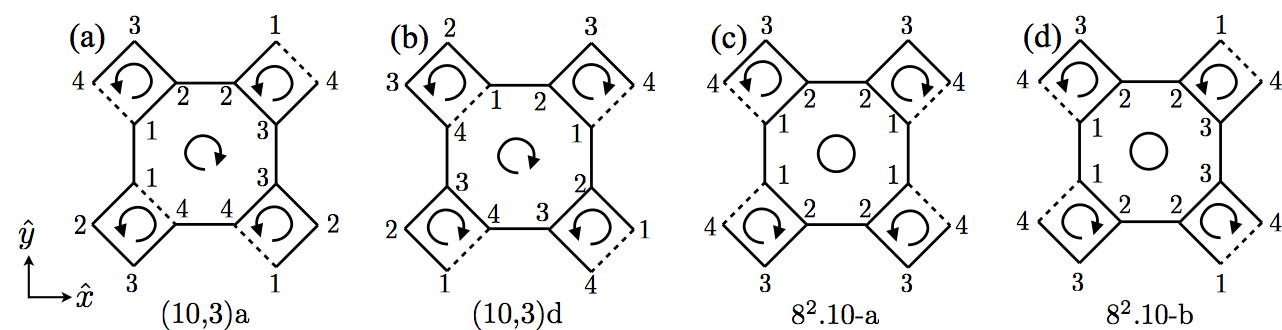}
	\caption{Projection along the crystallographic $c$-axis ($\hat z$ in the following)  for the (a) (10,3)a, (b) (10,3)d,  (c) \latname, and (d) \latnameb~lattice. The numbers indicate the height (in multiples of $1/4$) along the $\hat z$ axis and the lines denote the nearest-neighbor bonds. Dashed lines are used to indicate that the `squares'/`octagons' form spirals. The first lattice is chiral, while the other three are inversion symmetric and contain square-spirals of both chiralities. Note that the octagons in (a) and (b) are spirals, while they close for (c) and (d).}
	\label{lattice_comparison}
\end{figure*}

In this article, we show that non-symmorphic lattice symmetries can also enrich KSL phases. 
We study the Kitaev model on two closely related inversion symmetric tricoordinated lattice structures, termed \latname~and (10,3)d~\cite{Wells1977}.\footnote{In this article, we prefer to use the terminology introduced by Wells~\cite{Wells1977} to distinguish the lattices. The \latname~ lattice is also known as LiGe or \textbf{lig} net, and (10,3)d as  \textbf{utp} net in O'Keeffe's three-letter codes~\cite{OKeeffe2003regular,OKeeffe2003semiregular} which is intensively used in the chemistry literature and database.}
Both lattices are closely related to the hyperoctagon or (10,3)a lattice \footnote{The (10,3)a lattice is also known as SrSi$_2$ or \textbf{srs} net~\cite{OKeeffe2003regular,OKeeffe2003semiregular,Batten2009review}, Laves graph~\cite{laves}, or K$_4$ crystal~\cite{K4}}, as shown in Fig.~\ref{lattice_comparison}, but show very different KSL phases. 
The \latname~lattice harbors a gapless KSL, where the  dispersion exhibits two 3D \emph{Dirac cones} in the bulk Brillouin zone, protected by a combination of the 4-fold screw and glide symmetry. The (10,3)d lattice exhibits a set of touching nodal lines --- a nodal chain~\cite{Bzdusek2016nodal} --- in the bulk Brillouin zone, where the touching points are protected by the combination of time-reversal and a glide symmetry.

Generically, nodal lines in KSLs are protected  by time-reversal symmetry. Breaking this symmetry gaps the lines, leaving behind an even number of Weyl nodes~\cite{Hermanns2015weyl, Obrien2016classification}. In the Kitaev model on (10,3)d, however, one of the nodal lines remains stable even when the time-reversal symmetry is broken, owing to the glide symmetry~\cite{Yang2017topological}. 
Thus, the presence of non-symmorphic lattice symmetries leads to interesting gapless topological phases, not anticipated by the earlier classification based on the time-reversal and inversion symmetries.
This indicates that there are a lot of phases protected by space group symmetries yet to be discovered in Kitaev spin liquids.

Although the Kitaev model looks like a theoretical toy model, the strongly anisotropic spin interactions of the Kitaev model can arise in real materials from the spin-orbit entangled nature of $j=1/2$ magnetic moments in 4$d$ or 5$d$ transition metal ions with a strong spin-orbit coupling~\cite{Khaliullin2005orbital,Jackeli2009}.
Besides conventional materials like iridates~\cite{Singh2012relevance,Takayama2015hyperhoneycomb,Kimchi2014three} and $\alpha$-RuCl$_3$~\cite{Banerjee2016proximate}, a new avenue of material candidates has been opened by \emph{metal-organic frameworks} (MOFs)~\cite{Yamada2017MOF}. In essence, they consist of metal ions bonded to organic ligands, and can form highly complex networks. This is an exciting avenue to realize crystal structures and, thus, types of KSLs that cannot be stabilized in  conventional materials. A prominent example is the hyperoctagon lattice~\cite{Hermanns2014quantum}, which occurs naturally in MOFs~\cite{Oehrstrom2003} and possibly harbors a gapless Kitaev spin liquid with a Majorana Fermi surface.
 
\begin{figure*}
	\centering
	\includegraphics[width=2\columnwidth]{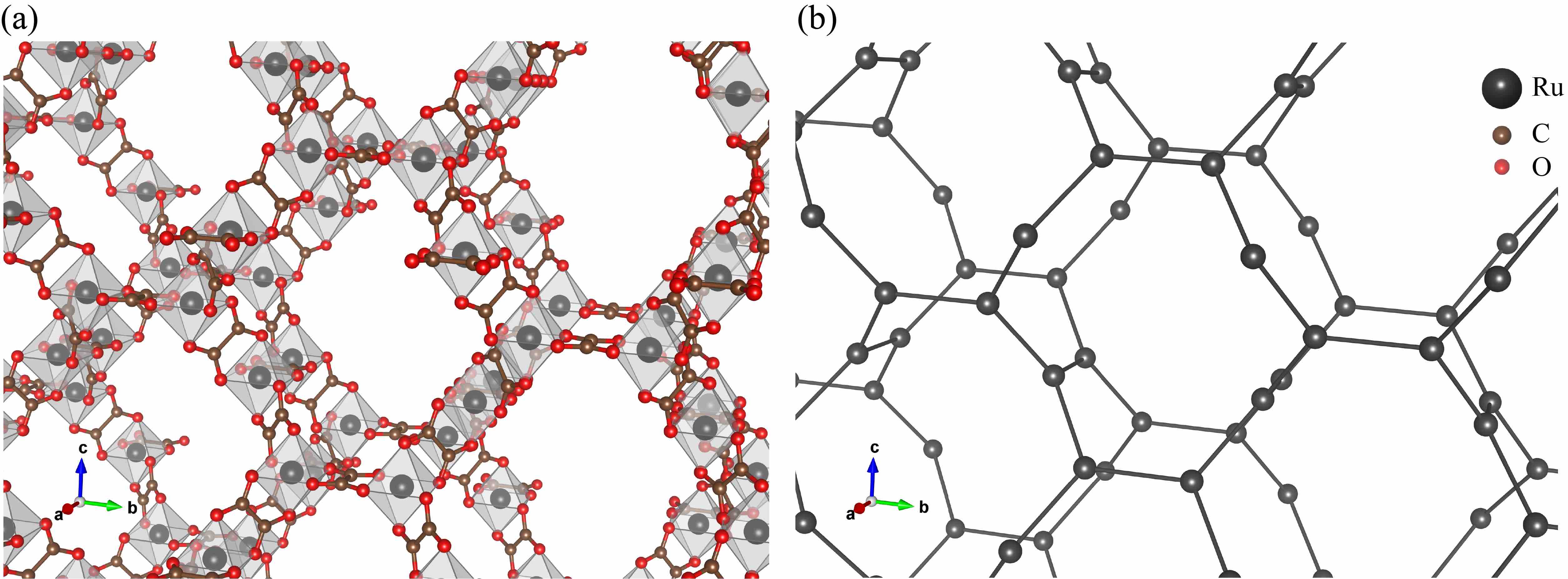}
	\caption{(a) Idealized structure of \latname~MOF, Ru$_2$(C$_2$O$_4$)$_3$ with an $I4_1/amd$ symmetry.
		The $abc$-axes are crystallographic cubic axes.  Gray spheres inside the oxygen octahedra represent Ru atoms
		with a $j=1/2$ magnetic moment due to the octahedral ligand field in combination
		with a strong SOC of Ru$^{3+}.$
		(b) Skeletal structure of the corresponding MOF.  Only Ru atoms are shown here and they
		form the complete \latname~lattice.}
	\label{mof}
\end{figure*}

In fact,  the \latname~lattice has also been realized~\cite{ClementeLeon2013}, albeit in a somewhat distorted form and with the `wrong' metal ions. 
An undistorted version of this lattice can be embedded in a network of edge-sharing octahedra (visualized in Fig.~\ref{mof}), which leads us to expect that this lattice topology with Ir or Ru as the metal ion will be a good candidate for the realization of dominant Kitaev interactions.

The rest of this paper is organized as follows: In Sec.~\ref{sec:mat}, we discuss the MOFs as a potential realization of the 3D lattices of interest.  In Sec.~\ref{sec:ksl}, we summarize the basic physics associated with Kitaev spin liquids on tricoordinated lattices, as well as discuss their classification. We introduce the details of our lattices in Sec.~\ref{sec:lat}, and study the physics of the Kitaev models on the \latname~and (10,3)d lattices in Sec.~\ref{sec:8_3_x} and \ref{sec:10_3_d}, respectively. We conclude our discussions in Sec.~\ref{sec:conc}. 

A point about notation: in the rest of this article, we shall use $\vk = (k_x, k_y, k_z)$ to denote the lattice momenta with components along the Cartesian directions. We shall also use $\vq_j$ to denote the reciprocal lattice vectors corresponding to the lattice translation vectors $\va_j$ and define $k_j = \vk\cdot\va_j, \, j = 1,2,3$, so that $\vk = \sum_j k_j \, \vq_j$.


\section{Metal-organic frameworks}    \label{sec:mat}
In this section, we provide a brief introduction to metal-organic frameworks (MOFs), a class of 
materials which can potentially realize Kitaev physics on certain tricoordinated lattices. 
We also discuss the mechanism that generates the Kitaev interactions, and some possible candidates for  materials that realize the 3D nets studied in the rest of this article.

\subsection{Structure and construction}

An MOF~\cite{Hendon2017MOF}, also known as a porous coordination polymer~\cite{Kitagawa2004PCP}, is a \emph{coordination polymer} consisting of metal ions
and organic ligands, which are connected by \emph{coordinate bonds}, 
i.e, the organic ligands donate a pair of electrons to the metal ion to form chemical bonds. 
The metal ions are typically referred to as \emph{nodes}, while the ligands are often referred
to as \emph{struts} or \emph{linkers}.  During the last few decades, these materials have been a 
subject of much interest, owing to their diverse applications as well as the possibility 
of realizing diverse structural topologies in 1, 2 or 3 dimensions~\cite{Batten2009review}.
Their applications range from catalysis~\cite{Lee2009catalyst} and gas storage~\cite{Murray2009review} to usage in electronic devices~\cite{Stassen2017device}, spanning chemistry and physics. 

A guiding principle in determining the structure of MOFs is the ligand field experienced by the transition metal ions.  
The originally degenerate $d$ or $f$-orbitals of the metal ion split in the presence of ligands, and the resulting energy gain leads to the chemical bonding via coordinate bonds. 
There are several possible coordinations --- octahedral, tetrahedral, tetragonal, etc. --- and for a given metal-ligand combination, one can be energetically favorable over the others. This leads to a `default' structure of the ligands around the metal ions and opens up a way to `design' MOFs corresponding to a given lattice structure by a suitable choice of metal-ligand combination. 
In the resulting MOF, the metal ions form the nodes of the corresponding 3D lattice, while the organic ligands form the links. 
This provides a bottom-up approach for synthesis of MOFs by self-assembly~\cite{Norbert2012review}.

Based on the node-linker construction of MOFs, many tricoordinated nets have been synthesized, including (but not limited to) the layered honeycomb lattice, as well as (10,3)a (hyperoctagon), (10,3)b (hyperhoneycomb), (10,3)d, and the \latname~lattice, see~\cite{Oehrstrom2003} for a more complete list.
However, if one wants to realize the Kitaev model on these lattices, one needs to consider a special subclass of MOFs, as discussed in the following section. 

\begin{figure}
\centering
\includegraphics[width=0.6\columnwidth]{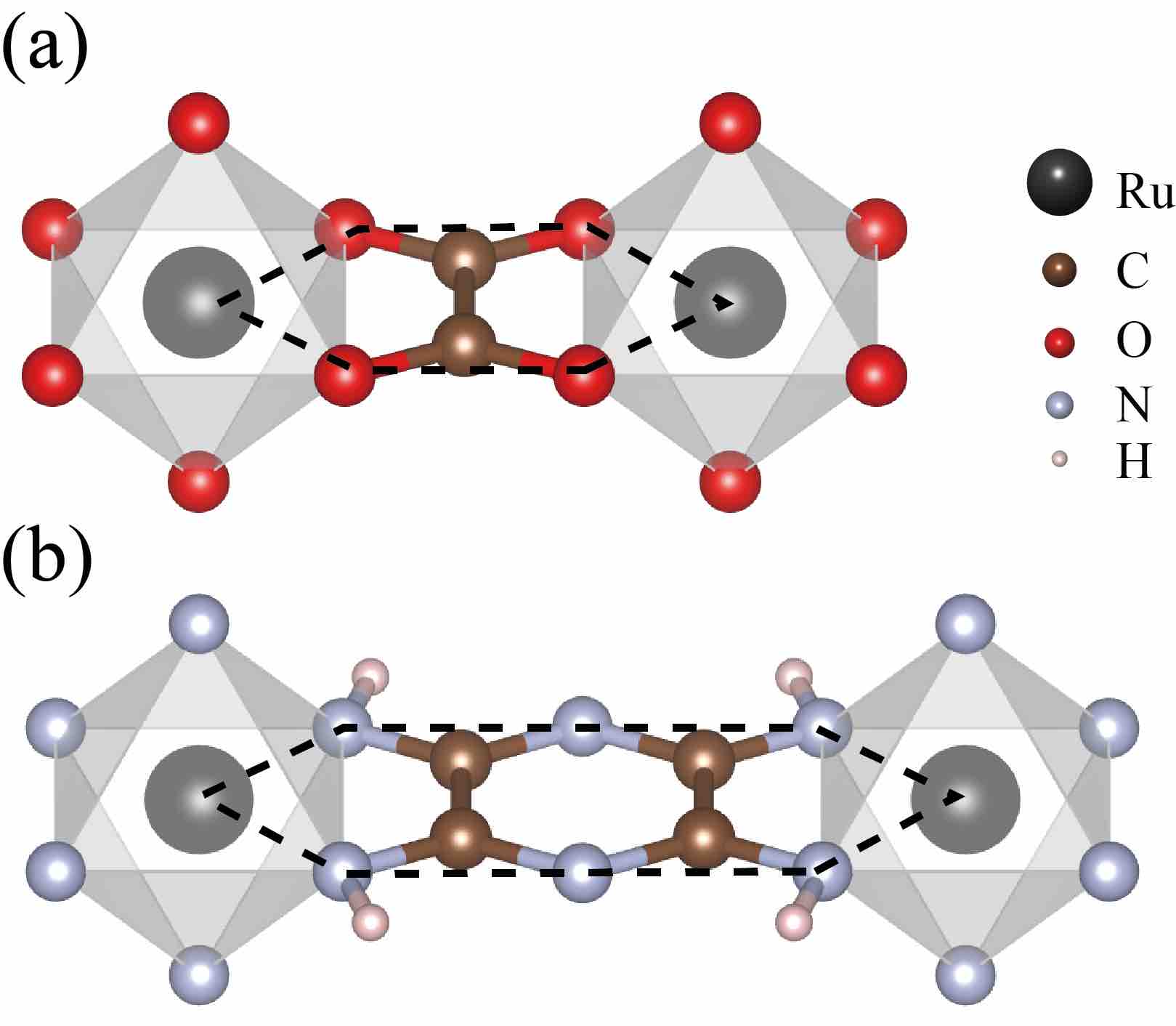}
\caption{Superexchange paths for the (a) Ru-oxalate-Ru and (b) Ru-tetraaminopyrazine-Ru structure, where the Ru atoms are surrounded by (a) oxygen or (b) nitrogen octahedra.
In each figure, the black dashed lines represent the two exchange paths whose destructive interference suppresses the Heisenberg interaction.  
}
\label{super}
\end{figure}

\subsection{Realization of Kitaev physics}

Let us first briefly discuss, how Kitaev interactions can arise in materials. 
That the Kitaev model is not just an interesting toy model, but can be relevant for actual materials, was first pointed out by Jackeli and Khaliullin~\cite{Khaliullin2005orbital,Jackeli2009}.
They showed that Kitaev interactions can be the most dominant interaction in certain transition metal compounds, where the ions are in a $d^5$ electronic configuration --- the experimentally realized materials contain Ir$^{4+}$ (5$d^5$) and Ru$^{3+}$ (4$d^5$), but one would also expect the same physics for osmates, rhenates~\cite{TrebstReview}, and rhodates~\cite{rhodates}.
The central ingredients needed for dominant Kitaev interactions are oxygen octahedra, spin-orbit-coupling (SOC) and multiple superexchange paths. 
The crystal field created by the oxygen octahedron splits the degenerate $d$-orbitals into two (high-energy) $e_g$-orbitals and three (low-energy) $t_{2g}$-orbitals. 
The latter multiplet is further split by the strong SOC into a completely filled $j=3/2$ quartet and a half-filled $j=1/2$ Kramers doublet. 
Even moderately strong electronic correlations can now drive the system into a Mott insulator regime where \emph{spin-orbit-entangled} $j=1/2$ moments are localized on the transition metal sites. 

The interaction between these local moments depends on the microscopic arrangement of the oxygen octahedra. 
Because of the large distance between the magnetic sites, the main contribution comes from the superexchange via the oxygen atoms.  For an edge-sharing~\cite{Khaliullin2005orbital,Jackeli2009} or a parallel-edge-sharing~\cite{Becker2015spin} configuration, there are two exchange path that interfere destructively.
For ideal octahedra, this mechanism suppresses the Heisenberg exchange almost completely (a small contribution always comes from the direct exchange, which however is suppressed exponentially in the distance between magnetic ions) and leaves the Kitaev exchange as the dominant interaction.   

Using the bottom-up (self-assembly) approach discussed in the previous subsection, one can propose a sub-class of MOFs where Kitaev interactions should be dominant~\cite{Yamada2016MOF}. 
If we choose the node to be a transition metal ion $M$ favoring an octahedral coordination, and the linker to be an organic ligand $L$ with \emph{two} oxygen atoms at each end, then they would self-assemble into the structure $M_2 L_3$, where $M$ is surrounded by six oxygen atoms forming an octahedron and the bond angle between the three ligands $L$ is 120 degrees.  
As a concrete example, let us consider the case where $M$ is Ru and $L$ is oxalate. 
As evident from Fig.~\ref{super}, the local structure around the Ru ion is very similar to that of the conventional materials that were considered in Ref.~\cite{Jackeli2009}, and we expect the existence of local, spin-orbit entangled Kramers doublets at the metal ion sites.
The two exchange paths are longer than those in the conventional materials because of the larger size of the organic ligand, but they still interfere destructively (see Fig.~\ref{super}(a)), leading to a dominant Kitaev interaction~\cite{Yamada2016MOF}. 
In fact, the longer exchange path is an advantage, because the direct overlap between the orbitals of neighboring ions is very strongly suppressed. 
There still are other possible interactions arising from the superexchange, e.g. Heisenberg interactions. 
Their relative strengths can be calculated for each individual case starting from the microscopic description of the system in the framework of fragment molecular orbital theory~\cite{Yamada2016MOF}.

Another advantage of MOFs over iridates (and other conventional systems) is that we can `tune' the interaction strength between neighboring atoms without deforming the oxygen octahedra. 
For instance, we can (i) replace the oxygen in oxalate with other elements, such as sulphur (tetrathiooxalate (C$_2$S$_4$)$^{2-}$) or the amide group
(tetraaminooxalate (C$_2$(NH)$_4$)$^{2-}$), or (ii) use ligands based on other organic motifs, such as the tetraaminopyrazine ((C$_4$N$_6$H$_4$)$^{2-}$) family [see Fig.~\ref{super}(b)].
This freedom allows us to explore the phase diagram of the Kitaev model on a given lattice. For instance, using identical ligands for all three bond types, we expect to realize the isotropic Kitaev model with $J_x=J_y=J_z.$ However, using different ligands --- e.g. the tetraaminooxalate-based ligands for the $z$-bonds and tetraaminopyrazine-based ligands for the $x$- and $y$-bonds --- we can tune the value of $J_z$ with respect to $J_x=J_y.$

The octahedral coordination is favorable for most transition metals, so we can expect a variety of MOFs where the metal ions form tricoordinated lattices. 
In fact, $M_2 L_3$ MOFs forming the (10,3)b lattice~\cite{Zhang2012hyperhoneycomb} as well as the layered honeycomb lattice~\cite{Zhang2014honeycomb} have already been reported.
Another important 3D tricoordinated net is the hyperoctagon, or (10,3)a, lattice~\cite{Hermanns2014quantum}.
This lattice  is difficult to realize in inorganic materials due to its chirality, but it can naturally be realized in MOFs~\cite{Coronado2001hyperoctagon}. 
Also the \latname~lattice can be realized as an $M_2 L_3$ MOF, as shown in Fig.~\ref{mof}, even though the current MOF realization --- Mn/Cr-oxalate framework (MnCr(C$_2$O$_4$)$_3$) --- is still quite distorted~\cite{ClementeLeon2013}. 
To the best of our knowledge, there is no realization of (10,3)d MOF which can potentially
be used to realize the Kitaev model on the (10,3)d lattice.  Further materials search is
necessary in this direction.


\section{Kitaev spin liquids}    \label{sec:ksl}
In this section, we briefly review the analytical solution of the Kitaev model and the classification of Kitaev spin liquids based on time-reversal and inversion symmetries. This discussion is intended merely to orient the reader and set up our notation; for more details, we refer the reader to Ref.~\cite{Obrien2016classification}.

\subsection{Solution of Kitaev model}
The original solution~\cite{Kitaev2006anyons} of the Kitaev model of \eq{eq:kitaev}
hinges on the existence of an infinite number of conserved quantities associated with the loops (``\emph{plaquettes}'') of the lattice. For each such loop $\ell$, one defines a loop operator 
\beq 
\label{eq:loop}
  W_\ell = \prod_{s\in\ell} \sigma_s^{\gamma_s} \sigma_{s-1}^{\gamma_s}, 
\eeq 
where the $(s-1, s)$ bond is of type $\gamma_s$. It has eigenvalues $\pm 1$ for loops of even length and $\pm i$ for loops of odd length. 
All the loop operators $W_\ell$ commute with the Hamiltonian and with each other and, thus, can be diagonalized simultaneously. 
Consequently, the Hilbert space decomposes into subspaces --- one for each particular flux configuration of the emergent $\intg_2$ gauge field --- and we can restrict the discussion to a single `flux sector'. 

For 3D lattices, the set of eigenvalues of $W_\ell$ are not all independent; instead, they are restricted by a set of \emph{volume constraints}~\cite{Obrien2016classification}. 
These arise because for a set of loops $\ell_j, \, j = 1, \dots N_\ell$ forming a closed volume, the product of loop operators must satisfy $\prod_j W_{\ell_j} =  \id$, since each edge forming the closed volume appears in an even number of loop operators. 
This is equivalent to the statement that there are no monopoles in the emergent $\intg_2$ gauge field. 

To expose this gauge structure and solve the model, Kitaev wrote the spin operators in terms of four Majorana operators $a^x_j,\,a^y_j,\,a^z_j,\,c_j$ for each $j$ as 
\beq   
  \sigma_j^\gamma = i a_j^\gamma c_j,
\eeq 
with the anticommutation relations 
\[
 \big\{ a_j^\gamma, a_{k}^{\delta} \big\} = 2 \delta^{\gamma\delta} \delta_{jk},  \quad  \big\{ c_j, c_{k} \big\} = 2 \delta_{jk}, \quad  \big\{ a_j^\gamma, c_{k} \big\} = 0.
\]
Since four Majorana fermions are equivalent to two complex fermions, this prescription extends the Hilbert space at each lattice site from two dimensional to four dimensional. Thus, to recover the original spin Hilbert space, we project down to the one (complex) fermion sector of this Hilbert space by defining the operator $D_j = a_j^x a_j^y a_j^z c_j$ and demanding that the physical states satisfy $D_j\ket{\psi} = \ket{\psi} \, \forall \, j$. 

The $\intg_2$ gauge field can be written as the operators $\hat{u}_{jk} \equiv i a_j^\gamma a_k^\gamma$, (where $\gamma$ is the type of the $j$-$k$ bond), whose eigenvalues are $\pm 1$ in the extended Hilbert space. 
As the bond operators, $\hat{u}_{jk}$, commute with the Hamiltonian and each other, we can simply fix their eigenvalues (`fix a gauge') for any fixed flux configuration of the loop operators \eqref{eq:loop}.
We, thereby, obtain a noninteracting hopping Hamiltonian in terms of $c_j$'s, which we shall term the \emph{Kitaev-Majorana Hamiltonian}. 
The latter can be studied using the techniques conventionally employed for the noninteracting fermionic systems.

The remaining, essential classical, problem is to determine the flux sector that minimizes the ground state energy, i.e. the ground state flux configuration of the $\intg_2$ gauge field.
Most generally, one can compute the lowest energy sector by exhaustively searching in the space of all configurations by numerically diagonalizing the Hamiltonian on a finite lattice. However, for a few lattices, we can use Lieb's theorem, which, predicts the ground state flux configuration for certain plaquettes posessing a mirror symmetry~\cite{Lieb1994flux}.

\subsection{Symmetries}
Symmetries play an important role in studying the various phases of the Kitaev spin liquids.  While the original symmetry operators, acting on the spins, form an \emph{ordinary} representation of the associated symmetry group, the corresponding operators acting on the Majorana fermions form a \emph{projective} representation of the symmetry group~\cite{Wen2002quantum}. The essential reason for this is the presence of a $\intg_2$ gauge symmetry, so that the symmetry operations can be (and often are) complemented by a gauge transformation, which corresponds to the ``projective'' part of the representation.

The most basic ``symmetry'' of the Kitaev-Majorana Hamiltonian is the particle-hole symmetry, which is essentially a redundancy in the description arising from the enlargement of the Hilbert space when we represent the spins in terms of the Majorana operators. 
Besides, the original spin model possesses a time-reversal symmetry, which turns out to be closely related to the sublattice (or chiral) symmetry in the effective Majorana Hamiltonian~\cite{Kitaev2006anyons,Obrien2016classification}.

For bipartite lattices, the Bloch Hamiltonian can always be recast in the form
\beq 
  \hlt(\vk) = 
  \begin{pmatrix}
   0 & A(\vk) \\ 
   A^\dagger(\vk) & 0 
  \end{pmatrix},    \label{eq:hlt_chiral}
\eeq 
This representation is convenient since $\det \hlt(\vk) = \abs{\det A(\vk)}^2$, which we can use to efficiently determine zero modes of the Hamiltonian. Furthermore, the matrix $A$ can be used to conveniently write the Berry phase associated with a curve $\curve$ in the momentum space and compute the chiral invariant associated with nodal lines as
\beq 
  \theta = \frac{1}{4\pi i} \oint_\curve \trenv{ A^{-1} dA - \left( A^\dagger \right)^{-1} d A^\dagger }.    \label{eq:chiral_inv}
\eeq

In order to study the interplay of KSLs with symmetries, we also seek to break those symmetries in a controllable fashion. The lattice symmetries can naturally be broken by various distortions of the lattice, which basically amounts to changing the hopping parameters. 
The only relevant \emph{intrinsic} symmetry, {\viz} the time-reversal symmetry, can be broken by addition of a Zeeman term ($\vh\cdot\bsg$) to the original Kitaev Hamiltonian. The model is not exactly solvable anymore; however, for small magnetic field, the effect of this term can be analyzed by a perturbative expansion about the exactly solvable $\vh = \nullv$ point~\cite{Kitaev2006anyons}. The first nontrivial contribution occurs at the third order in perturbation theory, and can be written as 
\beq 
  \hlt_{\text{eff}} \propto \sum_{j,l,k} \sigma_j^\alpha \sigma_k^\beta \sigma_l^\gamma, 
\eeq 
where $j,k,l$ are three lattice sites such that $(j,l)$ and $(l,k)$ are nearest neighbors with bond types $\alpha$ and $\beta$, respectively, and $\gamma \neq \alpha, \beta$ is the remaining bond label. This Hamiltonian can again be rewritten in terms of the Majorana operators, resulting in a next-nearest-neighbor hopping term.

\begin{figure*}[tb]
	\centering  
	\includegraphics[width=0.65\columnwidth]{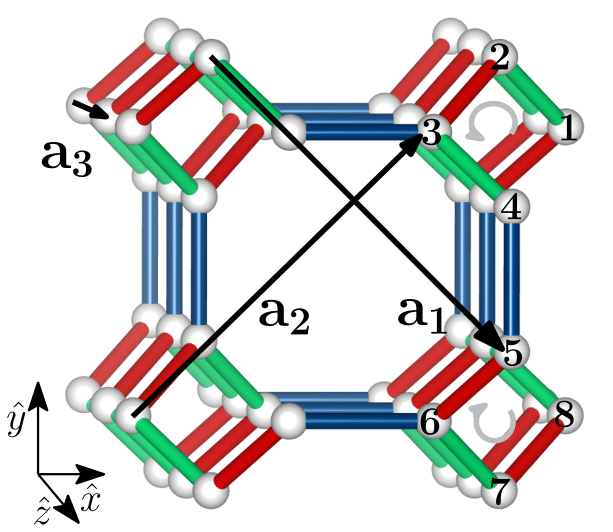} $\quad\quad$
	\includegraphics[width=0.45\columnwidth]{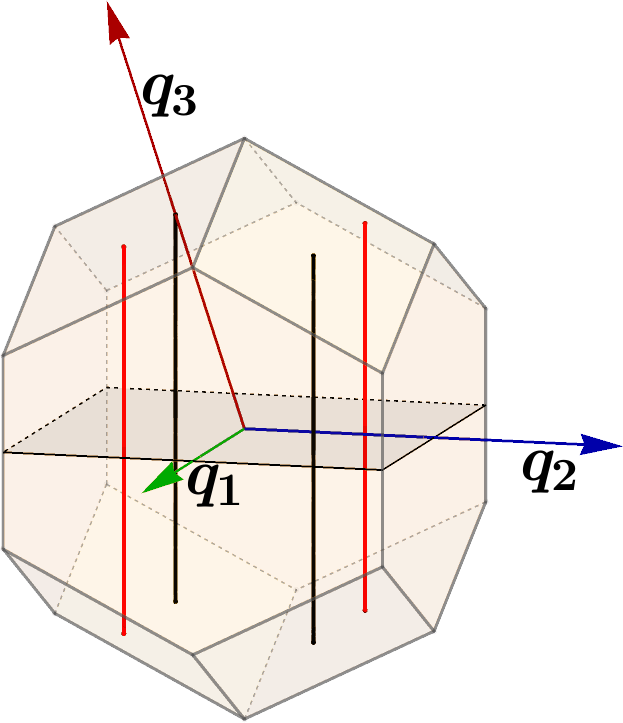}   $\quad\quad$
	\includegraphics[width=0.65\columnwidth]{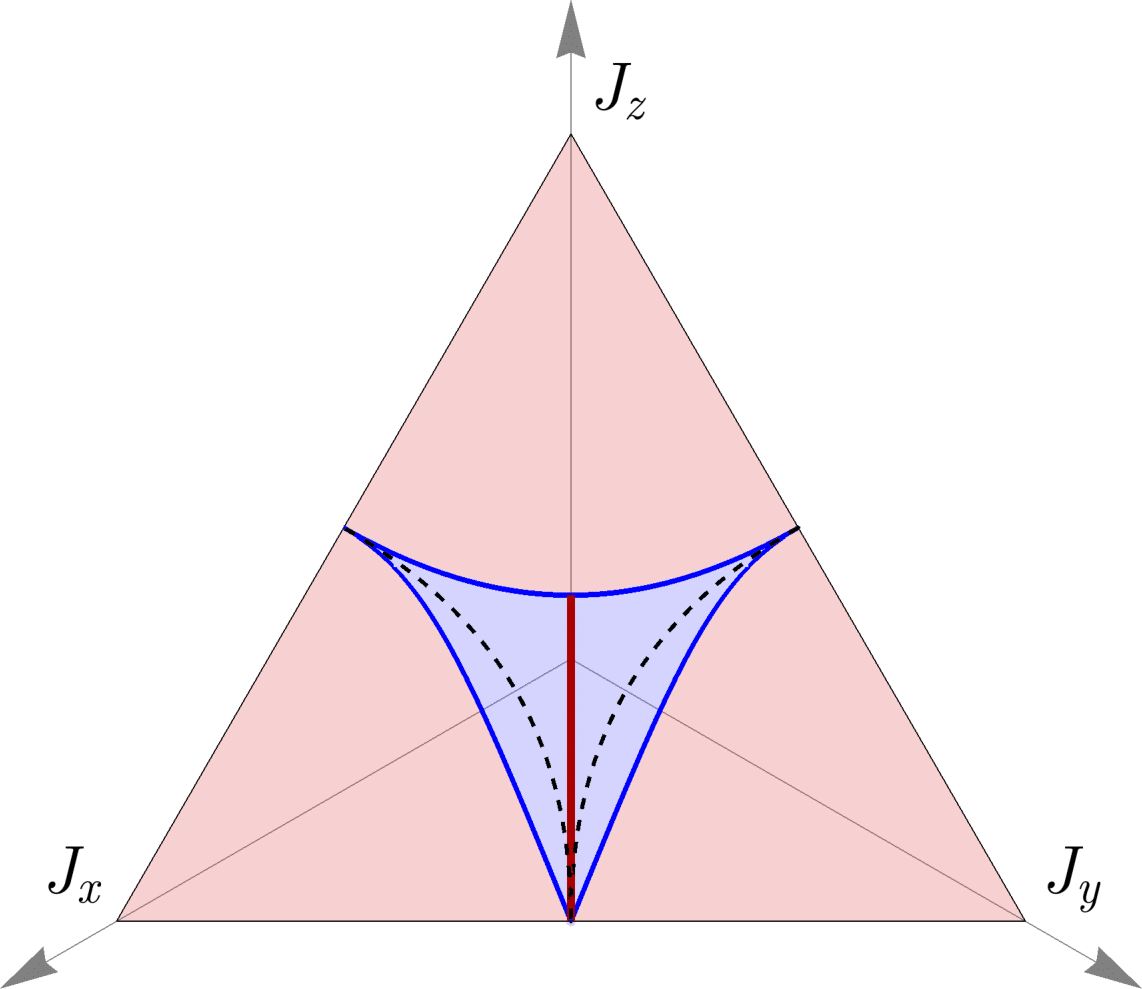} 
	\caption{
		(Left) The unit cell and translation vectors for the \latname~lattice, with the $x$-,$y$- and $z$-bonds colored by green, red, blue respectively. 
		(Middle) The Brillouin zone for the \latname~lattice. We also depict the lines in the Brillouin zone symmetric under fourfold screw (red) and twofold screw (black).
		(Right) The phase diagram for the Kitaev model on \latname.  The parameter region shaded blue corresponds to the gapless regime, while the region shaded pink represents the gapped regime. The solid red line corresponds to $J_x = J_y$, where the system exhibits 3D Dirac nodes in the bulk, while the dashed black line denotes the parameter values for which the two nodal lines touch (see Fig.~\ref{fig:8x_nodes}). 
	} 
	\label{fig:8x}
\end{figure*}

\subsection{Classification of Kitaev spin liquids} 
When classifying KSLs we are interested in the projective representation of the symmetry group. 
For the projective symmetry of the Majorana system, the associated gauge transformation may enlarge the unit cell, resulting in the implementation of the symmetry with an additional translation in the momentum space.
This  makes the symmetry classification of KSL richer than the corresponding classification for free fermions~\cite{Chiu2016classification}; in particular, the Majorana system may realize phases of matter that are  forbidden for complex fermions. 
A systematic classification of KSLs based on time-reversal and inversion symmetry was presented in  Ref.~\cite{Obrien2016classification}. 
Here we only discuss the main results in order to contrast the generic behavior of KSLs to the (partially) atypical behavior of the KSL on both  (10,3)d and \latname. 

Let us first discuss time-reversal symmetry for bipartite lattices. If the sublattices have the same translation vectors as the full lattice --- that is true e.g. for the honeycomb or hyperhoneycomb lattice --- then  time-reversal is implemented trivially, i.e. without any additional translation in momentum space. This is the most common behavior for KSLs, and both the (10,3)d and the \latname~lattice fall in this category. 
Such systems generically harbor gapless spin liquid phases with one or more nodal lines, which are protected by time-reversal symmetry. Breaking the time-reversal symmetry generically gaps the nodal line(s) up to an even number of Weyl points. 
In the following sections, we show that neither \latname~ nor (10,3)d follow this generic behavior, mostly due to additional symmetries present in the systems. 

If the sublattice is not invariant under the translation vectors of the full lattice --- i.e. the sublattice gauge transformations enlarge the unit cell --- then time-reversal is implemented with an additional translation in momentum space. This is e.g. the case for the hyperoctagon and the hyperhexagon lattices. 
The resulting KSLs generically harbor an even number of Fermi surfaces that  are often topological --- i.e. they surround a Weyl node that lies below the chemical potential. 
If the Majorana Hamiltonian is, in addition,  inversion symmetric, then these Weyl nodes are constrained to lie exactly at the chemical potential, and the system becomes a Weyl spin liquid. 
The latter is a prominent example of how projective symmetries may lead to a richer variety of phases: Weyl nodes are  forbidden in free fermion systems with both inversion and time-reversal symmetry, but they can exist in KSLs, e.g. for the Kitaev model on the hyperhexagon lattice~\cite{Obrien2016classification}. 
Breaking the time-reversal symmetry usually has very little effect on this class of KSLs --- the Majorana Fermi surfaces deform and the Weyl nodes move in momentum space, but the overall nature of the KSL remains unchanged. 
Weyl spin liquids are, in fact, stable against arbitrary perturbations (e.g. a small Heisenberg exchange) away from the exactly solvable Kitaev model, whereas the Majorana Fermi surface undergoes a Spin-Peierls-BCS transition that gaps the Fermi surfaces up to an (odd) number of nodal lines~\cite{Hermanns2015spin-peierls}.


\section{Lattice descriptions}   \label{sec:lat} 
In this section, we describe the \latname~and (10,3)d lattices and their symmetries relevant to Kitaev physics. 
They  belong to a class of 3D lattices whose projection along the crystallographic $c$ direction (chosen as $\hat z$ in the following) is a square-octagon lattice, as shown in Fig.~\ref{lattice_comparison}.  In the following, we use the terms `square' and `octagon' to describe the lattices. However, the reader should keep in mind that the `squares' do not close, but instead form spirals along $\hat z$. The nonplanar octagons close for \latname, but not for (10,3)d.

\subsection{The \latname~lattice}

The \latname~lattice can be described as a body-centered tetragonal lattice with 8 sites per unit cell. Within the unit cell centered at origin, the positions of these sites are given by 
\begin{align}
  & \vr_1 = \frac{1}{4} \left(  a,  b, 	1  \right), & 
  & \vr_2 = \frac{1}{4} \left(  0,  a+b,2  \right), & \nonumber \\ 
  & \vr_3 = \frac{1}{4} \left( -a,  b, 	3  \right), & 
  & \vr_4 = \frac{1}{4} \left(  0, -a+b,4 \right), & \nonumber \\ 
  & \vr_5 = \frac{1}{4} \left(  0, a-b, 4  \right), & 
  & \vr_6 = \frac{1}{4} \left( -a, -b, 	3  \right), & \nonumber \\ 
  & \vr_7 = \frac{1}{4} \left(	0,-a-b, 2   \right), & 
  & \vr_8 = \frac{1}{4} \left(  a,-b, 	1 \right), & 
\end{align}
where $a = 1/\sqrt{2}$ and $b = 2a = \sqrt{2}$ (see Fig.~\ref{fig:8x}). 
The lattice is bipartite, and we choose the A sublattice as all the odd numbered sites, and the B sublattice as all the even numbered sites.  
We take the lattice vectors as 
\begin{align}
  & \va_1 = \; \frac{1}{2} \left( b,-b, 1 \right),  
  & \va_2 =  \;\frac{1}{2} \left( b, b, 1 \right), \nonumber \\ 
  & \va_3 = \left( 0, 0, 1 \right),  & 
\end{align}
and the corresponding reciprocal lattice vectors are given by 
\begin{align}
  & \vq_1 = \; \left( \frac{2\pi}{b},-\frac{2\pi}{b}, 0 \right),
  & \vq_2 = \; \left( \frac{2\pi}{b}, \frac{2\pi}{b}, 0 \right),  \nonumber \\ 
  & \vq_3 = \; \left( -\frac{2\pi}{b}, 0, 2\pi \right).  & 
\end{align}
The first Brillouin zone is illustrated in the middle panel of Fig.~\ref{fig:8x}.

\begin{figure}
	\centering
	\includegraphics[width=\columnwidth]{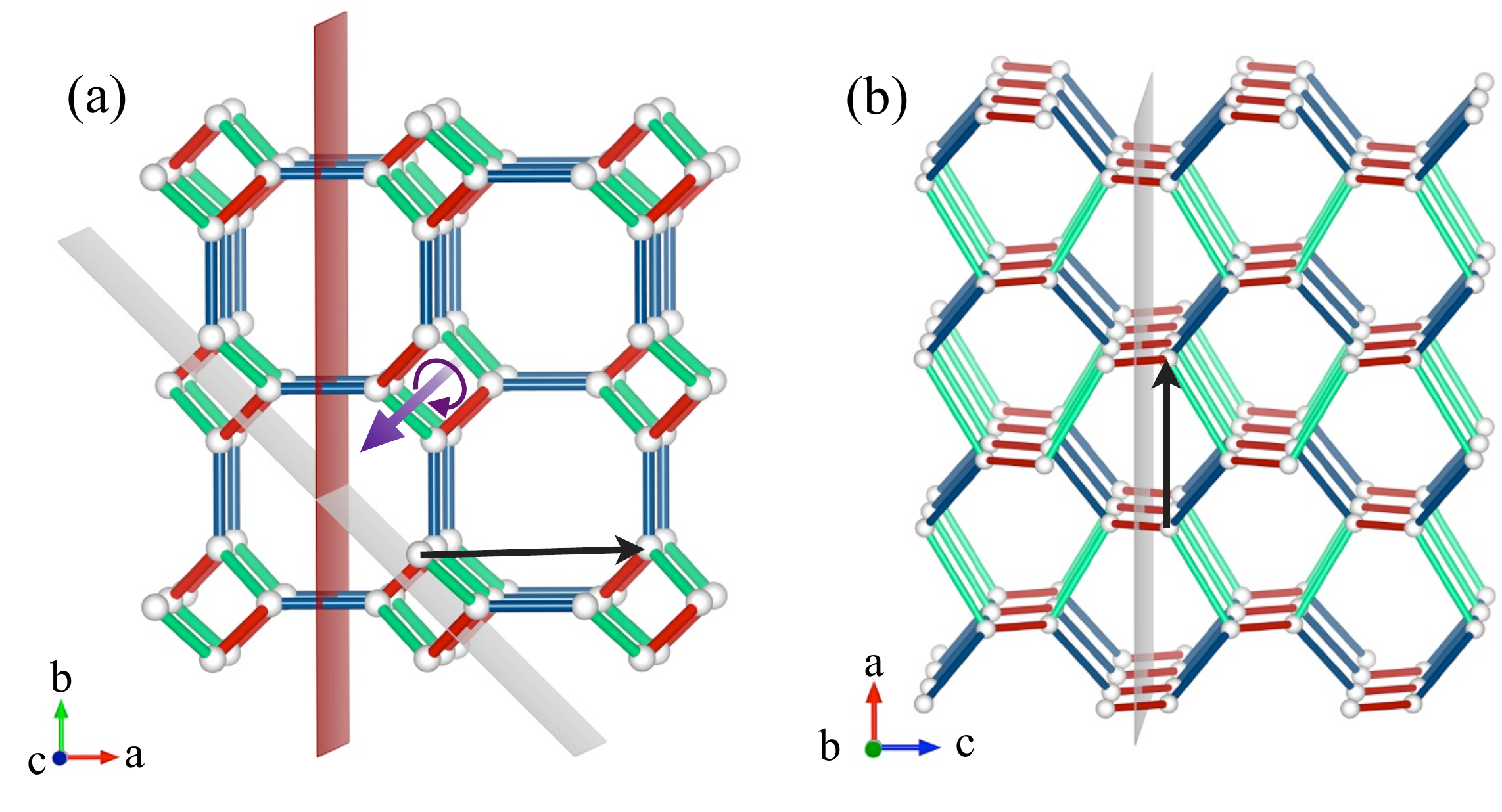}
	\caption{(a) Symmetries of the \latname~lattice: The purple arrow shows  the rotation axis of the 2- and 4-fold screw rotation. The gray plane indicates one of the glide mirror planes, with the corresponding translation indicated by the black arrow. The red plane indicates one of the mirror planes that are relevant for Lieb's theorem. (b) Glide symmetry of the (10,3)d lattice. The black arrow indicates the corresponding  translation.}
	\label{Fig:symmetries}
\end{figure}

The symmetries of the \latname~lattice correspond to the space group $I4_1/amd$ (No. \textbf{141}). For the Kitaev physics, the most relevant symmetries are the inversion, mirror, screw  and glide symmetries. 
The inversion and mirror symmetries are defined with respect to the centers of the octagons. 
The screw  symmetries involve a fourfold or twofold rotation around a rotation axis along $\hat z$ that is located at the center of the square spirals, followed by a translation along the rotation axis. The glide symmetries involve a reflection about planes cutting through the square spirals followed by a translation, which is visualized in Fig.~\ref{Fig:symmetries}(a).

\begin{figure}
	\centering
	\includegraphics[width=\columnwidth]{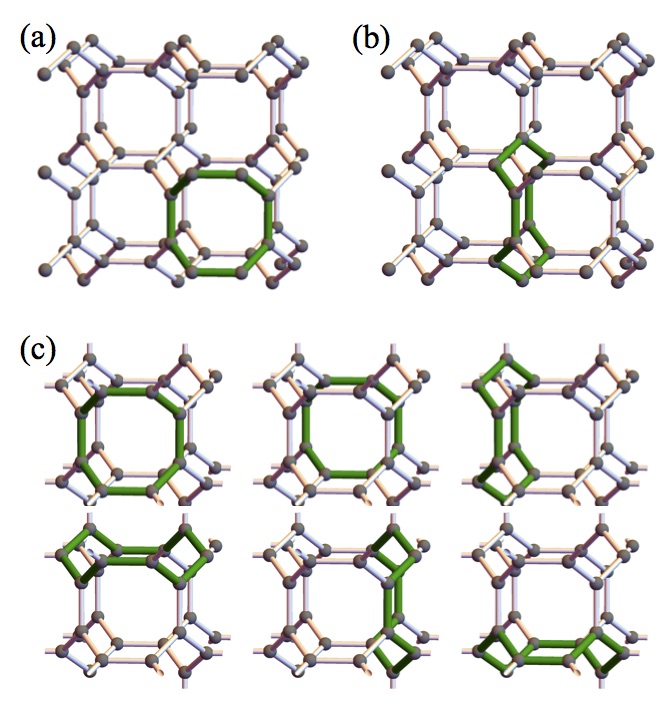}
	\caption{
		There are two distinct types of loops in \latname: (a) 8-loops that run along the nonplanar octagons, and (b) 10-loops that spiral up along one square and down along a neighboring square. 
		Closed volumes are formed by six loops, as shown in (c).  
	}
	\label{fig:810loops}
\end{figure}
The \latname~lattice is an example of a net, where not all the elementary loops are of the same length. There are six elementary loops associated with each unit cell: two of length 8 (8-loops) and four of length 10 (10-loops), as shown in Fig.~\ref{fig:810loops}. The two 8-loops correspond to the two distinct (nonplanar) octagons, while the remaining four 10-loops involve spiraling up along one square and then down along an adjacent one.

\subsection{The {\normalfont (10,3)d} lattice}  
\begin{figure*}[htb]
	\centering  
	\includegraphics[width=0.7\columnwidth]{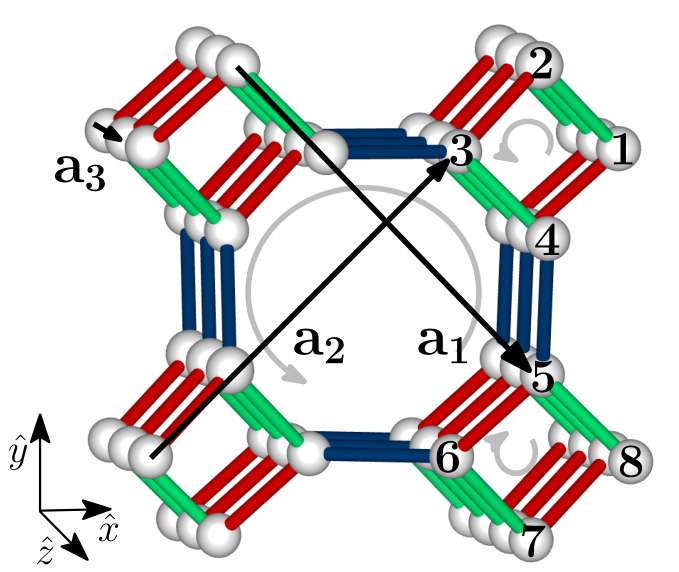} $\quad\quad$
	\includegraphics[width=0.4\columnwidth]{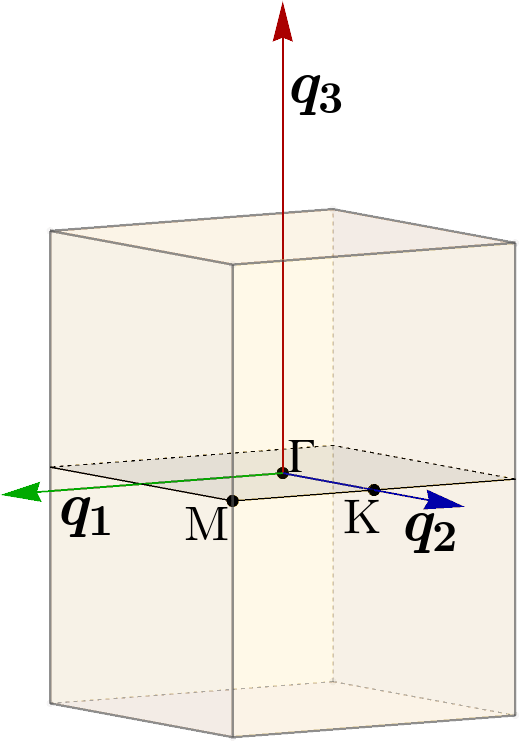}   $\quad\quad$
	\includegraphics[width=0.65\columnwidth]{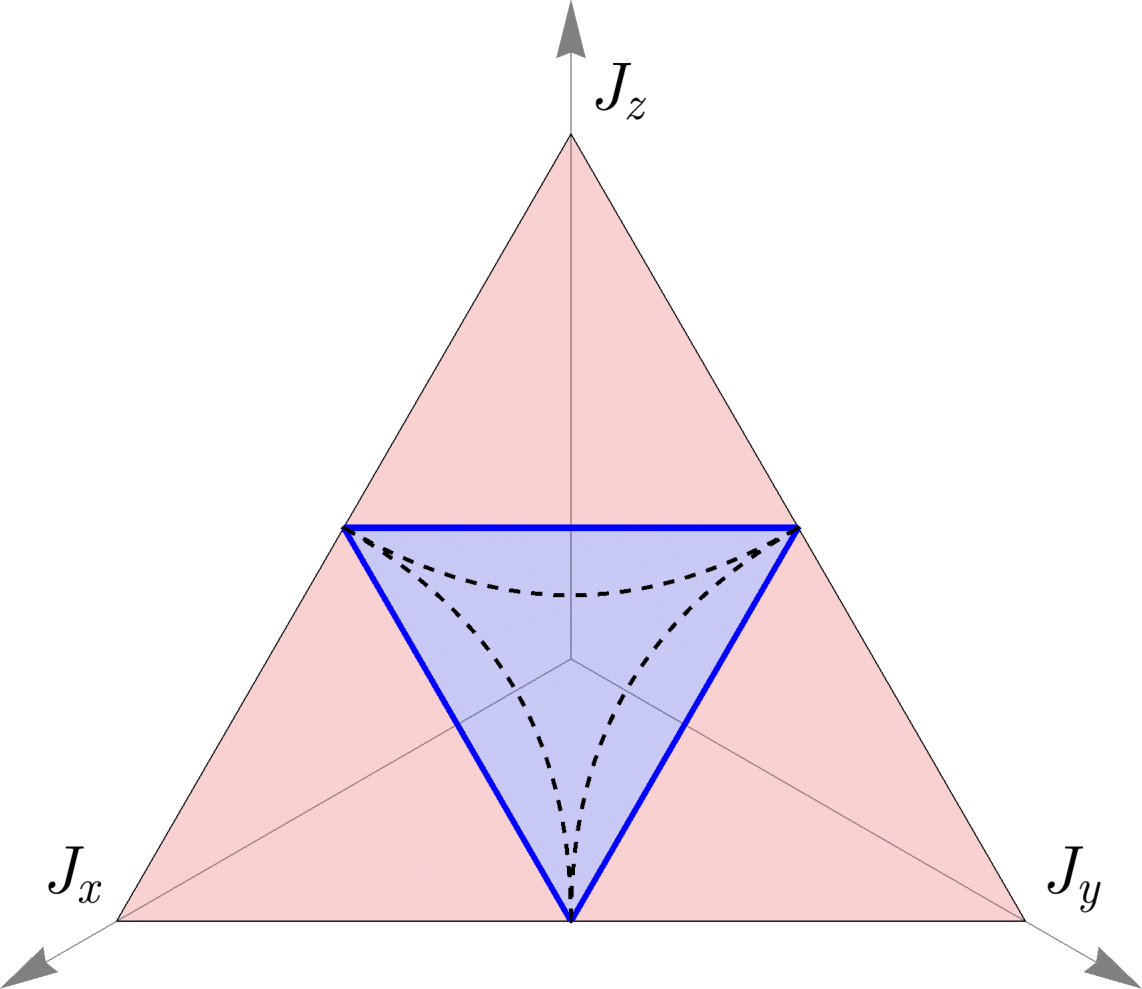} 
  \caption{
	(Left) The unit cell and translation vectors for the (10,3)d lattice, with the $x$-,$y$- and $z$-bonds colored by green, red, blue respectively. 
	(Middle) The Brillouin zone with some of the high-symmetry points indicated for  (10,3)d.
	(Right) The phase diagram for the Kitaev model on (10,3)d.  The parameter region shaded blue corresponds to the gapless regime, while the region shaded pink represents the gapped regime. The dashed black line denotes the parameter values for which the two of the nodal lines gap out, leaving a single nodal line located on the $k_z=0$ plane (see Fig.~\ref{fig:10d_nodes}). 
} 
	\label{fig:10d}
\end{figure*}

The (10,3)d lattice can be described as a primitive orthorhombic lattice with 8 sites per unit cell. Within the unit cell centered at origin, the positions of these sites are given by 
\begin{align}
  & \vr_1 = \frac{1}{4} \left(  a,  b, 1  \right), & 
  & \vr_2 = \frac{1}{4} \left(  0,  a+b, 2  \right), & \nonumber \\ 
  & \vr_3 = \frac{1}{4} \left( -a,  b, 3  \right), & 
  & \vr_4 = \frac{1}{4} \left(  0, -a+b, 4 \right), & \nonumber \\ 
  & \vr_5 = \frac{1}{4} \left(  0, a-b, 3  \right), & 
  & \vr_6 = \frac{1}{4} \left( -a, -b, 2  \right), & \nonumber \\ 
  & \vr_7 = \frac{1}{4} \left(	0,-a-b, 1   \right), & 
  & \vr_8 = \frac{1}{4} \left(  a,-b, 4 \right), & 
\end{align}
where $a = 4 - 2\sqrt{2}$ and $b = 2$ (see Fig.~\ref{fig:10d}). This lattice is also bipartite, and similar to the case of \latname, we choose the A sublattice as all the odd numbered sites, and the B sublattice as all the even numbered sites.  We define the lattice translation vectors as 
\begin{align}
  & \va_1 = \;  \frac{1}{2}\left( b, -b, 0 \right),  
  & \va_2 =  \; \frac{1}{2}\left( b, b, 0 \right), \nonumber \\ 
  & \va_3 =  \left( 0, 0, 1 \right),  & 
\end{align}
and the corresponding reciprocal lattice vectors are given by 
\begin{align}
  & \vq_1 = \; \left( \frac{2\pi}{b}, -\frac{2\pi}{b}, 0 \right),
  & \vq_2 = \; \left( \frac{2\pi}{b}, \frac{2\pi}{b}, 0 \right),  \nonumber \\ 
  & \vq_3 = \; \left( 0, 0, 2\pi \right).  & 
\end{align}
The first Brillouin zone, along with some of the high-symmetry points, is illustrated in the middle panel of Fig.~\ref{fig:10d}. 

The symmetries of the (10,3)d lattice correspond to the space group $Pnna$ (No. \textbf{52}). For the Kitaev physics, the most relevant symmetries are the inversion symmetry about the center of the bonds connecting the squares, and the glide mirror symmetry, which involves a reflection about the $xy$-plane followed by a translation, as shown in Fig.~\ref{Fig:symmetries}(b). 

The (10,3)d lattice has eight elementary loops associated with each unit cell. Four of these are essentially identical to the 10-loops on the \latname~lattice --- spiraling up along one square and down along any of the four neighboring squares --- while the remaining four involve spiraling up along a square and then down along the octagon, as depicted in Fig.~\ref{fig:10dloops}.


\section{Kitaev spin liquid on \latname}    \label{sec:8_3_x}
In this section, we analyze the Kitaev model on the \latname~lattice, based on the Majorana representation discussed in Sec.~\ref{sec:ksl}.

\subsection{Kitaev model and symmetries}
We study the Kitaev model on the tricoordinated \latname~lattice with a bond coloring that is commensurate with the unit cell and preserves the inversion symmetry of the lattice, as shown in Fig.~\ref{fig:8x}. This choice of bond coloring is the one that should be realized in MOFs, in accordance with Ref.~\cite{Jackeli2009}. Thus, explicitly, we consider the spin Hamiltonian
\begin{align}
  \hlt = - & \sum_{\vR} \left\{ J_x \left[ \sigma_1^x(\vR) \sigma_2^x(\vR) + \sigma_3^x(\vR) \sigma_4^x(\vR) \right. \right. \nonumber \\ 
  & \left. \;\; \left. + \sigma_5^x(\vR) \sigma_8^x(\vR+\va_3)  + \sigma_6^x(\vR) \sigma_7^x(\vR) \right] \right. \nonumber \\ 
  & \left. + J_y \left[ \sigma_1^y(\vR) \sigma_4^y(\vR-\va_3) + \sigma_2^y(\vR) \sigma_3^y(\vR) \right. \right. \phantom{\sum} \nonumber \\ 
  & \left. \;\; \left. + \sigma_5^y(\vR) \sigma_6^y(\vR) + \sigma_7^y(\vR) \sigma_8^y(\vR) \right] \right. \nonumber \\ 
  & \left. + J_z \left[ \sigma_1^z(\vR) \sigma_6^z(\vR+\va_2-\va_3) \right. \right. \phantom{\sum} \nonumber \\ 
  & \left. \;\; \left. + \sigma_2^z(\vR) \sigma_7^z(\vR-\va_1+\va_2)  \right. \right. \nonumber \\ 
  & \left. \;\; \left. + \sigma_3^z(\vR) \sigma_8^z(\vR-\va_1+\va_3) + \sigma_4^z(\vR) \sigma_5^z(\vR) \right] \right\}
\end{align}

To analyze the Hamiltonian, we follow the discussion in Sec.~\ref{sec:ksl} to derive an effective hopping model for Majorana fermions with a fixed background $\intg_2$ gauge field in the ground state sector. To fix the fluxes associated with loops, we note that \latname~has six elementary loops per unit cell. Furthermore, we need six elementary loops to enclose a volume, so that the volume constraint demands that for any flux configuration, the number of loops for each bounded volume carrying a $\pi$ flux must be \emph{even}. 
In fact, one can actually determine the ground state flux configuration rigorously using Lieb's theorem~\cite{Lieb1994flux}, since \latname~ possesses mirror planes (100) and (010) passing through the centers of the octagons, which do not pass through any of the lattice sites [see Fig.~\ref{Fig:symmetries}(a)]. Thus, each loop of length 8 is threaded by a $\pi$ flux, while the 10-loops have no flux. 

\begin{figure}
	\centering
	\includegraphics[width=\columnwidth]{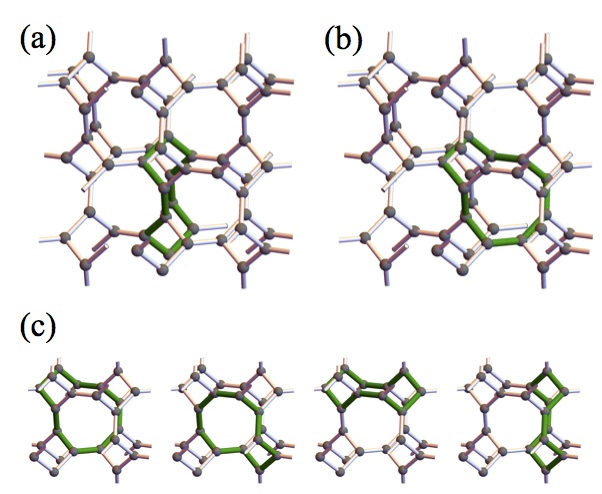}
	\caption{
		There are two distinct types of 10-loops in (10,3)d: loops that spiral up along one square and down along (a) a neighboring square and (b) a neighboring octagon.
		Closed volumes are formed by four loops, as shown in (c).  
	}
	\label{fig:10dloops}
\end{figure}

To define a gauge field configuration that realizes this flux configuration, we need to choose a direction for each bond. We use the bipartite nature of the \latname~lattice to define the positive bond directions from sublattice A to B for all bonds. Then, the desired flux configuration can be realized by setting the bond operators $u_{jk}=+1$ ($-1$) if $j$ is odd (even), except for the 1-6 bond where  $u_{16}=-1$. 
After a permutation of the sites, the resulting  Kitaev-Majorana Hamiltonian can then be expressed in the form of Eq.~\eqref{eq:hlt_chiral} with
\beq
  A = i \begin{pmatrix}
   0 & 				-J_z e^{2\pi ik_{23}} & 	J_x & 			J_y e^{-2\pi ik_3} 	\\
   J_z e^{2\pi i k_{31}} & 	0 & 			J_y & 			J_x 		\\
   J_x e^{2\pi ik_3} & 		J_y & 			0 & 			J_z 	  	\\ 
   J_y & 			J_x & 			J_z e^{2\pi ik_{12}} & 	0 			 
  \end{pmatrix},
\eeq
where $k_{mn}=k_m-k_n.$ 

We start by noting that since the sublattice transformation and inversion do not enlarge the unit cell, both time-reversal and inversion symmetries are implemented trivially. Thus,  based on the general classification approach, one would expect the system to exhibit nodal lines in the bulk. However, the Kitaev-Majorana Hamiltonian is symmetric under many of the lattice symmetries, which leads to a richer set of bulk nodes. In particular, the screw and glide symmetries are implemented projectively, which we now describe in more detail. 

\paragraph*{Screw  symmetry:}
The \latname~lattice possesses four-fold screw symmetries, which map $x$-bonds to $y$-bonds and vice versa. Thus, they are symmetries of the Hamiltonian only when $J_x = J_y$.
They can be written as 
\beq \label{eq:screw}
  \hlt(S_4\vk) = \screwF(\vk) \hlt(\vk) \screwF^\dagger(\vk), \quad \ve(S_4\vk) = -\ve(\vk), 
\eeq 
where 
\[ 
  S_4\left( k_1,k_2,k_3 \right) = \left( k_3-k_2, k_1+\frac{1}{2}, k_3 \right),  
\]
and the unitary matrix $\screwF(\vk)$ depends on the rotation axis. Here and in the following, we suppress the band index when writing the energy relations.
However, the reader should keep in mind that the symmetries generically relate different bands to each other. 
The locus of momenta invariant under the fourfold screw symmetries is a line in the Brillouin zone (depicted in red in the middle panel of Fig.~\ref{fig:8x}). It can be parametrized as 
\beq 
  \lsym_t = t \, \vq_1 + \left( t + \frac{1}{2} \right) \vq_2 + \left( 2t + \frac{1}{2} \right) \vq_3,    \label{eq:scr4_inv_axis}
\eeq 
and is periodic under $t \mapsto t + 1$.

The twofold screw, which is nothing but two successive fourfold screws, maps each bond type onto itself, so that it is a symmetry throughout the phase diagram. It can be written as 
\beq 
  \hlt(S_2\vk) = \screwT(\vk) \hlt(\vk) \screwT^\dagger(\vk), \quad \ve(S_2\vk) = \ve(\vk), 
\eeq 
where 
\[ 
  S_2\left( k_1,k_2,k_3 \right) = \left( k_3-k_1+\frac{1}{2}, k_3-k_2+\frac{1}{2}, k_3 \right),  
\]
where $\screwT(\vk) = \screwF^2(\vk)$. The locus of lattice momenta invariant under the twofold screw symmetry is a set of two lines in the Brillouin zone, \viz, the fourfold axis of \eq{eq:scr4_inv_axis}, as well as the line
\beq
  \wt{\lsym}_t = t \, \vq_1 + t \vq_2 + \left( 2t + \frac{1}{2} \right) \vq_3.    \label{eq:scr2_inv_axis}
\eeq 
The latter is depicted in black in the middle panel of Fig.~\ref{fig:8x}. 

\paragraph*{Glide  symmetry:} Inspecting Fig.~\ref{Fig:symmetries}, it becomes clear that the glide symmetries map each bond type onto itself, so that they are  symmetries of the model throughout the phase diagram. They can be implemented as 
\beq 
  \hlt(G \vk) = \glide(\vk) \hlt(\vk) \glide^\dagger(\vk), \quad \ve(G \vk) = \ve(\vk), 
\eeq 
where 
\[ 
  G \left( k_1,k_2,k_3 \right) = \left( k_1, k_3-k_2+\frac{1}{2}, k_3 \right),  
\]
and the unitary matrix $\glide(\vk)$ depends on the mirror plane. 

We note that the screw symmetric lines of \eq{eq:scr4_inv_axis} and \eq{eq:scr2_inv_axis} are also symmetric under a glide-mirror. On these lines, the interplay of the glide and the screw symmetries leads to an interesting situation, since both $\glide(\vk)$ and $\screwT(\vk)$ commute with the Hamiltonian for $\vk = \lsym_t, \wt{\lsym}_t$, and furthermore, 
\begin{align}\label{eq:non_comm}
\screwT(\lsym_t)\glide(\lsym_t) = & \; - \glide(\lsym_t)\screwT(\lsym_t),  \nonumber \\ 
\screwT(\wt{\lsym}_t)\glide(\wt{\lsym}_t) = & \;   \glide(\wt{\lsym}_t)\screwT(\wt{\lsym}_t).
\end{align} 
Thus, for $\wt{\lsym}$, the three operators, \viz, $\hlt, \screwT$ and $\glide$, can be diagonalized simultaneously, so that each band carries a unique glide and twofold screw quantum number. However, for $\lsym$, since the glide and the screw operators do not commute, they cannot be simultaneously diagonalized. Thus, each eigenstate of the Hamiltonian at these momenta must be doubly degenerate for all values of the parameters.

Let us note a subtlety regarding the screw symmetries: it turns out that only the twofold screw remains a symmetry when time-reversal is broken, but not the fourfold screw. This is a consequence of the close relation between the time-reversal and sublattice symmetries  --- they are actually equivalent symmetries in KSLs. Breaking the sublattice symmetry destroys the fourfold screw symmetry, as the latter maps the sublattice A onto B. The other symmetries are unaffected as they only map sublattice A to A and B to B.

\subsection{Phase diagram}
We next study the various phases of the Kitaev model on \latname~by diagonalizing the Kitaev-Majorana Hamiltonian and studying its spectrum. Since the phases are invariant under a rescaling of all $J_\gamma$'s, we restrict ourself to the plane $J_x + J_y + J_z = 1$. We observe a gapless phase with two Dirac cones on the line $J_x=J_y$. In addition, we observe two gapless phases with nodal lines near the isotropic point. For $J_a \gtrsim J_b + J_c \, \forall \, a \neq b \neq c$, the KSL is gapped, as shown in Fig.~\ref{fig:8x}.

\subsubsection{3D Dirac nodes}
The system exhibits a pair of 3D Dirac nodes for $J_x = J_y$, located on the fourfold screw symmetric line of \eq{eq:scr4_inv_axis} with 
\beq
  t_0 = \pm \frac{1}{2\pi} \cos^{-1} \left( \frac{J_z^2}{2 J^2} \right), \quad J = J_x = J_y. \label{eq:dirac_pos}
\eeq
For $J_z \geq J \sqrt{2}$, the two Dirac nodes collide at $\vk = \frac{1}{2} \left( \vq_2 + \vq_3 \right)$ and gap out.

\begin{figure}
	\centering
	\includegraphics[width=0.95\columnwidth]{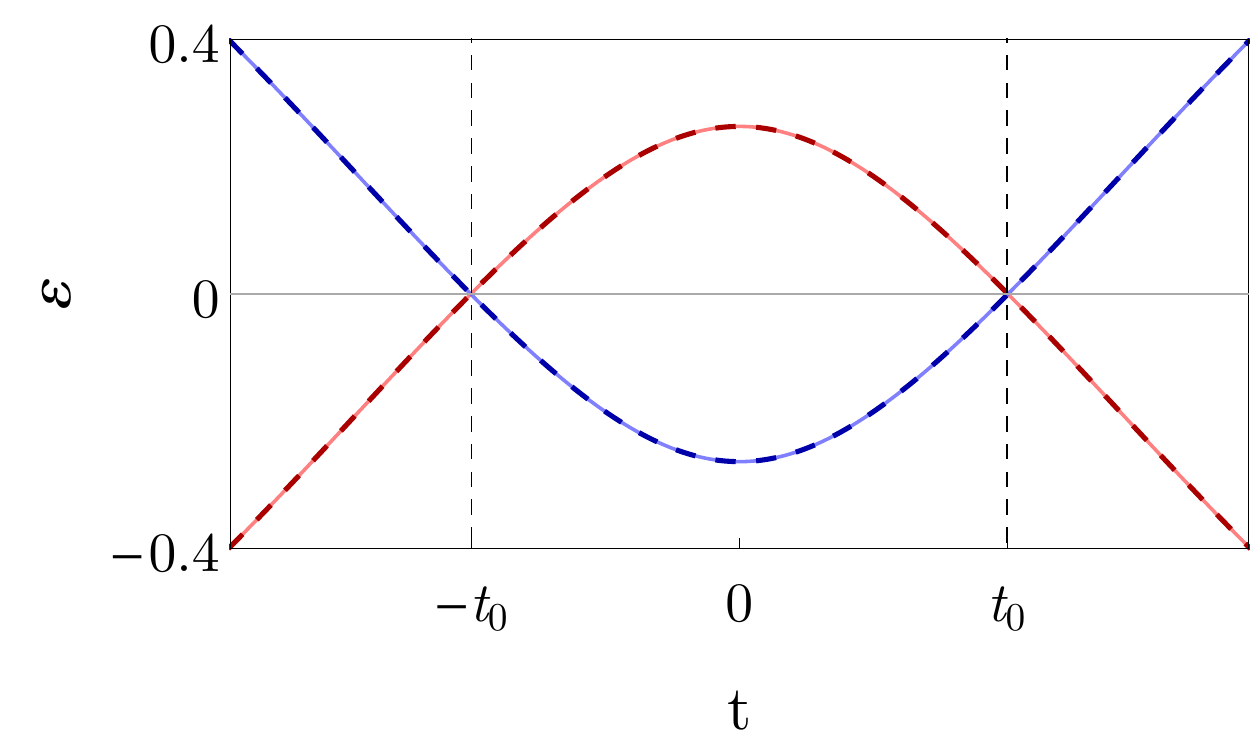} 
	\caption{
		Spectrum of the four bulk bands that are closest to $\varepsilon=0$ and that form the Dirac nodes, plotted along the screw symmetric line of \eq{eq:scr4_inv_axis}. The parameters are given by $J_x=J_y=0.37, \; J_z=0.26$, with the Dirac nodes corresponding to $t_0 = \pm 0.42\pi$ [Eq.~\eqref{eq:dirac_pos}]. The bands are labeled by their screw eigenvalues $\rho_n e^{- 3 \, i\pi t}$ with $\rho_n = e^{ i (2n+1)\pi / 4}$. The bands with eigenvalues $\rho_0, \rho_1$ are denoted by red solid and dashed lines,  and those with eigenvalues $\rho_2, \rho_3$ by blue solid and dashed lines, respectively. 
	} 
	\label{fig:screw_inv}
\end{figure}

Na\"ively, one might think that the 3D Dirac nodes can be trivially gapped out by the addition of a local operator to the Hamiltonian, corresponding to the fermionic ``mass''. However, for the case at hand, the Dirac nodes are protected by the combination of the glide and the  fourfold screw rotation --- the latter is a symmetry of the Hamiltonian only on the $J_x = J_y$ line. 
Explicitly, we note that for any momentum $\gamma_t$ on the symmetric line $\lsym$, 
\beq 
   [\screwF(\lsym_t), \hlt(\lsym_t)] = 0,
\eeq 
and  $\left[ \screwF(\lsym_t) \right]^4 = -e^{- 12 \, i \pi t} \id_8$. 
Thus, the eigenstates $\ket{\varphi_t}$ of $\hlt(\lsym_t)$ satisfy 
\beq
  \screwF(\lsym_t) \ket{\varphi_t} = \rho_n e^{- 3 \, i\pi t} \ket{\varphi_t},    \label{eq:screw_eig}
\eeq
with $\rho_n = \expn{i \frac{2n+1}{4} \pi}; \, n = 0,1,2,3$, and to each bulk band we associate a screw eigenvalue corresponding to $\rho_n$. 
From the discussion below Eq.~\eqref{eq:non_comm}, we know that each of the bulk bands must be (at least) doubly degenerate. 
In Fig.~\ref{fig:screw_inv}, we plot the bulk bands on the screw symmetric line with the corresponding screw eigenvalue identified, and note that the bands are inverted at the Dirac point. Since the $\rho_n$ cannot change continuously, the Dirac nodes are topologically protected  in the presence of  fourfold screw and glide symmetry, and can only be gapped out by colliding with another Dirac node.

\begin{figure*}[htb]
\centering  
  \includegraphics[width=2\columnwidth]{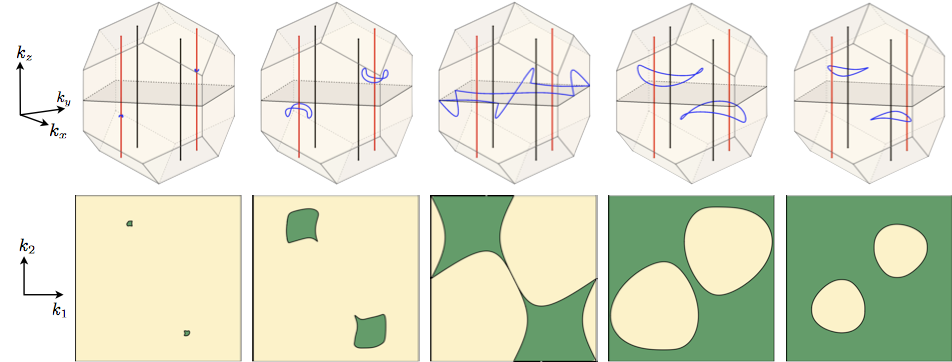} 
  \caption{
	(Top) The bulk nodes for the \latname~lattice in the first Brillouin zone for parameters $J_x = \frac{1}{3} + \Delta J, J_{y,z} = \frac{1}{3} - \frac{1}{2} \Delta J$, with $ \Delta J = 0.01, 0.05, 0.0808, 0.10, 0.12$ (from left to right).
	(Bottom) The chiral invariant computed for the loops along the $k_3$ axis as a function of $(k_1, k_2)$, where the values 0, and $-1$ are represented by yellow and green, respectively. The black solid line depicts the projection of the bulk nodal line along $k_3$. 
}
  \label{fig:8x_nodes}
\end{figure*}

\subsubsection{Nodal lines}
As one moves away from the $J_x=J_y$ line, the Dirac nodes turn into nodal lines formed by the intersection of two bands, while the remaining two bands that formed the Dirac nodes simply gap out. As one moves away from the isotropic point in the phase diagram, the two nodal lines touch at four points and flip inside out. Moving further away from the isotropic point, the nodal lines finally gap out.  This progression of bulk nodes is depicted in Fig.~\ref{fig:8x_nodes}. 

Using the fact that the gapless points are given by $\abs{\det A(\vk)} = 0$, we can obtain closed form expressions for the various phase boundaries. In particular, the nodes always touch in the $k_3=0$ plane at the time-reversal invariant momenta for parameters along the lines given by 
\beq 
 J_z= \sqrt{\abs{J_x^2 - J_y^2}} , 
\eeq
which form rectangular hyperbolae in the plane given by $J_x + J_y + J_z = $ constant. Furthermore, since the twofold screw  is a symmetry of the Hamiltonian throughout the phase diagram, if $\vk = \vk_0$ is a gapless point, then so is $\vk = S_2 \vk_0$. Thus, the point of contraction of the line nodes must satisfy $S_2\vk_0 = \vk_0$, i.e, it must lie on the twofold screw symmetric line given by \eq{eq:scr2_inv_axis}. Using this fact, the phase boundaries between the gapless and the gapped phases can be computed to be
\begin{align}
  2J_x &= \abs{J_z \pm \sqrt{J_z^2 - 4 i J_x J_y}},\nonumber\\
  J_z & = \sqrt{J_x^2+J_y^2}\,,
\end{align}
as depicted in Fig.~\ref{fig:8x}. 
The nodal lines are associated with  the chiral invariant defined in Eq.~\eqref{eq:chiral_inv}, which is plotted in  Fig.~\ref{fig:8x_nodes} as a function of $k_1, k_2$, i.e. the invariant is computed for (noncontractible) loops along $k_3$. 
A nonzero value of the chiral invariant implies that there is symmetry-protected zero-energy mode on the surface. 
Consequently, we find a flat zero-energy surface band in the gapped phase with $J_x \gg J_y,J_z$.
Note that there is no \emph{strong} topological index that can distinguish different gapped KSLs, because there are no topological phases in symmetry class D for noninteracting fermions~\cite{Schnyder2008classification,Kitaev2009topological}.

\subsection{Breaking the time-reversal symmetry}
By breaking the time-reversal symmetry (turning on an external magnetic field), the two nodal lines gap out completely,  in stark contrast to the generic behavior. 
In particular, in Ref.~\cite{Obrien2016classification} it was argued that Weyl nodes have to appear when breaking the time-reversal symmetry in a KSL with a nodal line.  
This argument relies on interpreting the 3D model as a 2D one with one momentum, e.g. $k_3$, regarded as a `tunable parameter'. 
Depending on the value of $k_3$, the corresponding 2D system is either a trivial insulator or harbors 2D Dirac nodes. Breaking the time-reversal symmetry leaves the trivial insulator intact, but gaps out the Dirac nodes, resulting in a Chern insulator. 
At the interface between these two insulators resides a gapless node --- the Weyl node.

So why do Weyl nodes not occur in \latname?
One obvious caveat of the argument in Ref.~\cite{Obrien2016classification} is that it only applies to KSLs with an odd number of nodal lines --- i.e. in cases where there is one `special' nodal line that is invariant under the particle-hole symmetry. 
This implies that in the corresponding 2D model, Dirac cones always occur in odd number of pairs, and  breaking the time-reversal symmetry results in an odd Chern number. 
Applying above argument to \latname, we always encounter an even pair of Dirac nodes, and the corresponding Chern number is also even. In particular, there is nothing that prevents the Chern number to be zero, in which case there will be no Weyl nodes. One should be able to make this argument more rigorous using the particle-hole symmetry.


\section{Kitaev spin liquid on {\normalfont (10,3)\lowercase{d}}}    \label{sec:10_3_d}
In this section, we analyze the Kitaev model on the (10,3)d lattice.

\subsection{Kitaev model and symmetries}

We assign the bond coloring that respects the inversion symmetry as well as the lattice periodicity, as shown in Fig.~\ref{fig:10d}, and again define all bond to be directed from sublattice A to B. Thus, we consider the spin Hamiltonian 
\begin{align}
  \hlt = - & \sum_{\vR} \left\{ J_x \left[ \sigma_1^x(\vR) \sigma_2^x(\vR) + \sigma_3^x(\vR) \sigma_4^x(\vR) \right. \right. \nonumber \\ 
  & \left. \;\; \left. + \sigma_5^x(\vR) \sigma_8^x(\vR)  + \sigma_6^x(\vR) \sigma_7^x(\vR) \right] \right. \nonumber \\ 
  & \left. + J_y \left[ \sigma_1^y(\vR) \sigma_4^y(\vR-\va_3) + \sigma_2^y(\vR) \sigma_3^y(\vR) \right. \right. \phantom{\sum} \nonumber \\ 
  & \left. \;\; \left. + \sigma_5^y(\vR) \sigma_6^y(\vR) + \sigma_7^y(\vR) \sigma_8^y(\vR-\va_3) \right] \right. \nonumber \\ 
  & \left. + J_z \left[ \sigma_1^z(\vR) \sigma_6^z(\vR+\va_2) + \sigma_2^z(\vR) \sigma_7^z(\vR-\va_1+\va_2) \right. \right. \phantom{\sum} \nonumber \\ 
  & \left. \;\; \left. + \sigma_3^z(\vR) \sigma_8^z(\vR-\va_1) + \sigma_4^z(\vR) \sigma_5^z(\vR) \right] \right\}. 
\end{align}

Next, we need to assign the $\intg_2$ fluxes to the bonds, subject to the volume constraints for each closed surface consisting of four loops, as shown in Fig.~\ref{fig:10dloops}(c). Since we do not have mirror planes for (10,3)d, strictly speaking, we cannot resort to Lieb's theorem to deduce the ground states sector. However, if Lieb's theorem were to apply, the ground states would have $0$ flux through each closed loop. 
This is indeed the correct ground state flux configuration~\cite{QMCunpublished}, as can be verified using the quantum Monte Carlo techniques introduced in Ref.~\cite{Nasu2014vaporization}. 

 Since (10,3)d is bipartite, the Kitaev-Majorana Hamiltonian, after a permutation of the sites, can be expressed in the form of \eq{eq:hlt_chiral}, with
\beq 
A = i
  \begin{pmatrix}
   0   & 		J_z e^{2\pi i k_2} & 	J_x &			J_y e^{-2\pi i k_3} 		\\ 
   J_z e^{-2\pi ik_1} & 	0 & 			J_y & 			J_x		 	\\ 
   J_x &		J_y & 			0 & 			J_z	 		\\
   J_y e^{-2\pi i k_3} & 	J_x & 			J_z e^{2\pi i k_{12}} & 	0 
  \end{pmatrix}.
\eeq 

For (10,3)d, as is the case of \latname, both time-reversal and inversion symmetries are implemented trivially, and we expect nodal lines in the bulk. However, the existence of additional lattice symmetries again leads to richer physics, as discussed below. In particular, the glide-mirror symmetry of (10,3)d, implemented trivially, is given by 
\beq 
  \hlt(G \vk) = \glide(\vk) \hlt(\vk) \glide^\dagger(\vk), \quad \ve(G \vk) = \ve(\vk), 
\eeq 
where 
\[ 
  G \left( k_1, k_2, k_3 \right) = \left( k_1, k_2, -k_3 \right),  
\]
and the unitary matrix $\glide(\vk)$ depends on the mirror plane. Thus, the planes $k_3=0,1/2$ are invariant under this glide-mirror.

\begin{figure*}[htb]
\centering  
	\includegraphics[width=2\columnwidth]{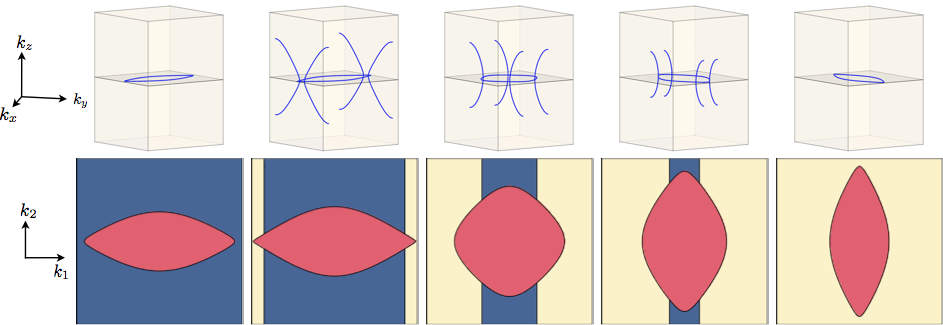} 
   \caption{
 	(Top) The bulk nodes for the (10,3)d~lattice in the first Brillouin zone for parameters $J_z = \frac{1}{3}, J_{x,,y} = \frac{1}{3} \pm \Delta J$, with $ \Delta J = -0.1, -0.08, 0, 0.05, 0.1$ (from left to right).     
 	(Bottom) The chiral invariant computed for the loops along the $k_3$ axis as a function of $(k_1, k_2)$, where the values 0, 1, 2 are represented by yellow, red and blue, respectively. The black solid line depicts the projection of the bulk nodal line along $k_3$. 
 }
  \label{fig:10d_nodes}
\end{figure*}

\subsection{Phase diagram}
The system exhibits a gapless Majorana phase with nodal lines for  $J_a + J_b \geq J_c, \; a \neq b \neq c$, and a gapped insulating phase otherwise, as plotted in Fig.~\ref{fig:10d}. Near the isotropic point, we observe three touching nodal lines, which lie on orthogonal planes. As we move towards the phase boundary, two of the three nodes are gapped out at the dotted line in the phase diagram, either by contracting to a point or colliding with each other. We are left with a single nodal line, which finally contracts to a point and gaps out at the phase boundary between the gapped and gapless phases	.

Explicitly, there is a nodal line in the $k_3=0$ plane, given by 
\beq
  J_x^2 \cos (2\pi k_1) + J_y^2 \cos (2\pi k_2) = \frac{\left( J_x^2 - J_y^2 \right)^2 + J_z^4}{2J_z^2}.
\eeq 
It can be checked that this equation has a solution iff 
\begin{align}
\abs{J_x-J_y}\leq J_z \leq J_x+J_y.
\end{align}
 The other set of nodal lines lie on the $k_2$--$k_3$ planes defined by
\beq 
  \cos \left( 2\pi k_1 \right)  = \frac{J_x^4 - J_y^4 + J_z^4}{2 J_x^2 J_z^2},
\eeq 
which has a solution iff
\beq 
 -2 J_x^2 J_z^2 \leq J_x^4 - J_y^4 + J_z^4 \leq 2 J_x^2 J_z^2.
\eeq 
The nodal lines are described by the equation
\beq 
  J_y^2 \cos (2\pi k_3) - J_z^2 \cos (2\pi k_2) = J_x^2, 
\eeq 
which has a solution iff $J_x^2 \leq \abs{J_y^2 - J_z^2}$. From these conditions, we deduce that the phase boundaries are given by $J_a^2 + J_b^2 = J_c^2, \quad a \neq b \neq c$, which form a set of rectangular hyperbolae in the phase diagram.

The physics of these nodal lines can be captured by a simple 2-band model, defined as
\beq 
\hlt(\vk) = \eta_x(\vk) \sigma_x + \eta_y(\vk) \sigma_y,  
\eeq 
whose spectrum is given by $\ve = \pm \sqrt{\eta_x^2 + \eta_y^2}$, so that the bulk spectrum exhibits nodal lines given by 
\[
\eta_x(\vk) = \eta_y(\vk) = 0.
\]
For (10,3)d in the gapless regime, using the explicit expressions for the nodal lines, we set
\begin{align}
\eta_x(\vk) = & \; \varphi_1(2\pi k_1) \sin (2\pi k_3), \nonumber \\ 
\eta_x(\vk) = & \; \sum_{i=1}^3 \varphi_i(2 \pi k_i) + \gamma \varphi_1(2 \pi k_1) \varphi_3(k_3), 
,
\end{align}
with $\varphi_i(q) = \alpha_i \cos q - \beta_i$. The parameters are given by 
\begin{align*}
\bal = & \; \left( J_x^2, \;\; -J_z^2, \;\; J_y^2 \right),  \\ 
\bbe = & \;  \left( \frac{J_x^4 - J_y^4 + J_z^4}{2J_z^2}, \;\;J_x^2 + J_y^2, \;\; -J_y^2  \right),  \\ 
\gamma =  & \; - \frac{J_y^2 + J_z^2}{2 J_y^4},
\end{align*}
with $\bal = \left( \alpha_1, \alpha_2, \alpha_3 \right)$, etc. This model  captures the interesting behavior of the nodal lines, and owing to its simplicity, it can be further used to compute interesting quantities associated with the physics of the (10,3)d Kitaev model.\footnote{Note that the value of the chiral invariant is shifted by 1 compared to the full model.}

\begin{figure}
  \centering
  \includegraphics[width=0.8\columnwidth]{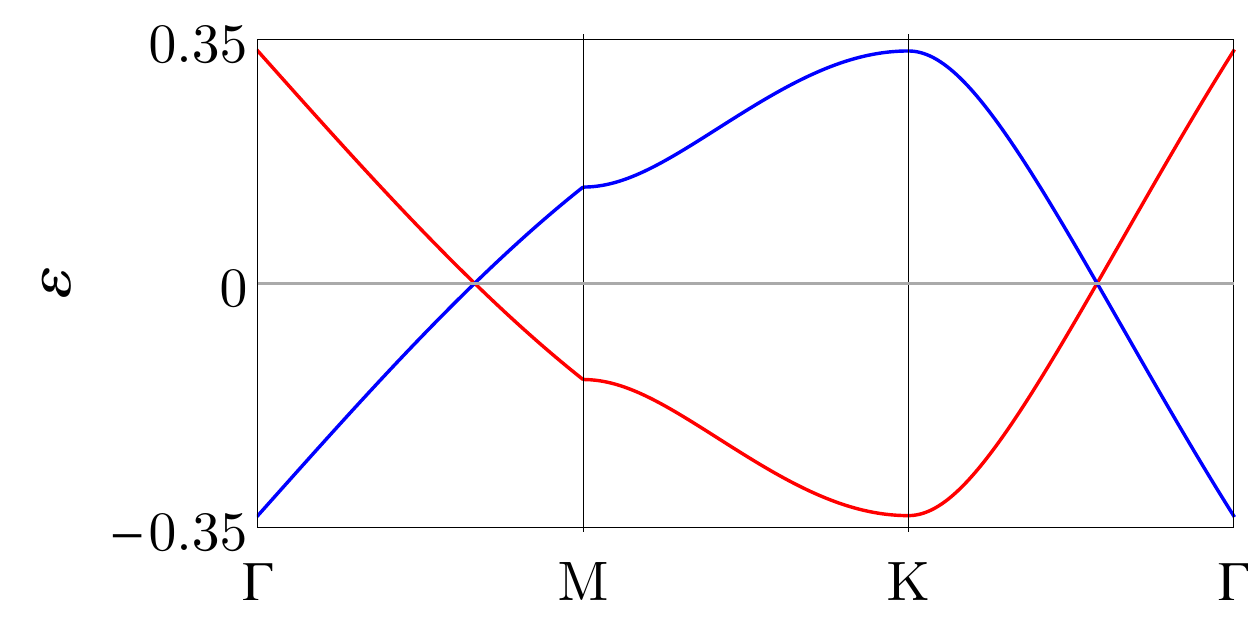} 
  \caption{
    Spectrum of the two  bulk bands that are closest to $\varepsilon=0$ and that form the nodal lines, plotted  along the high symmetry line in the glide-mirror symmetric plane of the Brillouin zone and for isotropic couplings $J_x=J_y=J_z=1/3$. The bands are labeled by their glide eigenvalues $\rho_n e^{i \pi (k_1 - k_2)}$ with $\rho_n = e^{ i (2n+1)\pi / 2}$. The bands with eigenvalues $\rho_0, \rho_1$ are denoted by red and blue solid lines, respectively. 
  } 
  \label{fig:glide_inv}
\end{figure}

\subsection{Breaking the time-reversal symmetry}
By breaking the time-reversal symmetry, the two vertical nodal lines (see Fig.~\ref{fig:10d_nodes}) gap out completely, but the nodal line in the $k_3 = 0$ plane remains intact. 
To explain this, we recall that the $k_3=0$ plane is invariant under the glide-mirror symmetry, so that for $\vk_{\parallel} =  k_1 \vq_1+ k_2 \vq_2$, the Hamiltonian commutes with $\glide(\vk_{\parallel})$, and the bands can be labeled by their glide eigenvalues. Furthermore, $\left[ \glide(\vk_\parallel) \right]^2 = -e^{2\pi i (k_1 - k_2)}$, so that its eigenvalues are  
\beq 
  \lambda_n = \rho_n e^{i\pi (k_1 - k_2)}, \quad \rho_n = e^{ i \frac{2n+1}{2} \pi}, \; n = 0,1.
\eeq 
Plotting these eigenvalues associated with the conduction and valence bands in the Brillouin zone in Fig.~\ref{fig:glide_inv}, we clearly see that they are inverted across the nodal line. Thus, the nodal line in the $k_3 = 0$ plane is topologically protected, even in the absence of the time-reversal symmetry. 
Upon breaking also the glide mirror symmetry, the nodal line gaps out, leaving two Weyl nodes behind. These are then stable  against any other perturbation that is  local in momentum space.


\section{Conclusions}    \label{sec:conc}
We have studied the Kitaev model on two 3D lattices, where the existence of lattice symmetries lead to a richer physics than what would be expected from a classification based only on the time-reversal and inversion symmetries. 
In particular, we found that non-symmorphic symmetries can stabilize 3D Dirac cones as well as nodal chains in Kitaev spin liquids. 
Our results suggest that there is still a host of interesting  spin liquid phases to be discovered that rely on the interplay of fractionalized excitations with lattice symmetries. 
We also give a concrete proposal for realizing these quantum spin liquids in metal-organic frameworks.

The tunability and bottom-up construction of MOFs allows for the realization of a wide variety of lattices. 
The four-fold spiral --- a common feature among the lattices in Fig.~\ref{lattice_comparison} --- occurs naturally in MOFs~\cite{Gruselle2006enantioselective}. 
This makes studying Kitaev physics on these lattices  interesting because they can potentially be realized experimentally. 
Consider, for instance, the \latnameb~lattice in Fig.~\ref{lattice_comparison}(d).  
Like the \latname~lattice, it is inversion symmetric, has equal-length bonds with 120$\degree$ bond angles, and can be embedded in a network of edge-sharing octahedra. The Kitaev model on this lattice exhibits a Weyl spin liquid~\cite{Hermanns2015weyl},  and shows qualitatively the same behavior as the Kitaev model on the (8,3)b  (hyperhexagon) lattice~\cite{Obrien2016classification}. Although MOFs with the (8,3)b net topology  have been realized~\cite{Tranchemontagne2008room,Rosi2005rod}, the synthesis of an (8,3)b metal-oxalate framework has, to the best of our knowledge, not yet been achieved. Thus, the \latnameb~lattice provides an alternative route for realizing this fascinating KSL phase.

Admittedly, in real materials, Kitaev interactions may be dominant, but other (symmetry-allowed) interactions will always be present. 
Due to the finite gap of the $\intg_2$ flux loop excitations, the Kitaev spin liquid phase is stable against small perturbations away from the exactly solvable Kitaev model. Unfortunately, in three dimensions, it is hard to accurately determine the extent of the spin liquid phase~\cite{Kimchi2014three}, but numerical simulations in two dimensions show that the Kitaev coupling must be larger than the Heisenberg coupling by a factor of 8 in order to stabilize the  quantum spin liquid ground state~\cite{Chaloupka2010,Jiang2011possible}, but it may in fact be less for generic interactions~\cite{Rousochatzakis2015phase}.

For perfect crystal structures, non-Kitaev-like interactions will be suppressed exponentially because of the small direct overlap of the magnetic ions. 
However, for  metal-oxalate MOF, there generically exists a trigonal compression in the $M$O$_6$ octahedra. A recent study~\cite{WinterValenti2016} discovered that this may in fact be beneficial to the realization of KSLs. They considered materials of the form $A_2$IrO$_3$ and showed that Kitaev interactions are most dominant when the O--Ir--O angle is about 80$\degree$ (instead of 90$\degree$ as depicted in Fig.~\ref{super}).  
The observed bond angles in Ru-based MOFs~\cite{Dikhtiarenko2016} range from 81\degree~to 83\degree, which indicates that this trigonal distortion in MOFs should support our proposal, assuming that the effect of trigonal distortion is similar for RuO$_6$ and IrO$_6$~\cite{Yamada2016MOF}. 
Moreover, this trigonal distortion does not break any space group symmetries, including glide and screw symmetries, so our symmetry analysis should remain valid. 

Smoking-gun experimental signatures for quantum spin liquids have so far proven elusive. 
Nevertheless, there are several experimental probes that can detect remnants of the spin fractionalization in such systems. The simplest is the specific heat, which clearly shows the dichotomy present in these phases --- a \emph{metallic} specific heat behavior occurring in a Mott \emph{insulator}. For nodal lines, the specific heat behaves as $C(T) \propto T^2$ at low temperatures~\cite{Lee2014heisenberg}. In contrast to other nodal line Kitaev spin liquids, this behavior persists even for small magnetic fields in (10,3)d. 
The specific heat for the isotropic Kitaev model on \latname~ behaves as $C(T) \propto  T^3$, and tuning the model away from the isotropic point changes the behavior to that of a nodal line spin liquid. 
The exact solvability of these models also make them amenable to the computation of other experimental signatures, such as the spin-structure factor~\cite{Knolle2014dynamics,Smith2015neutron} accessible in neutron scattering~\cite{Banerjee2016neutron}, as well as Raman scattering~\cite{Perreault2015theory,Sandilands2015scattering,Nasu2016fermionic} and resonant inelastic scattering (RIXS)~\cite{RIXS3D} signatures.
The combination of all these signatures should be able to give fairly conclusive evidence of whether the experimental system is in (or at least close to) a KSL phase. 

Our results emphasize that the interplay between fractionalized excitations and symmetries may lead to new types of quantum spin liquids.  
In principle, one could construct counterparts of the whole zoo of symmetry-protected noninteracting fermionic phases in Kitaev spin liquids. 
Furthermore, MOFs may prove to be the ideal platform to design and investigate this type of physics. 

\acknowledgments
We thank V. Chua, H.~Fujita, G.~Jackeli, I.~Kimchi, Y.~Motome, M.~Oshikawa, R.~Takahashi,  M.~Takigawa, and S.~Trebst for insightful discussions, and in particular T.~Eschmann and S.~Trebst for providing numerical confirmation for the ground state flux sector of (10,3)d.
A part of this work was completed at KITP, UCSB, supported by the US National Science Foundation under Grant No. NSF PHY11-25915.
This work was supported by JSPS KAKENHI Grant Number JP17J05736, and by JSPS Strategic International Networks Program No. R2604 ``TopoNet''.
M.G.Y. is supported by the Materials Education program for the future leaders in Research, Industry, and Technology (MERIT), and by JSPS.
V.D. is partially funded by the SFB 1238 grant of the Deutsche Forschungsgemeinschaft. 
M.H. is partially funded by the Deutsche Forschungsgemeinschaft under the Emmy Noether  grant no HE 7267/1-1 and the SFB 1238.

\bibliography{8_3_X_paper}

\begin{thebibliography}{104}%
\makeatletter
\providecommand \@ifxundefined [1]{%
 \@ifx{#1\undefined}
}%
\providecommand \@ifnum [1]{%
 \ifnum #1\expandafter \@firstoftwo
 \else \expandafter \@secondoftwo
 \fi
}%
\providecommand \@ifx [1]{%
 \ifx #1\expandafter \@firstoftwo
 \else \expandafter \@secondoftwo
 \fi
}%
\providecommand \natexlab [1]{#1}%
\providecommand \enquote  [1]{``#1''}%
\providecommand \bibnamefont  [1]{#1}%
\providecommand \bibfnamefont [1]{#1}%
\providecommand \citenamefont [1]{#1}%
\providecommand \href@noop [0]{\@secondoftwo}%
\providecommand \href [0]{\begingroup \@sanitize@url \@href}%
\providecommand \@href[1]{\@@startlink{#1}\@@href}%
\providecommand \@@href[1]{\endgroup#1\@@endlink}%
\providecommand \@sanitize@url [0]{\catcode `\\12\catcode `\$12\catcode
  `\&12\catcode `\#12\catcode `\^12\catcode `\_12\catcode `\%12\relax}%
\providecommand \@@startlink[1]{}%
\providecommand \@@endlink[0]{}%
\providecommand \url  [0]{\begingroup\@sanitize@url \@url }%
\providecommand \@url [1]{\endgroup\@href {#1}{\urlprefix }}%
\providecommand \urlprefix  [0]{URL }%
\providecommand \Eprint [0]{\href }%
\providecommand \doibase [0]{http://dx.doi.org/}%
\providecommand \selectlanguage [0]{\@gobble}%
\providecommand \bibinfo  [0]{\@secondoftwo}%
\providecommand \bibfield  [0]{\@secondoftwo}%
\providecommand \translation [1]{[#1]}%
\providecommand \BibitemOpen [0]{}%
\providecommand \bibitemStop [0]{}%
\providecommand \bibitemNoStop [0]{.\EOS\space}%
\providecommand \EOS [0]{\spacefactor3000\relax}%
\providecommand \BibitemShut  [1]{\csname bibitem#1\endcsname}%
\let\auto@bib@innerbib\@empty
\bibitem [{\citenamefont {Balents}(2010)}]{Balents2010spin}%
  \BibitemOpen
  \bibfield  {author} {\bibinfo {author} {\bibfnamefont {Leon}\ \bibnamefont
  {Balents}},\ }\bibfield  {title} {\enquote {\bibinfo {title} {{Spin liquids
  in frustrated magnets}},}\ }\href@noop {} {\bibfield  {journal} {\bibinfo
  {journal} {Nature}\ }\textbf {\bibinfo {volume} {464}},\ \bibinfo {pages}
  {199--208} (\bibinfo {year} {2010})}\BibitemShut {NoStop}%
\bibitem [{\citenamefont {Witczak-Krempa}\ \emph {et~al.}(2014)\citenamefont
  {Witczak-Krempa}, \citenamefont {Chen}, \citenamefont {Kim},\ and\
  \citenamefont {Balents}}]{WitczakKrempa2014correlatedAnnualReview}%
  \BibitemOpen
  \bibfield  {author} {\bibinfo {author} {\bibfnamefont {William}\ \bibnamefont
  {Witczak-Krempa}}, \bibinfo {author} {\bibfnamefont {Gang}\ \bibnamefont
  {Chen}}, \bibinfo {author} {\bibfnamefont {Yong~Baek}\ \bibnamefont {Kim}}, \
  and\ \bibinfo {author} {\bibfnamefont {Leon}\ \bibnamefont {Balents}},\
  }\bibfield  {title} {\enquote {\bibinfo {title} {{Correlated Quantum
  Phenomena in the Strong Spin-Orbit Regime}},}\ }\href {\doibase
  10.1146/annurev-conmatphys-020911-125138} {\bibfield  {journal} {\bibinfo
  {journal} {Annual Review of Condensed Matter Physics}\ }\textbf {\bibinfo
  {volume} {5}},\ \bibinfo {pages} {57--82} (\bibinfo {year}
  {2014})}\BibitemShut {NoStop}%
\bibitem [{\citenamefont {{Savary}}\ and\ \citenamefont
  {{Balents}}(2017)}]{Savary2017quantum}%
  \BibitemOpen
  \bibfield  {author} {\bibinfo {author} {\bibfnamefont {L.}~\bibnamefont
  {{Savary}}}\ and\ \bibinfo {author} {\bibfnamefont {L.}~\bibnamefont
  {{Balents}}},\ }\bibfield  {title} {\enquote {\bibinfo {title} {{Quantum spin
  liquids: a review}},}\ }\href {\doibase 10.1088/0034-4885/80/1/016502}
  {\bibfield  {journal} {\bibinfo  {journal} {Reports on Progress in Physics}\
  }\textbf {\bibinfo {volume} {80}},\ \bibinfo {eid} {016502} (\bibinfo {year}
  {2017})}\BibitemShut {NoStop}%
\bibitem [{\citenamefont {Misguich}(2010)}]{Misguich2010qsl}%
  \BibitemOpen
  \bibfield  {author} {\bibinfo {author} {\bibfnamefont {G.}~\bibnamefont
  {Misguich}},\ }\enquote {\bibinfo {title} {Quantum spin liquids and
  fractionalization},}\ in\ \href@noop {} {\emph {\bibinfo {booktitle}
  {Introduction to frustrated magnetism}}}\ (\bibinfo  {publisher} {Springer},\
  \bibinfo {address} {Heidelberg},\ \bibinfo {year} {2010})\BibitemShut
  {NoStop}%
\bibitem [{\citenamefont {Kitaev}\ and\ \citenamefont
  {Preskill}(2006)}]{KitaevPreskill}%
  \BibitemOpen
  \bibfield  {author} {\bibinfo {author} {\bibfnamefont {Alexei}\ \bibnamefont
  {Kitaev}}\ and\ \bibinfo {author} {\bibfnamefont {John}\ \bibnamefont
  {Preskill}},\ }\bibfield  {title} {\enquote {\bibinfo {title} {Topological
  entanglement entropy},}\ }\href {\doibase 10.1103/PhysRevLett.96.110404}
  {\bibfield  {journal} {\bibinfo  {journal} {Phys. Rev. Lett.}\ }\textbf
  {\bibinfo {volume} {96}},\ \bibinfo {pages} {110404} (\bibinfo {year}
  {2006})}\BibitemShut {NoStop}%
\bibitem [{\citenamefont {Levin}\ and\ \citenamefont {Wen}(2006)}]{LevinWen}%
  \BibitemOpen
  \bibfield  {author} {\bibinfo {author} {\bibfnamefont {Michael}\ \bibnamefont
  {Levin}}\ and\ \bibinfo {author} {\bibfnamefont {Xiao-Gang}\ \bibnamefont
  {Wen}},\ }\bibfield  {title} {\enquote {\bibinfo {title} {Detecting
  topological order in a ground state wave function},}\ }\href {\doibase
  10.1103/PhysRevLett.96.110405} {\bibfield  {journal} {\bibinfo  {journal}
  {Phys. Rev. Lett.}\ }\textbf {\bibinfo {volume} {96}},\ \bibinfo {pages}
  {110405} (\bibinfo {year} {2006})}\BibitemShut {NoStop}%
\bibitem [{\citenamefont {Jiang}\ \emph {et~al.}(2012)\citenamefont {Jiang},
  \citenamefont {Wang},\ and\ \citenamefont {Balents}}]{Jiang2012identifying}%
  \BibitemOpen
  \bibfield  {author} {\bibinfo {author} {\bibfnamefont {Hong-Chen}\
  \bibnamefont {Jiang}}, \bibinfo {author} {\bibfnamefont {Zhenghan}\
  \bibnamefont {Wang}}, \ and\ \bibinfo {author} {\bibfnamefont {Leon}\
  \bibnamefont {Balents}},\ }\bibfield  {title} {\enquote {\bibinfo {title}
  {Identifying topological order by entanglement entropy},}\ }\href {\doibase
  doi:10.1038/nphys2465} {\bibfield  {journal} {\bibinfo  {journal} {Nature
  Physics}\ }\textbf {\bibinfo {volume} {8}},\ \bibinfo {pages} {902} (\bibinfo
  {year} {2012})}\BibitemShut {NoStop}%
\bibitem [{\citenamefont {Hermanns}\ \emph {et~al.}()\citenamefont {Hermanns},
  \citenamefont {Kimchi},\ and\ \citenamefont {Knolle}}]{Hermanns2017physics}%
  \BibitemOpen
  \bibfield  {author} {\bibinfo {author} {\bibfnamefont {Maria}\ \bibnamefont
  {Hermanns}}, \bibinfo {author} {\bibfnamefont {Itamar}\ \bibnamefont
  {Kimchi}}, \ and\ \bibinfo {author} {\bibfnamefont {Johannes}\ \bibnamefont
  {Knolle}},\ }\href@noop {} {\enquote {\bibinfo {title} {{Physics of the
  Kitaev model: fractionalization, dynamical correlations, and material
  connections}},}\ }\bibinfo {howpublished} {{arXiv:1705.01740, to be published
  in Ann. Rev. Cond. Mat. 9 (2018) }}\BibitemShut {NoStop}%
\bibitem [{\citenamefont {Trebst}(2017)}]{TrebstReview}%
  \BibitemOpen
  \bibfield  {author} {\bibinfo {author} {\bibfnamefont {Simon}\ \bibnamefont
  {Trebst}},\ }\bibfield  {title} {\enquote {\bibinfo {title} {Kitaev
  materials},}\ }\href@noop {} {\bibfield  {journal} {\bibinfo  {journal}
  {arXiv:1701.07056}\ } (\bibinfo {year} {2017})}\BibitemShut {NoStop}%
\bibitem [{\citenamefont {Winter}\ \emph {et~al.}(2017)\citenamefont {Winter},
  \citenamefont {Tsirlin}, \citenamefont {Daghofer}, \citenamefont {van~den
  Brink}, \citenamefont {Singh}, \citenamefont {Gegenwart},\ and\ \citenamefont
  {Valenti}}]{RoserReview}%
  \BibitemOpen
  \bibfield  {author} {\bibinfo {author} {\bibfnamefont {Stephen~M.}\
  \bibnamefont {Winter}}, \bibinfo {author} {\bibfnamefont {Alexander~A.}\
  \bibnamefont {Tsirlin}}, \bibinfo {author} {\bibfnamefont {Maria}\
  \bibnamefont {Daghofer}}, \bibinfo {author} {\bibfnamefont {Jeroen}\
  \bibnamefont {van~den Brink}}, \bibinfo {author} {\bibfnamefont {Yogesh}\
  \bibnamefont {Singh}}, \bibinfo {author} {\bibfnamefont {Philipp}\
  \bibnamefont {Gegenwart}}, \ and\ \bibinfo {author} {\bibfnamefont {Roser}\
  \bibnamefont {Valenti}},\ }\bibfield  {title} {\enquote {\bibinfo {title}
  {Models and materials for generalized kitaev magnetism},}\ }\href@noop {}
  {\bibfield  {journal} {\bibinfo  {journal} {arXiv:1706.06113}\ } (\bibinfo
  {year} {2017})}\BibitemShut {NoStop}%
\bibitem [{\citenamefont {Kitaev}(2006)}]{Kitaev2006anyons}%
  \BibitemOpen
  \bibfield  {author} {\bibinfo {author} {\bibfnamefont {Alexei}\ \bibnamefont
  {Kitaev}},\ }\bibfield  {title} {\enquote {\bibinfo {title} {{Anyons in an
  exactly solved model and beyond}},}\ }\href {\doibase
  10.1016/j.aop.2005.10.005} {\bibfield  {journal} {\bibinfo  {journal} {Ann.
  Phys.}\ }\textbf {\bibinfo {volume} {321}},\ \bibinfo {pages} {2--111}
  (\bibinfo {year} {2006})},\ \bibinfo {note} {january Special
  Issue}\BibitemShut {NoStop}%
\bibitem [{\citenamefont {Yao}\ and\ \citenamefont
  {Kivelson}(2007)}]{Yao2007exact}%
  \BibitemOpen
  \bibfield  {author} {\bibinfo {author} {\bibfnamefont {Hong}\ \bibnamefont
  {Yao}}\ and\ \bibinfo {author} {\bibfnamefont {Steven~A.}\ \bibnamefont
  {Kivelson}},\ }\bibfield  {title} {\enquote {\bibinfo {title} {{Exact Chiral
  Spin Liquid with Non-Abelian Anyons}},}\ }\href {\doibase
  10.1103/PhysRevLett.99.247203} {\bibfield  {journal} {\bibinfo  {journal}
  {Phys. Rev. Lett.}\ }\textbf {\bibinfo {volume} {99}},\ \bibinfo {pages}
  {247203} (\bibinfo {year} {2007})}\BibitemShut {NoStop}%
\bibitem [{\citenamefont {Yang}\ \emph {et~al.}(2007)\citenamefont {Yang},
  \citenamefont {Zhou},\ and\ \citenamefont {Sun}}]{Yang2007mosaic}%
  \BibitemOpen
  \bibfield  {author} {\bibinfo {author} {\bibfnamefont {S.}~\bibnamefont
  {Yang}}, \bibinfo {author} {\bibfnamefont {D.~L.}\ \bibnamefont {Zhou}}, \
  and\ \bibinfo {author} {\bibfnamefont {C.~P.}\ \bibnamefont {Sun}},\
  }\bibfield  {title} {\enquote {\bibinfo {title} {{Mosaic spin models with
  topological order}},}\ }\href {\doibase 10.1103/PhysRevB.76.180404}
  {\bibfield  {journal} {\bibinfo  {journal} {Phys. Rev. B}\ }\textbf {\bibinfo
  {volume} {76}},\ \bibinfo {pages} {180404} (\bibinfo {year}
  {2007})}\BibitemShut {NoStop}%
\bibitem [{\citenamefont {Kamfor}\ \emph {et~al.}(2010)\citenamefont {Kamfor},
  \citenamefont {Dusuel}, \citenamefont {Vidal},\ and\ \citenamefont
  {Schmidt}}]{Kamfor2010kitaev}%
  \BibitemOpen
  \bibfield  {author} {\bibinfo {author} {\bibfnamefont {Michael}\ \bibnamefont
  {Kamfor}}, \bibinfo {author} {\bibfnamefont {S{\'e}bastien}\ \bibnamefont
  {Dusuel}}, \bibinfo {author} {\bibfnamefont {Julien}\ \bibnamefont {Vidal}},
  \ and\ \bibinfo {author} {\bibfnamefont {Kai~Phillip}\ \bibnamefont
  {Schmidt}},\ }\bibfield  {title} {\enquote {\bibinfo {title} {Kitaev model
  and dimer coverings on the honeycomb lattice},}\ }\href@noop {} {\bibfield
  {journal} {\bibinfo  {journal} {Journal of Statistical Mechanics: Theory and
  Experiment}\ }\textbf {\bibinfo {volume} {2010}},\ \bibinfo {pages} {P08010}
  (\bibinfo {year} {2010})}\BibitemShut {NoStop}%
\bibitem [{\citenamefont {Rachel}\ \emph {et~al.}(2016)\citenamefont {Rachel},
  \citenamefont {Fritz},\ and\ \citenamefont {Vojta}}]{Rachel2016landau}%
  \BibitemOpen
  \bibfield  {author} {\bibinfo {author} {\bibfnamefont {Stephan}\ \bibnamefont
  {Rachel}}, \bibinfo {author} {\bibfnamefont {Lars}\ \bibnamefont {Fritz}}, \
  and\ \bibinfo {author} {\bibfnamefont {Matthias}\ \bibnamefont {Vojta}},\
  }\bibfield  {title} {\enquote {\bibinfo {title} {{Landau Levels of Majorana
  Fermions in a Spin Liquid}},}\ }\href {\doibase
  10.1103/PhysRevLett.116.167201} {\bibfield  {journal} {\bibinfo  {journal}
  {Phys. Rev. Lett.}\ }\textbf {\bibinfo {volume} {116}},\ \bibinfo {pages}
  {167201} (\bibinfo {year} {2016})}\BibitemShut {NoStop}%
\bibitem [{\citenamefont {Si}\ and\ \citenamefont {Yu}(2008)}]{Si2008anyonic}%
  \BibitemOpen
  \bibfield  {author} {\bibinfo {author} {\bibfnamefont {Tieyan}\ \bibnamefont
  {Si}}\ and\ \bibinfo {author} {\bibfnamefont {Yue}\ \bibnamefont {Yu}},\
  }\bibfield  {title} {\enquote {\bibinfo {title} {{Anyonic loops in
  three-dimensional spin liquid and chiral spin liquid}},}\ }\href {\doibase
  10.1016/j.nuclphysb.2008.06.009} {\bibfield  {journal} {\bibinfo  {journal}
  {Nuclear Physics B}\ }\textbf {\bibinfo {volume} {803}},\ \bibinfo {pages}
  {428--449} (\bibinfo {year} {2008})}\BibitemShut {NoStop}%
\bibitem [{\citenamefont {Mandal}\ and\ \citenamefont
  {Surendran}(2009)}]{Mandal2009exactly}%
  \BibitemOpen
  \bibfield  {author} {\bibinfo {author} {\bibfnamefont {Saptarshi}\
  \bibnamefont {Mandal}}\ and\ \bibinfo {author} {\bibfnamefont {Naveen}\
  \bibnamefont {Surendran}},\ }\bibfield  {title} {\enquote {\bibinfo {title}
  {{Exactly solvable Kitaev model in three dimensions}},}\ }\href {\doibase
  10.1103/PhysRevB.79.024426} {\bibfield  {journal} {\bibinfo  {journal} {Phys.
  Rev. B}\ }\textbf {\bibinfo {volume} {79}},\ \bibinfo {pages} {024426}
  (\bibinfo {year} {2009})}\BibitemShut {NoStop}%
\bibitem [{\citenamefont {Hermanns}\ and\ \citenamefont
  {Trebst}(2014)}]{Hermanns2014quantum}%
  \BibitemOpen
  \bibfield  {author} {\bibinfo {author} {\bibfnamefont {M.}~\bibnamefont
  {Hermanns}}\ and\ \bibinfo {author} {\bibfnamefont {S.}~\bibnamefont
  {Trebst}},\ }\bibfield  {title} {\enquote {\bibinfo {title} {{Quantum spin
  liquid with a Majorana Fermi surface on the three-dimensional hyperoctagon
  lattice}},}\ }\href {\doibase 10.1103/PhysRevB.89.235102} {\bibfield
  {journal} {\bibinfo  {journal} {Phys. Rev. B}\ }\textbf {\bibinfo {volume}
  {89}},\ \bibinfo {pages} {235102} (\bibinfo {year} {2014})}\BibitemShut
  {NoStop}%
\bibitem [{\citenamefont {Hermanns}\ \emph
  {et~al.}(2015{\natexlab{a}})\citenamefont {Hermanns}, \citenamefont
  {O'Brien},\ and\ \citenamefont {Trebst}}]{Hermanns2015weyl}%
  \BibitemOpen
  \bibfield  {author} {\bibinfo {author} {\bibfnamefont {M.}~\bibnamefont
  {Hermanns}}, \bibinfo {author} {\bibfnamefont {K.}~\bibnamefont {O'Brien}}, \
  and\ \bibinfo {author} {\bibfnamefont {S.}~\bibnamefont {Trebst}},\
  }\bibfield  {title} {\enquote {\bibinfo {title} {{Weyl Spin Liquids}},}\
  }\href {\doibase 10.1103/PhysRevLett.114.157202} {\bibfield  {journal}
  {\bibinfo  {journal} {Phys. Rev. Lett.}\ }\textbf {\bibinfo {volume} {114}},\
  \bibinfo {pages} {157202} (\bibinfo {year} {2015}{\natexlab{a}})}\BibitemShut
  {NoStop}%
\bibitem [{\citenamefont {Hermanns}\ \emph
  {et~al.}(2015{\natexlab{b}})\citenamefont {Hermanns}, \citenamefont
  {Trebst},\ and\ \citenamefont {Rosch}}]{Hermanns2015spin-peierls}%
  \BibitemOpen
  \bibfield  {author} {\bibinfo {author} {\bibfnamefont {Maria}\ \bibnamefont
  {Hermanns}}, \bibinfo {author} {\bibfnamefont {Simon}\ \bibnamefont
  {Trebst}}, \ and\ \bibinfo {author} {\bibfnamefont {Achim}\ \bibnamefont
  {Rosch}},\ }\bibfield  {title} {\enquote {\bibinfo {title} {{Spin-Peierls
  Instability of Three-Dimensional Spin Liquids with Majorana Fermi
  Surfaces}},}\ }\href {\doibase 10.1103/PhysRevLett.115.177205} {\bibfield
  {journal} {\bibinfo  {journal} {Phys. Rev. Lett.}\ }\textbf {\bibinfo
  {volume} {115}},\ \bibinfo {pages} {177205} (\bibinfo {year}
  {2015}{\natexlab{b}})}\BibitemShut {NoStop}%
\bibitem [{\citenamefont {O'Brien}\ \emph {et~al.}(2016)\citenamefont
  {O'Brien}, \citenamefont {Hermanns},\ and\ \citenamefont
  {Trebst}}]{Obrien2016classification}%
  \BibitemOpen
  \bibfield  {author} {\bibinfo {author} {\bibfnamefont {Kevin}\ \bibnamefont
  {O'Brien}}, \bibinfo {author} {\bibfnamefont {Maria}\ \bibnamefont
  {Hermanns}}, \ and\ \bibinfo {author} {\bibfnamefont {Simon}\ \bibnamefont
  {Trebst}},\ }\bibfield  {title} {\enquote {\bibinfo {title} {{Classification
  of gapless ${\mathbb{Z}}_{2}$ spin liquids in three-dimensional Kitaev
  models}},}\ }\href {\doibase 10.1103/PhysRevB.93.085101} {\bibfield
  {journal} {\bibinfo  {journal} {Phys. Rev. B}\ }\textbf {\bibinfo {volume}
  {93}},\ \bibinfo {pages} {085101} (\bibinfo {year} {2016})}\BibitemShut
  {NoStop}%
\bibitem [{\citenamefont {Chiu}\ \emph {et~al.}(2016)\citenamefont {Chiu},
  \citenamefont {Teo}, \citenamefont {Schnyder},\ and\ \citenamefont
  {Ryu}}]{Chiu2016classification}%
  \BibitemOpen
  \bibfield  {author} {\bibinfo {author} {\bibfnamefont {Ching-Kai}\
  \bibnamefont {Chiu}}, \bibinfo {author} {\bibfnamefont {Jeffrey C.~Y.}\
  \bibnamefont {Teo}}, \bibinfo {author} {\bibfnamefont {Andreas~P.}\
  \bibnamefont {Schnyder}}, \ and\ \bibinfo {author} {\bibfnamefont {Shinsei}\
  \bibnamefont {Ryu}},\ }\bibfield  {title} {\enquote {\bibinfo {title}
  {Classification of topological quantum matter with symmetries},}\ }\href
  {\doibase 10.1103/RevModPhys.88.035005} {\bibfield  {journal} {\bibinfo
  {journal} {Rev. Mod. Phys.}\ }\textbf {\bibinfo {volume} {88}},\ \bibinfo
  {pages} {035005} (\bibinfo {year} {2016})}\BibitemShut {NoStop}%
\bibitem [{\citenamefont {Lee}\ \emph {et~al.}(2014)\citenamefont {Lee},
  \citenamefont {Schaffer}, \citenamefont {Bhattacharjee},\ and\ \citenamefont
  {Kim}}]{Lee2014heisenberg}%
  \BibitemOpen
  \bibfield  {author} {\bibinfo {author} {\bibfnamefont {Eric Kin-Ho}\
  \bibnamefont {Lee}}, \bibinfo {author} {\bibfnamefont {Robert}\ \bibnamefont
  {Schaffer}}, \bibinfo {author} {\bibfnamefont {Subhro}\ \bibnamefont
  {Bhattacharjee}}, \ and\ \bibinfo {author} {\bibfnamefont {Yong~Baek}\
  \bibnamefont {Kim}},\ }\bibfield  {title} {\enquote {\bibinfo {title}
  {{Heisenberg-Kitaev model on the hyperhoneycomb lattice}},}\ }\href {\doibase
  10.1103/PhysRevB.89.045117} {\bibfield  {journal} {\bibinfo  {journal} {Phys.
  Rev. B}\ }\textbf {\bibinfo {volume} {89}},\ \bibinfo {pages} {045117}
  (\bibinfo {year} {2014})}\BibitemShut {NoStop}%
\bibitem [{\citenamefont {Kimchi}\ \emph {et~al.}(2014)\citenamefont {Kimchi},
  \citenamefont {Analytis},\ and\ \citenamefont
  {Vishwanath}}]{Kimchi2014three}%
  \BibitemOpen
  \bibfield  {author} {\bibinfo {author} {\bibfnamefont {Itamar}\ \bibnamefont
  {Kimchi}}, \bibinfo {author} {\bibfnamefont {James~G.}\ \bibnamefont
  {Analytis}}, \ and\ \bibinfo {author} {\bibfnamefont {Ashvin}\ \bibnamefont
  {Vishwanath}},\ }\bibfield  {title} {\enquote {\bibinfo {title}
  {{Three-dimensional quantum spin liquids in models of harmonic-honeycomb
  iridates and phase diagram in an infinite-$D$ approximation}},}\ }\href
  {\doibase 10.1103/PhysRevB.90.205126} {\bibfield  {journal} {\bibinfo
  {journal} {Phys. Rev. B}\ }\textbf {\bibinfo {volume} {90}},\ \bibinfo
  {pages} {205126} (\bibinfo {year} {2014})}\BibitemShut {NoStop}%
\bibitem [{\citenamefont {{Slager Robert-Jan}}\ \emph
  {et~al.}(2013)\citenamefont {{Slager Robert-Jan}}, \citenamefont {{Mesaros
  Andrej}}, \citenamefont {{Juri\v{c}i\'{c} Vladimir}},\ and\ \citenamefont
  {{Zaanen Jan}}}]{Slager2013}%
  \BibitemOpen
  \bibfield  {author} {\bibinfo {author} {\bibnamefont {{Slager Robert-Jan}}},
  \bibinfo {author} {\bibnamefont {{Mesaros Andrej}}}, \bibinfo {author}
  {\bibnamefont {{Juri\v{c}i\'{c} Vladimir}}}, \ and\ \bibinfo {author}
  {\bibnamefont {{Zaanen Jan}}},\ }\bibfield  {title} {\enquote {\bibinfo
  {title} {{The space group classification of topological band-insulators}},}\
  }\href {\doibase http://dx.doi.org/10.1038/nphys2513} {\bibfield  {journal}
  {\bibinfo  {journal} {Nat Phys}\ }\textbf {\bibinfo {volume} {9}},\ \bibinfo
  {pages} {98--102} (\bibinfo {year} {2013})},\ \bibinfo {note}
  {10.1038/nphys2513}\BibitemShut {NoStop}%
\bibitem [{\citenamefont {Morimoto}\ and\ \citenamefont
  {Furusaki}(2013)}]{PhysRevB.88.125129}%
  \BibitemOpen
  \bibfield  {author} {\bibinfo {author} {\bibfnamefont {Takahiro}\
  \bibnamefont {Morimoto}}\ and\ \bibinfo {author} {\bibfnamefont {Akira}\
  \bibnamefont {Furusaki}},\ }\bibfield  {title} {\enquote {\bibinfo {title}
  {{Topological classification with additional symmetries from Clifford
  algebras}},}\ }\href {\doibase 10.1103/PhysRevB.88.125129} {\bibfield
  {journal} {\bibinfo  {journal} {Phys. Rev. B}\ }\textbf {\bibinfo {volume}
  {88}},\ \bibinfo {pages} {125129} (\bibinfo {year} {2013})}\BibitemShut
  {NoStop}%
\bibitem [{\citenamefont {Jadaun}\ \emph {et~al.}(2013)\citenamefont {Jadaun},
  \citenamefont {Xiao}, \citenamefont {Niu},\ and\ \citenamefont
  {Banerjee}}]{PhysRevB.88.085110}%
  \BibitemOpen
  \bibfield  {author} {\bibinfo {author} {\bibfnamefont {Priyamvada}\
  \bibnamefont {Jadaun}}, \bibinfo {author} {\bibfnamefont {Di}~\bibnamefont
  {Xiao}}, \bibinfo {author} {\bibfnamefont {Qian}\ \bibnamefont {Niu}}, \ and\
  \bibinfo {author} {\bibfnamefont {Sanjay~K.}\ \bibnamefont {Banerjee}},\
  }\bibfield  {title} {\enquote {\bibinfo {title} {Topological classification
  of crystalline insulators with space group symmetry},}\ }\href {\doibase
  10.1103/PhysRevB.88.085110} {\bibfield  {journal} {\bibinfo  {journal} {Phys.
  Rev. B}\ }\textbf {\bibinfo {volume} {88}},\ \bibinfo {pages} {085110}
  (\bibinfo {year} {2013})}\BibitemShut {NoStop}%
\bibitem [{\citenamefont {Shiozaki}\ and\ \citenamefont
  {Sato}(2014)}]{Shiozaki2014topology}%
  \BibitemOpen
  \bibfield  {author} {\bibinfo {author} {\bibfnamefont {Ken}\ \bibnamefont
  {Shiozaki}}\ and\ \bibinfo {author} {\bibfnamefont {Masatoshi}\ \bibnamefont
  {Sato}},\ }\bibfield  {title} {\enquote {\bibinfo {title} {Topology of
  crystalline insulators and superconductors},}\ }\href {\doibase
  10.1103/PhysRevB.90.165114} {\bibfield  {journal} {\bibinfo  {journal} {Phys.
  Rev. B}\ }\textbf {\bibinfo {volume} {90}},\ \bibinfo {pages} {165114}
  (\bibinfo {year} {2014})}\BibitemShut {NoStop}%
\bibitem [{\citenamefont {Altland}\ and\ \citenamefont
  {Zirnbauer}(1997)}]{Altland1997classification}%
  \BibitemOpen
  \bibfield  {author} {\bibinfo {author} {\bibfnamefont {Alexander}\
  \bibnamefont {Altland}}\ and\ \bibinfo {author} {\bibfnamefont {Martin~R.}\
  \bibnamefont {Zirnbauer}},\ }\bibfield  {title} {\enquote {\bibinfo {title}
  {Nonstandard symmetry classes in mesoscopic normal-superconducting hybrid
  structures},}\ }\href {\doibase 10.1103/PhysRevB.55.1142} {\bibfield
  {journal} {\bibinfo  {journal} {Phys. Rev. B}\ }\textbf {\bibinfo {volume}
  {55}},\ \bibinfo {pages} {1142--1161} (\bibinfo {year} {1997})}\BibitemShut
  {NoStop}%
\bibitem [{\citenamefont {Schnyder}\ \emph {et~al.}(2008)\citenamefont
  {Schnyder}, \citenamefont {Ryu}, \citenamefont {Furusaki},\ and\
  \citenamefont {Ludwig}}]{Schnyder2008classification}%
  \BibitemOpen
  \bibfield  {author} {\bibinfo {author} {\bibfnamefont {Andreas~P.}\
  \bibnamefont {Schnyder}}, \bibinfo {author} {\bibfnamefont {Shinsei}\
  \bibnamefont {Ryu}}, \bibinfo {author} {\bibfnamefont {Akira}\ \bibnamefont
  {Furusaki}}, \ and\ \bibinfo {author} {\bibfnamefont {Andreas W.~W.}\
  \bibnamefont {Ludwig}},\ }\bibfield  {title} {\enquote {\bibinfo {title}
  {{Classification of topological insulators and superconductors in three
  spatial dimensions}},}\ }\href {\doibase 10.1103/PhysRevB.78.195125}
  {\bibfield  {journal} {\bibinfo  {journal} {Phys. Rev. B}\ }\textbf {\bibinfo
  {volume} {78}},\ \bibinfo {pages} {195125} (\bibinfo {year}
  {2008})}\BibitemShut {NoStop}%
\bibitem [{\citenamefont {Kitaev}(2009)}]{Kitaev2009periodic}%
  \BibitemOpen
  \bibfield  {author} {\bibinfo {author} {\bibfnamefont {Alexei}\ \bibnamefont
  {Kitaev}},\ }\bibfield  {title} {\enquote {\bibinfo {title} {{Periodic table
  for topological insulators and superconductors}},}\ }\href {\doibase
  10.1063/1.3149495} {\bibfield  {journal} {\bibinfo  {journal} {AIP Conference
  Proceedings}\ }\textbf {\bibinfo {volume} {1134}},\ \bibinfo {pages} {22--30}
  (\bibinfo {year} {2009})}\BibitemShut {NoStop}%
\bibitem [{\citenamefont {Fu}(2011)}]{Fu2011tci}%
  \BibitemOpen
  \bibfield  {author} {\bibinfo {author} {\bibfnamefont {Liang}\ \bibnamefont
  {Fu}},\ }\bibfield  {title} {\enquote {\bibinfo {title} {Topological
  crystalline insulators},}\ }\href {\doibase 10.1103/PhysRevLett.106.106802}
  {\bibfield  {journal} {\bibinfo  {journal} {Phys. Rev. Lett.}\ }\textbf
  {\bibinfo {volume} {106}},\ \bibinfo {pages} {106802} (\bibinfo {year}
  {2011})}\BibitemShut {NoStop}%
\bibitem [{\citenamefont {Hsieh}\ \emph {et~al.}(2012)\citenamefont {Hsieh},
  \citenamefont {Lin}, \citenamefont {Liu}, \citenamefont {Duan}, \citenamefont
  {Bansil},\ and\ \citenamefont {Fu}}]{Hsieh2012tci}%
  \BibitemOpen
  \bibfield  {author} {\bibinfo {author} {\bibfnamefont {Timothy~H}\
  \bibnamefont {Hsieh}}, \bibinfo {author} {\bibfnamefont {Hsin}\ \bibnamefont
  {Lin}}, \bibinfo {author} {\bibfnamefont {Junwei}\ \bibnamefont {Liu}},
  \bibinfo {author} {\bibfnamefont {Wenhui}\ \bibnamefont {Duan}}, \bibinfo
  {author} {\bibfnamefont {Arun}\ \bibnamefont {Bansil}}, \ and\ \bibinfo
  {author} {\bibfnamefont {Liang}\ \bibnamefont {Fu}},\ }\bibfield  {title}
  {\enquote {\bibinfo {title} {{Topological crystalline insulators in the SnTe
  material class}},}\ }\href@noop {} {\bibfield  {journal} {\bibinfo  {journal}
  {Nat. Commun.}\ }\textbf {\bibinfo {volume} {3}},\ \bibinfo {pages} {982}
  (\bibinfo {year} {2012})}\BibitemShut {NoStop}%
\bibitem [{\citenamefont {Fang}\ \emph {et~al.}(2012)\citenamefont {Fang},
  \citenamefont {Gilbert},\ and\ \citenamefont
  {Bernevig}}]{PhysRevB.86.115112}%
  \BibitemOpen
  \bibfield  {author} {\bibinfo {author} {\bibfnamefont {Chen}\ \bibnamefont
  {Fang}}, \bibinfo {author} {\bibfnamefont {Matthew~J.}\ \bibnamefont
  {Gilbert}}, \ and\ \bibinfo {author} {\bibfnamefont {B.~A.}\ \bibnamefont
  {Bernevig}},\ }\bibfield  {title} {\enquote {\bibinfo {title} {Bulk
  topological invariants in noninteracting point group symmetric insulators},}\
  }\href {\doibase 10.1103/PhysRevB.86.115112} {\bibfield  {journal} {\bibinfo
  {journal} {Phys. Rev. B}\ }\textbf {\bibinfo {volume} {86}},\ \bibinfo
  {pages} {115112} (\bibinfo {year} {2012})}\BibitemShut {NoStop}%
\bibitem [{\citenamefont {Teo}\ and\ \citenamefont
  {Hughes}(2013)}]{PhysRevLett.111.047006}%
  \BibitemOpen
  \bibfield  {author} {\bibinfo {author} {\bibfnamefont {Jeffrey C.~Y.}\
  \bibnamefont {Teo}}\ and\ \bibinfo {author} {\bibfnamefont {Taylor~L.}\
  \bibnamefont {Hughes}},\ }\bibfield  {title} {\enquote {\bibinfo {title}
  {{Existence of Majorana-Fermion Bound States on Disclinations and the
  Classification of Topological Crystalline Superconductors in Two
  Dimensions}},}\ }\href {\doibase 10.1103/PhysRevLett.111.047006} {\bibfield
  {journal} {\bibinfo  {journal} {Phys. Rev. Lett.}\ }\textbf {\bibinfo
  {volume} {111}},\ \bibinfo {pages} {047006} (\bibinfo {year}
  {2013})}\BibitemShut {NoStop}%
\bibitem [{\citenamefont {Fu}\ and\ \citenamefont
  {Kane}(2007)}]{FuKane2007tci}%
  \BibitemOpen
  \bibfield  {author} {\bibinfo {author} {\bibfnamefont {Liang}\ \bibnamefont
  {Fu}}\ and\ \bibinfo {author} {\bibfnamefont {C.~L.}\ \bibnamefont {Kane}},\
  }\bibfield  {title} {\enquote {\bibinfo {title} {Topological insulators with
  inversion symmetry},}\ }\href {\doibase 10.1103/PhysRevB.76.045302}
  {\bibfield  {journal} {\bibinfo  {journal} {Phys. Rev. B}\ }\textbf {\bibinfo
  {volume} {76}},\ \bibinfo {pages} {045302} (\bibinfo {year}
  {2007})}\BibitemShut {NoStop}%
\bibitem [{\citenamefont {Turner}\ \emph {et~al.}(2010)\citenamefont {Turner},
  \citenamefont {Zhang},\ and\ \citenamefont
  {Vishwanath}}]{Turner2010entanglement}%
  \BibitemOpen
  \bibfield  {author} {\bibinfo {author} {\bibfnamefont {Ari~M.}\ \bibnamefont
  {Turner}}, \bibinfo {author} {\bibfnamefont {Yi}~\bibnamefont {Zhang}}, \
  and\ \bibinfo {author} {\bibfnamefont {Ashvin}\ \bibnamefont {Vishwanath}},\
  }\bibfield  {title} {\enquote {\bibinfo {title} {Entanglement and inversion
  symmetry in topological insulators},}\ }\href {\doibase
  10.1103/PhysRevB.82.241102} {\bibfield  {journal} {\bibinfo  {journal} {Phys.
  Rev. B}\ }\textbf {\bibinfo {volume} {82}},\ \bibinfo {pages} {241102}
  (\bibinfo {year} {2010})}\BibitemShut {NoStop}%
\bibitem [{\citenamefont {Hughes}\ \emph {et~al.}(2011)\citenamefont {Hughes},
  \citenamefont {Prodan},\ and\ \citenamefont
  {Bernevig}}]{Hughes2011inversion}%
  \BibitemOpen
  \bibfield  {author} {\bibinfo {author} {\bibfnamefont {Taylor~L.}\
  \bibnamefont {Hughes}}, \bibinfo {author} {\bibfnamefont {Emil}\ \bibnamefont
  {Prodan}}, \ and\ \bibinfo {author} {\bibfnamefont {B.~A.}\ \bibnamefont
  {Bernevig}},\ }\bibfield  {title} {\enquote {\bibinfo {title}
  {Inversion-symmetric topological insulators},}\ }\href {\doibase
  10.1103/PhysRevB.83.245132} {\bibfield  {journal} {\bibinfo  {journal} {Phys.
  Rev. B}\ }\textbf {\bibinfo {volume} {83}},\ \bibinfo {pages} {245132}
  (\bibinfo {year} {2011})}\BibitemShut {NoStop}%
\bibitem [{Note1()}]{Note1}%
  \BibitemOpen
  \bibinfo {note} {Even though lattice symmetries are necessarily broken in
  real materials due to imperfections, it is sufficient that the symmetry is
  realized `on average' to see experimental signatures of topological
  crystalline insulators~\cite
  {Hsieh2012tci,Tanaka2012SnTe,Yang2012SnTe,Sziawa2012SnTe}.}\BibitemShut
  {Stop}%
\bibitem [{\citenamefont {Volovik}(2009)}]{Volovik2003}%
  \BibitemOpen
  \bibfield  {author} {\bibinfo {author} {\bibfnamefont {Grigory~E.}\
  \bibnamefont {Volovik}},\ }\href@noop {} {\emph {\bibinfo {title} {{The
  Universe in a Helium Droplet}}}}\ (\bibinfo  {publisher} {Oxford University
  Press, Oxford, UK},\ \bibinfo {year} {2009})\BibitemShut {NoStop}%
\bibitem [{\citenamefont {Ho\ifmmode~\check{r}\else
  \v{r}\fi{}ava}(2005)}]{Horava2005stability}%
  \BibitemOpen
  \bibfield  {author} {\bibinfo {author} {\bibfnamefont {Petr}\ \bibnamefont
  {Ho\ifmmode~\check{r}\else \v{r}\fi{}ava}},\ }\bibfield  {title} {\enquote
  {\bibinfo {title} {{Stability of Fermi Surfaces and $K$ Theory}},}\ }\href
  {\doibase 10.1103/PhysRevLett.95.016405} {\bibfield  {journal} {\bibinfo
  {journal} {Phys. Rev. Lett.}\ }\textbf {\bibinfo {volume} {95}},\ \bibinfo
  {pages} {016405} (\bibinfo {year} {2005})}\BibitemShut {NoStop}%
\bibitem [{\citenamefont {Matsuura}\ \emph {et~al.}(2013)\citenamefont
  {Matsuura}, \citenamefont {Chang}, \citenamefont {Schnyder},\ and\
  \citenamefont {Ryu}}]{Matsuura2013protected}%
  \BibitemOpen
  \bibfield  {author} {\bibinfo {author} {\bibfnamefont {Shunji}\ \bibnamefont
  {Matsuura}}, \bibinfo {author} {\bibfnamefont {Po-Yao}\ \bibnamefont
  {Chang}}, \bibinfo {author} {\bibfnamefont {Andreas~P}\ \bibnamefont
  {Schnyder}}, \ and\ \bibinfo {author} {\bibfnamefont {Shinsei}\ \bibnamefont
  {Ryu}},\ }\bibfield  {title} {\enquote {\bibinfo {title} {Protected boundary
  states in gapless topological phases},}\ }\href
  {http://stacks.iop.org/1367-2630/15/i=6/a=065001} {\bibfield  {journal}
  {\bibinfo  {journal} {New Journal of Physics}\ }\textbf {\bibinfo {volume}
  {15}},\ \bibinfo {pages} {065001} (\bibinfo {year} {2013})}\BibitemShut
  {NoStop}%
\bibitem [{\citenamefont {Zhao}\ and\ \citenamefont
  {Wang}(2013)}]{Zhao2013topological}%
  \BibitemOpen
  \bibfield  {author} {\bibinfo {author} {\bibfnamefont {Y.~X.}\ \bibnamefont
  {Zhao}}\ and\ \bibinfo {author} {\bibfnamefont {Z.~D.}\ \bibnamefont
  {Wang}},\ }\bibfield  {title} {\enquote {\bibinfo {title} {{Topological
  Classification and Stability of Fermi Surfaces}},}\ }\href {\doibase
  10.1103/PhysRevLett.110.240404} {\bibfield  {journal} {\bibinfo  {journal}
  {Phys. Rev. Lett.}\ }\textbf {\bibinfo {volume} {110}},\ \bibinfo {pages}
  {240404} (\bibinfo {year} {2013})}\BibitemShut {NoStop}%
\bibitem [{\citenamefont {Chiu}\ and\ \citenamefont
  {Schnyder}(2014)}]{Chiu2014classification}%
  \BibitemOpen
  \bibfield  {author} {\bibinfo {author} {\bibfnamefont {Ching-Kai}\
  \bibnamefont {Chiu}}\ and\ \bibinfo {author} {\bibfnamefont {Andreas~P.}\
  \bibnamefont {Schnyder}},\ }\bibfield  {title} {\enquote {\bibinfo {title}
  {Classification of reflection-symmetry-protected topological semimetals and
  nodal superconductors},}\ }\href {\doibase 10.1103/PhysRevB.90.205136}
  {\bibfield  {journal} {\bibinfo  {journal} {Phys. Rev. B}\ }\textbf {\bibinfo
  {volume} {90}},\ \bibinfo {pages} {205136} (\bibinfo {year}
  {2014})}\BibitemShut {NoStop}%
\bibitem [{\citenamefont {Yang}\ \emph {et~al.}(2014)\citenamefont {Yang},
  \citenamefont {Pan},\ and\ \citenamefont {Zhang}}]{Yang2014dirac}%
  \BibitemOpen
  \bibfield  {author} {\bibinfo {author} {\bibfnamefont {Shengyuan~A.}\
  \bibnamefont {Yang}}, \bibinfo {author} {\bibfnamefont {Hui}\ \bibnamefont
  {Pan}}, \ and\ \bibinfo {author} {\bibfnamefont {Fan}\ \bibnamefont
  {Zhang}},\ }\bibfield  {title} {\enquote {\bibinfo {title} {{Dirac and Weyl
  Superconductors in Three Dimensions}},}\ }\href {\doibase
  10.1103/PhysRevLett.113.046401} {\bibfield  {journal} {\bibinfo  {journal}
  {Phys. Rev. Lett.}\ }\textbf {\bibinfo {volume} {113}},\ \bibinfo {pages}
  {046401} (\bibinfo {year} {2014})}\BibitemShut {NoStop}%
\bibitem [{\citenamefont {Parameswaran}\ \emph {et~al.}(2013)\citenamefont
  {Parameswaran}, \citenamefont {Turner}, \citenamefont {Arovas},\ and\
  \citenamefont {Vishwanath}}]{Parameswaran2013nonsymmorphic}%
  \BibitemOpen
  \bibfield  {author} {\bibinfo {author} {\bibfnamefont {Siddharth~A.}\
  \bibnamefont {Parameswaran}}, \bibinfo {author} {\bibfnamefont {Ari~M.}\
  \bibnamefont {Turner}}, \bibinfo {author} {\bibfnamefont {Daniel~P.}\
  \bibnamefont {Arovas}}, \ and\ \bibinfo {author} {\bibfnamefont {Ashvin}\
  \bibnamefont {Vishwanath}},\ }\bibfield  {title} {\enquote {\bibinfo {title}
  {{Topological order and absence of band insulators at integer filling in
  non-symmorphic crystals}},}\ }\href {\doibase
  http://dx.doi.org/10.1038/nphys2600 10.1038/nphys2600} {\bibfield  {journal}
  {\bibinfo  {journal} {Nat. Phys.}\ }\textbf {\bibinfo {volume} {9}},\
  \bibinfo {pages} {299--303} (\bibinfo {year} {2013})}\BibitemShut {NoStop}%
\bibitem [{\citenamefont {Fang}\ and\ \citenamefont {Fu}(2015)}]{Fag2015new}%
  \BibitemOpen
  \bibfield  {author} {\bibinfo {author} {\bibfnamefont {Chen}\ \bibnamefont
  {Fang}}\ and\ \bibinfo {author} {\bibfnamefont {Liang}\ \bibnamefont {Fu}},\
  }\bibfield  {title} {\enquote {\bibinfo {title} {New classes of
  three-dimensional topological crystalline insulators: Nonsymmorphic and
  magnetic},}\ }\href {\doibase 10.1103/PhysRevB.91.161105} {\bibfield
  {journal} {\bibinfo  {journal} {Phys. Rev. B}\ }\textbf {\bibinfo {volume}
  {91}},\ \bibinfo {pages} {161105} (\bibinfo {year} {2015})}\BibitemShut
  {NoStop}%
\bibitem [{\citenamefont {Zhao}\ and\ \citenamefont
  {Schnyder}(2016)}]{Zhao2016nonsymmorphic}%
  \BibitemOpen
  \bibfield  {author} {\bibinfo {author} {\bibfnamefont {Y.~X.}\ \bibnamefont
  {Zhao}}\ and\ \bibinfo {author} {\bibfnamefont {Andreas~P.}\ \bibnamefont
  {Schnyder}},\ }\bibfield  {title} {\enquote {\bibinfo {title} {Nonsymmorphic
  symmetry-required band crossings in topological semimetals},}\ }\href
  {\doibase 10.1103/PhysRevB.94.195109} {\bibfield  {journal} {\bibinfo
  {journal} {Phys. Rev. B}\ }\textbf {\bibinfo {volume} {94}},\ \bibinfo
  {pages} {195109} (\bibinfo {year} {2016})}\BibitemShut {NoStop}%
\bibitem [{\citenamefont {Wang}\ and\ \citenamefont
  {Liu}(2016)}]{Wang2016topological}%
  \BibitemOpen
  \bibfield  {author} {\bibinfo {author} {\bibfnamefont {Qing-Ze}\ \bibnamefont
  {Wang}}\ and\ \bibinfo {author} {\bibfnamefont {Chao-Xing}\ \bibnamefont
  {Liu}},\ }\bibfield  {title} {\enquote {\bibinfo {title} {Topological
  nonsymmorphic crystalline superconductors},}\ }\href {\doibase
  10.1103/PhysRevB.93.020505} {\bibfield  {journal} {\bibinfo  {journal} {Phys.
  Rev. B}\ }\textbf {\bibinfo {volume} {93}},\ \bibinfo {pages} {020505}
  (\bibinfo {year} {2016})}\BibitemShut {NoStop}%
\bibitem [{\citenamefont {{Wang Zhijun}}\ \emph {et~al.}(2016)\citenamefont
  {{Wang Zhijun}}, \citenamefont {{Alexandradinata A.}}, \citenamefont {{Cava
  R. J.}},\ and\ \citenamefont {{Bernevig B. A.}}}]{Zhijun2016}%
  \BibitemOpen
  \bibfield  {author} {\bibinfo {author} {\bibnamefont {{Wang Zhijun}}},
  \bibinfo {author} {\bibnamefont {{Alexandradinata A.}}}, \bibinfo {author}
  {\bibnamefont {{Cava R. J.}}}, \ and\ \bibinfo {author} {\bibnamefont
  {{Bernevig B. A.}}},\ }\bibfield  {title} {\enquote {\bibinfo {title}
  {{Hourglass fermions}},}\ }\href {\doibase
  http://dx.doi.org/10.1038/nature17410 10.1038/nature17410} {\bibfield
  {journal} {\bibinfo  {journal} {Nature}\ }\textbf {\bibinfo {volume} {532}},\
  \bibinfo {pages} {189--194} (\bibinfo {year} {2016})}\BibitemShut {NoStop}%
\bibitem [{\citenamefont {Takahashi}\ \emph {et~al.}(2017)\citenamefont
  {Takahashi}, \citenamefont {Hirayama},\ and\ \citenamefont
  {Murakami}}]{Takahashi2017topological}%
  \BibitemOpen
  \bibfield  {author} {\bibinfo {author} {\bibfnamefont {Ryo}\ \bibnamefont
  {Takahashi}}, \bibinfo {author} {\bibfnamefont {Motoaki}\ \bibnamefont
  {Hirayama}}, \ and\ \bibinfo {author} {\bibfnamefont {Shuichi}\ \bibnamefont
  {Murakami}},\ }\bibfield  {title} {\enquote {\bibinfo {title} {Topological
  nodal-line semimetals arising from crystal symmetry},}\ }\href@noop {}
  {\bibfield  {journal} {\bibinfo  {journal} {arXiv:1704.02151}\ } (\bibinfo
  {year} {2017})}\BibitemShut {NoStop}%
\bibitem [{\citenamefont {Wells}\ \emph {et~al.}(1977)\citenamefont {Wells}
  \emph {et~al.}}]{Wells1977}%
  \BibitemOpen
  \bibfield  {author} {\bibinfo {author} {\bibfnamefont {Alexander~Frank}\
  \bibnamefont {Wells}} \emph {et~al.},\ }\href@noop {} {\emph {\bibinfo
  {title} {Three dimensional nets and polyhedra}}}\ (\bibinfo  {publisher}
  {Wiley},\ \bibinfo {year} {1977})\BibitemShut {NoStop}%
\bibitem [{Note2()}]{Note2}%
  \BibitemOpen
  \bibinfo {note} {In this article, we prefer to use the terminology introduced
  by Wells~\cite {Wells1977} to distinguish the lattices. The $8^2.10$-$a$~
  lattice is also known as LiGe or \protect \textbf {lig} net, and (10,3)d as
  \protect \textbf {utp} net in O'Keeffe's three-letter codes~\cite
  {OKeeffe2003regular,OKeeffe2003semiregular} which is intensively used in the
  chemistry literature and database.}\BibitemShut {Stop}%
\bibitem [{Note3()}]{Note3}%
  \BibitemOpen
  \bibinfo {note} {The (10,3)a lattice is also known as SrSi$_2$ or \protect
  \textbf {srs} net~\cite
  {OKeeffe2003regular,OKeeffe2003semiregular,Batten2009review}, Laves
  graph~\cite {laves}, or K$_4$ crystal~\cite {K4}}\BibitemShut {NoStop}%
\bibitem [{\citenamefont {Bzdu{\v{s}}ek}\ \emph {et~al.}(2016)\citenamefont
  {Bzdu{\v{s}}ek}, \citenamefont {Wu}, \citenamefont {R{\"u}egg}, \citenamefont
  {Sigrist},\ and\ \citenamefont {Soluyanov}}]{Bzdusek2016nodal}%
  \BibitemOpen
  \bibfield  {author} {\bibinfo {author} {\bibfnamefont {Tom{\'a}{\v{s}}}\
  \bibnamefont {Bzdu{\v{s}}ek}}, \bibinfo {author} {\bibfnamefont {QuanSheng}\
  \bibnamefont {Wu}}, \bibinfo {author} {\bibfnamefont {Andreas}\ \bibnamefont
  {R{\"u}egg}}, \bibinfo {author} {\bibfnamefont {Manfred}\ \bibnamefont
  {Sigrist}}, \ and\ \bibinfo {author} {\bibfnamefont {Alexey~A}\ \bibnamefont
  {Soluyanov}},\ }\bibfield  {title} {\enquote {\bibinfo {title} {Nodal-chain
  metals},}\ }\href@noop {} {\bibfield  {journal} {\bibinfo  {journal}
  {Nature}\ }\textbf {\bibinfo {volume} {538}},\ \bibinfo {pages} {75–78}
  (\bibinfo {year} {2016})}\BibitemShut {NoStop}%
\bibitem [{\citenamefont {Yang}\ \emph {et~al.}(2017)\citenamefont {Yang},
  \citenamefont {Bojesen}, \citenamefont {Morimoto},\ and\ \citenamefont
  {Furusaki}}]{Yang2017topological}%
  \BibitemOpen
  \bibfield  {author} {\bibinfo {author} {\bibfnamefont {Bohm-Jung}\
  \bibnamefont {Yang}}, \bibinfo {author} {\bibfnamefont {Troels~Arnfred}\
  \bibnamefont {Bojesen}}, \bibinfo {author} {\bibfnamefont {Takahiro}\
  \bibnamefont {Morimoto}}, \ and\ \bibinfo {author} {\bibfnamefont {Akira}\
  \bibnamefont {Furusaki}},\ }\bibfield  {title} {\enquote {\bibinfo {title}
  {Topological semimetals protected by off-centered symmetries in nonsymmorphic
  crystals},}\ }\href {\doibase 10.1103/PhysRevB.95.075135} {\bibfield
  {journal} {\bibinfo  {journal} {Phys. Rev. B}\ }\textbf {\bibinfo {volume}
  {95}},\ \bibinfo {pages} {075135} (\bibinfo {year} {2017})}\BibitemShut
  {NoStop}%
\bibitem [{\citenamefont {Khaliullin}(2005)}]{Khaliullin2005orbital}%
  \BibitemOpen
  \bibfield  {author} {\bibinfo {author} {\bibfnamefont {Giniyat}\ \bibnamefont
  {Khaliullin}},\ }\bibfield  {title} {\enquote {\bibinfo {title} {{Orbital
  Order and Fluctuations in Mott Insulators}},}\ }\href {\doibase
  10.1143/PTPS.160.155} {\bibfield  {journal} {\bibinfo  {journal} {Progress of
  Theoretical Physics Supplement}\ }\textbf {\bibinfo {volume} {160}},\
  \bibinfo {pages} {155} (\bibinfo {year} {2005})}\BibitemShut {NoStop}%
\bibitem [{\citenamefont {Jackeli}\ and\ \citenamefont
  {Khaliullin}(2009)}]{Jackeli2009}%
  \BibitemOpen
  \bibfield  {author} {\bibinfo {author} {\bibfnamefont {G.}~\bibnamefont
  {Jackeli}}\ and\ \bibinfo {author} {\bibfnamefont {G.}~\bibnamefont
  {Khaliullin}},\ }\bibfield  {title} {\enquote {\bibinfo {title} {{Mott
  Insulators in the Strong Spin-Orbit Coupling Limit: From Heisenberg to a
  Quantum Compass and Kitaev Models}},}\ }\href {\doibase
  10.1103/PhysRevLett.102.017205} {\bibfield  {journal} {\bibinfo  {journal}
  {Phys. Rev. Lett.}\ }\textbf {\bibinfo {volume} {102}},\ \bibinfo {pages}
  {017205} (\bibinfo {year} {2009})}\BibitemShut {NoStop}%
\bibitem [{\citenamefont {Singh}\ \emph {et~al.}(2012)\citenamefont {Singh},
  \citenamefont {Manni}, \citenamefont {Reuther}, \citenamefont {Berlijn},
  \citenamefont {Thomale}, \citenamefont {Ku}, \citenamefont {Trebst},\ and\
  \citenamefont {Gegenwart}}]{Singh2012relevance}%
  \BibitemOpen
  \bibfield  {author} {\bibinfo {author} {\bibfnamefont {Yogesh}\ \bibnamefont
  {Singh}}, \bibinfo {author} {\bibfnamefont {S.}~\bibnamefont {Manni}},
  \bibinfo {author} {\bibfnamefont {J.}~\bibnamefont {Reuther}}, \bibinfo
  {author} {\bibfnamefont {T.}~\bibnamefont {Berlijn}}, \bibinfo {author}
  {\bibfnamefont {R.}~\bibnamefont {Thomale}}, \bibinfo {author} {\bibfnamefont
  {W.}~\bibnamefont {Ku}}, \bibinfo {author} {\bibfnamefont {S.}~\bibnamefont
  {Trebst}}, \ and\ \bibinfo {author} {\bibfnamefont {P.}~\bibnamefont
  {Gegenwart}},\ }\bibfield  {title} {\enquote {\bibinfo {title} {{Relevance of
  the Heisenberg-Kitaev Model for the Honeycomb Lattice Iridates
  A$_2$IrO$_3$}},}\ }\href {\doibase 10.1103/PhysRevLett.108.127203} {\bibfield
   {journal} {\bibinfo  {journal} {Phys. Rev. Lett.}\ }\textbf {\bibinfo
  {volume} {108}},\ \bibinfo {pages} {127203} (\bibinfo {year}
  {2012})}\BibitemShut {NoStop}%
\bibitem [{\citenamefont {Takayama}\ \emph {et~al.}(2015)\citenamefont
  {Takayama}, \citenamefont {Kato}, \citenamefont {Dinnebier}, \citenamefont
  {Nuss}, \citenamefont {Kono}, \citenamefont {Veiga}, \citenamefont {Fabbris},
  \citenamefont {Haskel},\ and\ \citenamefont
  {Takagi}}]{Takayama2015hyperhoneycomb}%
  \BibitemOpen
  \bibfield  {author} {\bibinfo {author} {\bibfnamefont {T.}~\bibnamefont
  {Takayama}}, \bibinfo {author} {\bibfnamefont {A.}~\bibnamefont {Kato}},
  \bibinfo {author} {\bibfnamefont {R.}~\bibnamefont {Dinnebier}}, \bibinfo
  {author} {\bibfnamefont {J.}~\bibnamefont {Nuss}}, \bibinfo {author}
  {\bibfnamefont {H.}~\bibnamefont {Kono}}, \bibinfo {author} {\bibfnamefont
  {L.~S.~I.}\ \bibnamefont {Veiga}}, \bibinfo {author} {\bibfnamefont
  {G.}~\bibnamefont {Fabbris}}, \bibinfo {author} {\bibfnamefont
  {D.}~\bibnamefont {Haskel}}, \ and\ \bibinfo {author} {\bibfnamefont
  {H.}~\bibnamefont {Takagi}},\ }\bibfield  {title} {\enquote {\bibinfo {title}
  {{Hyperhoneycomb Iridate
  $\ensuremath{\beta}\text{\ensuremath{-}}{\mathrm{Li}}_{2}{\mathrm{IrO}}_{3}$
  as a Platform for Kitaev Magnetism}},}\ }\href {\doibase
  10.1103/PhysRevLett.114.077202} {\bibfield  {journal} {\bibinfo  {journal}
  {Phys. Rev. Lett.}\ }\textbf {\bibinfo {volume} {114}},\ \bibinfo {pages}
  {077202} (\bibinfo {year} {2015})}\BibitemShut {NoStop}%
\bibitem [{\citenamefont {Banerjee}\ \emph {et~al.}(2016)\citenamefont
  {Banerjee}, \citenamefont {Bridges}, \citenamefont {Yan}, \citenamefont
  {Aczel}, \citenamefont {Li}, \citenamefont {Stone}, \citenamefont {Granroth},
  \citenamefont {Lumsden}, \citenamefont {Yiu}, \citenamefont {Knolle},
  \citenamefont {Bhattacharjee}, \citenamefont {Kovrizhin}, \citenamefont
  {Moessner}, \citenamefont {Tennant}, \citenamefont {Mandrus},\ and\
  \citenamefont {Nagler}}]{Banerjee2016proximate}%
  \BibitemOpen
  \bibfield  {author} {\bibinfo {author} {\bibfnamefont {A.}~\bibnamefont
  {Banerjee}}, \bibinfo {author} {\bibfnamefont {C.~A.}\ \bibnamefont
  {Bridges}}, \bibinfo {author} {\bibfnamefont {J.-Q.}\ \bibnamefont {Yan}},
  \bibinfo {author} {\bibfnamefont {A.~A.}\ \bibnamefont {Aczel}}, \bibinfo
  {author} {\bibfnamefont {L.}~\bibnamefont {Li}}, \bibinfo {author}
  {\bibfnamefont {M.~B.}\ \bibnamefont {Stone}}, \bibinfo {author}
  {\bibfnamefont {G.~E.}\ \bibnamefont {Granroth}}, \bibinfo {author}
  {\bibfnamefont {M.~D.}\ \bibnamefont {Lumsden}}, \bibinfo {author}
  {\bibfnamefont {Y.}~\bibnamefont {Yiu}}, \bibinfo {author} {\bibfnamefont
  {J.}~\bibnamefont {Knolle}}, \bibinfo {author} {\bibfnamefont
  {S.}~\bibnamefont {Bhattacharjee}}, \bibinfo {author} {\bibfnamefont {D.~L.}\
  \bibnamefont {Kovrizhin}}, \bibinfo {author} {\bibfnamefont {R.}~\bibnamefont
  {Moessner}}, \bibinfo {author} {\bibfnamefont {D.~A.}\ \bibnamefont
  {Tennant}}, \bibinfo {author} {\bibfnamefont {D.~G.}\ \bibnamefont
  {Mandrus}}, \ and\ \bibinfo {author} {\bibfnamefont {S.~E.}\ \bibnamefont
  {Nagler}},\ }\bibfield  {title} {\enquote {\bibinfo {title} {{Proximate
  Kitaev quantum spin liquid behaviour in a honeycomb magnet}},}\ }\href
  {\doibase 10.1038/nmat4604} {\bibfield  {journal} {\bibinfo  {journal}
  {Nature Materials}\ }\textbf {\bibinfo {volume} {15}},\ \bibinfo {pages}
  {733--740} (\bibinfo {year} {2016})}\BibitemShut {NoStop}%
\bibitem [{\citenamefont {Yamada}\ \emph {et~al.}(2017)\citenamefont {Yamada},
  \citenamefont {Fujita},\ and\ \citenamefont {Oshikawa}}]{Yamada2017MOF}%
  \BibitemOpen
  \bibfield  {author} {\bibinfo {author} {\bibfnamefont {Masahiko~G.}\
  \bibnamefont {Yamada}}, \bibinfo {author} {\bibfnamefont {Hiroyuki}\
  \bibnamefont {Fujita}}, \ and\ \bibinfo {author} {\bibfnamefont {Masaki}\
  \bibnamefont {Oshikawa}},\ }\bibfield  {title} {\enquote {\bibinfo {title}
  {Designing kitaev spin liquids in metal-organic frameworks},}\ }\href
  {\doibase 10.1103/PhysRevLett.119.057202} {\bibfield  {journal} {\bibinfo
  {journal} {Phys. Rev. Lett.}\ }\textbf {\bibinfo {volume} {119}},\ \bibinfo
  {pages} {057202} (\bibinfo {year} {2017})}\BibitemShut {NoStop}%
\bibitem [{\citenamefont {\"Ohrstr\"om}\ and\ \citenamefont
  {Larsson}(2004)}]{Oehrstrom2003}%
  \BibitemOpen
  \bibfield  {author} {\bibinfo {author} {\bibfnamefont {Lars}\ \bibnamefont
  {\"Ohrstr\"om}}\ and\ \bibinfo {author} {\bibfnamefont {Krister}\
  \bibnamefont {Larsson}},\ }\bibfield  {title} {\enquote {\bibinfo {title}
  {What kinds of three-dimensional nets are possible with tris-chelated metal
  complexes as building blocks?}}\ }\href {\doibase 10.1039/B310330G}
  {\bibfield  {journal} {\bibinfo  {journal} {Dalton Trans.}\ ,\ \bibinfo
  {pages} {347}} (\bibinfo {year} {2004})}\BibitemShut {NoStop}%
\bibitem [{\citenamefont {Clemente-Le\'on}\ \emph {et~al.}(2013)\citenamefont
  {Clemente-Le\'on}, \citenamefont {Coronado},\ and\ \citenamefont
  {L\'opez-Jord\`a}}]{ClementeLeon2013}%
  \BibitemOpen
  \bibfield  {author} {\bibinfo {author} {\bibfnamefont {Miguel}\ \bibnamefont
  {Clemente-Le\'on}}, \bibinfo {author} {\bibfnamefont {Eugenio}\ \bibnamefont
  {Coronado}}, \ and\ \bibinfo {author} {\bibfnamefont {Maurici}\ \bibnamefont
  {L\'opez-Jord\`a}},\ }\bibfield  {title} {\enquote {\bibinfo {title} {{2D and
  3D bimetallic oxalate-based ferromagnets prepared by insertion of MnIII-salen
  type complexes}},}\ }\href {\doibase 10.1039/C3DT32996H} {\bibfield
  {journal} {\bibinfo  {journal} {Dalton Trans.}\ }\textbf {\bibinfo {volume}
  {42}},\ \bibinfo {pages} {5100--5110} (\bibinfo {year} {2013})}\BibitemShut
  {NoStop}%
\bibitem [{\citenamefont {Hendon}\ \emph {et~al.}(2017)\citenamefont {Hendon},
  \citenamefont {Rieth}, \citenamefont {Korzy\'nski},\ and\ \citenamefont
  {Dinc\u{a}}}]{Hendon2017MOF}%
  \BibitemOpen
  \bibfield  {author} {\bibinfo {author} {\bibfnamefont {Christopher~H.}\
  \bibnamefont {Hendon}}, \bibinfo {author} {\bibfnamefont {Adam~J.}\
  \bibnamefont {Rieth}}, \bibinfo {author} {\bibfnamefont {Maciej~D.}\
  \bibnamefont {Korzy\'nski}}, \ and\ \bibinfo {author} {\bibfnamefont
  {Mircea}\ \bibnamefont {Dinc\u{a}}},\ }\bibfield  {title} {\enquote {\bibinfo
  {title} {{Grand Challenges and Future Opportunities for Metal–Organic
  Frameworks}},}\ }\href {\doibase 10.1021/acscentsci.7b00197} {\bibfield
  {journal} {\bibinfo  {journal} {ACS Cent. Sci.}\ }\textbf {\bibinfo {volume}
  {3}},\ \bibinfo {pages} {554} (\bibinfo {year} {2017})}\BibitemShut {NoStop}%
\bibitem [{\citenamefont {Kitagawa}\ \emph {et~al.}(2004)\citenamefont
  {Kitagawa}, \citenamefont {Kitaura},\ and\ \citenamefont
  {Noro}}]{Kitagawa2004PCP}%
  \BibitemOpen
  \bibfield  {author} {\bibinfo {author} {\bibfnamefont {Susumu}\ \bibnamefont
  {Kitagawa}}, \bibinfo {author} {\bibfnamefont {Ryo}\ \bibnamefont {Kitaura}},
  \ and\ \bibinfo {author} {\bibfnamefont {Shin-ichiro}\ \bibnamefont {Noro}},\
  }\bibfield  {title} {\enquote {\bibinfo {title} {Functional porous
  coordination polymers},}\ }\href {\doibase 10.1002/anie.200300610} {\bibfield
   {journal} {\bibinfo  {journal} {Angew. Chem. Int. Ed.}\ }\textbf {\bibinfo
  {volume} {43}},\ \bibinfo {pages} {2334--2375} (\bibinfo {year}
  {2004})}\BibitemShut {NoStop}%
\bibitem [{\citenamefont {Batten}\ \emph {et~al.}(2009)\citenamefont {Batten},
  \citenamefont {Neville},\ and\ \citenamefont {Turner}}]{Batten2009review}%
  \BibitemOpen
  \bibfield  {author} {\bibinfo {author} {\bibfnamefont {Stuart~R}\
  \bibnamefont {Batten}}, \bibinfo {author} {\bibfnamefont {Suzanne~M}\
  \bibnamefont {Neville}}, \ and\ \bibinfo {author} {\bibfnamefont {David~R}\
  \bibnamefont {Turner}},\ }\href {\doibase 10.1039/9781847558862} {\emph
  {\bibinfo {title} {Coordination Polymers}}}\ (\bibinfo  {publisher} {The
  Royal Society of Chemistry},\ \bibinfo {year} {2009})\BibitemShut {NoStop}%
\bibitem [{\citenamefont {Lee}\ \emph {et~al.}(2009)\citenamefont {Lee},
  \citenamefont {Farha}, \citenamefont {Roberts}, \citenamefont {Scheidt},
  \citenamefont {Nguyen},\ and\ \citenamefont {Hupp}}]{Lee2009catalyst}%
  \BibitemOpen
  \bibfield  {author} {\bibinfo {author} {\bibfnamefont {JeongYong}\
  \bibnamefont {Lee}}, \bibinfo {author} {\bibfnamefont {Omar~K.}\ \bibnamefont
  {Farha}}, \bibinfo {author} {\bibfnamefont {John}\ \bibnamefont {Roberts}},
  \bibinfo {author} {\bibfnamefont {Karl~A.}\ \bibnamefont {Scheidt}}, \bibinfo
  {author} {\bibfnamefont {SonBinh~T.}\ \bibnamefont {Nguyen}}, \ and\ \bibinfo
  {author} {\bibfnamefont {Joseph~T.}\ \bibnamefont {Hupp}},\ }\bibfield
  {title} {\enquote {\bibinfo {title} {{Metal-organic framework materials as
  catalysts}},}\ }\href {\doibase 10.1039/B807080F} {\bibfield  {journal}
  {\bibinfo  {journal} {Chem. Soc. Rev.}\ }\textbf {\bibinfo {volume} {38}},\
  \bibinfo {pages} {1450--1459} (\bibinfo {year} {2009})}\BibitemShut {NoStop}%
\bibitem [{\citenamefont {Murray}\ \emph {et~al.}(2009)\citenamefont {Murray},
  \citenamefont {Dinc\u{a}},\ and\ \citenamefont {Long}}]{Murray2009review}%
  \BibitemOpen
  \bibfield  {author} {\bibinfo {author} {\bibfnamefont {Leslie~J.}\
  \bibnamefont {Murray}}, \bibinfo {author} {\bibfnamefont {Mircea}\
  \bibnamefont {Dinc\u{a}}}, \ and\ \bibinfo {author} {\bibfnamefont
  {Jeffrey~R.}\ \bibnamefont {Long}},\ }\bibfield  {title} {\enquote {\bibinfo
  {title} {{Hydrogen storage in metal-organic frameworks}},}\ }\href {\doibase
  10.1039/B802256A} {\bibfield  {journal} {\bibinfo  {journal} {Chem. Soc.
  Rev.}\ }\textbf {\bibinfo {volume} {38}},\ \bibinfo {pages} {1294--1314}
  (\bibinfo {year} {2009})}\BibitemShut {NoStop}%
\bibitem [{\citenamefont {Stassen}\ \emph {et~al.}(2017)\citenamefont
  {Stassen}, \citenamefont {Burtch}, \citenamefont {Talin}, \citenamefont
  {Falcaro}, \citenamefont {Allendorf},\ and\ \citenamefont
  {Ameloot}}]{Stassen2017device}%
  \BibitemOpen
  \bibfield  {author} {\bibinfo {author} {\bibfnamefont {Ivo}\ \bibnamefont
  {Stassen}}, \bibinfo {author} {\bibfnamefont {Nicholas}\ \bibnamefont
  {Burtch}}, \bibinfo {author} {\bibfnamefont {Alec}\ \bibnamefont {Talin}},
  \bibinfo {author} {\bibfnamefont {Paolo}\ \bibnamefont {Falcaro}}, \bibinfo
  {author} {\bibfnamefont {Mark}\ \bibnamefont {Allendorf}}, \ and\ \bibinfo
  {author} {\bibfnamefont {Rob}\ \bibnamefont {Ameloot}},\ }\bibfield  {title}
  {\enquote {\bibinfo {title} {{An updated roadmap for the integration of
  metal-organic frameworks with electronic devices and chemical sensors}},}\
  }\href {\doibase 10.1039/C7CS00122C} {\bibfield  {journal} {\bibinfo
  {journal} {Chem. Soc. Rev.}\ }\textbf {\bibinfo {volume} {46}},\ \bibinfo
  {pages} {3185--3241} (\bibinfo {year} {2017})}\BibitemShut {NoStop}%
\bibitem [{\citenamefont {Stock}\ and\ \citenamefont
  {Biswas}(2012)}]{Norbert2012review}%
  \BibitemOpen
  \bibfield  {author} {\bibinfo {author} {\bibfnamefont {Norbert}\ \bibnamefont
  {Stock}}\ and\ \bibinfo {author} {\bibfnamefont {Shyam}\ \bibnamefont
  {Biswas}},\ }\bibfield  {title} {\enquote {\bibinfo {title} {{Synthesis of
  Metal-Organic Frameworks (MOFs): Routes to Various MOF Topologies,
  Morphologies, and Composites}},}\ }\href {\doibase 10.1021/cr200304e}
  {\bibfield  {journal} {\bibinfo  {journal} {Chem. Rev.}\ }\textbf {\bibinfo
  {volume} {112}},\ \bibinfo {pages} {933--969} (\bibinfo {year}
  {2012})}\BibitemShut {NoStop}%
\bibitem [{\citenamefont {Luo}\ \emph {et~al.}(2013)\citenamefont {Luo},
  \citenamefont {Cao}, \citenamefont {Si}, \citenamefont {Li}, \citenamefont
  {Bao}, \citenamefont {Guo}, \citenamefont {Yang}, \citenamefont {Shen},
  \citenamefont {Feng}, \citenamefont {Dai}, \citenamefont {Cao},\ and\
  \citenamefont {Xu}}]{rhodates}%
  \BibitemOpen
  \bibfield  {author} {\bibinfo {author} {\bibfnamefont {Yongkang}\
  \bibnamefont {Luo}}, \bibinfo {author} {\bibfnamefont {Chao}\ \bibnamefont
  {Cao}}, \bibinfo {author} {\bibfnamefont {Bingqi}\ \bibnamefont {Si}},
  \bibinfo {author} {\bibfnamefont {Yuke}\ \bibnamefont {Li}}, \bibinfo
  {author} {\bibfnamefont {Jinke}\ \bibnamefont {Bao}}, \bibinfo {author}
  {\bibfnamefont {Hanjie}\ \bibnamefont {Guo}}, \bibinfo {author}
  {\bibfnamefont {Xiaojun}\ \bibnamefont {Yang}}, \bibinfo {author}
  {\bibfnamefont {Chenyi}\ \bibnamefont {Shen}}, \bibinfo {author}
  {\bibfnamefont {Chunmu}\ \bibnamefont {Feng}}, \bibinfo {author}
  {\bibfnamefont {Jianhui}\ \bibnamefont {Dai}}, \bibinfo {author}
  {\bibfnamefont {Guanghan}\ \bibnamefont {Cao}}, \ and\ \bibinfo {author}
  {\bibfnamefont {Zhu-an}\ \bibnamefont {Xu}},\ }\bibfield  {title} {\enquote
  {\bibinfo {title} {{Li${}_{2}$RhO${}_{3}$: A spin-glassy relativistic Mott
  insulator}},}\ }\href {\doibase 10.1103/PhysRevB.87.161121} {\bibfield
  {journal} {\bibinfo  {journal} {Phys. Rev. B}\ }\textbf {\bibinfo {volume}
  {87}},\ \bibinfo {pages} {161121} (\bibinfo {year} {2013})}\BibitemShut
  {NoStop}%
\bibitem [{\citenamefont {Becker}\ \emph {et~al.}(2015)\citenamefont {Becker},
  \citenamefont {Hermanns}, \citenamefont {Bauer}, \citenamefont {Garst},\ and\
  \citenamefont {Trebst}}]{Becker2015spin}%
  \BibitemOpen
  \bibfield  {author} {\bibinfo {author} {\bibfnamefont {Michael}\ \bibnamefont
  {Becker}}, \bibinfo {author} {\bibfnamefont {Maria}\ \bibnamefont
  {Hermanns}}, \bibinfo {author} {\bibfnamefont {Bela}\ \bibnamefont {Bauer}},
  \bibinfo {author} {\bibfnamefont {Markus}\ \bibnamefont {Garst}}, \ and\
  \bibinfo {author} {\bibfnamefont {Simon}\ \bibnamefont {Trebst}},\ }\bibfield
   {title} {\enquote {\bibinfo {title} {{Spin-orbit physics of $j=\frac{1}{2}$
  Mott insulators on the triangular lattice}},}\ }\href {\doibase
  10.1103/PhysRevB.91.155135} {\bibfield  {journal} {\bibinfo  {journal} {Phys.
  Rev. B}\ }\textbf {\bibinfo {volume} {91}},\ \bibinfo {pages} {155135}
  (\bibinfo {year} {2015})}\BibitemShut {NoStop}%
\bibitem [{\citenamefont {Zhang}\ \emph {et~al.}(2012)\citenamefont {Zhang},
  \citenamefont {Zhang},\ and\ \citenamefont {Zhu}}]{Zhang2012hyperhoneycomb}%
  \BibitemOpen
  \bibfield  {author} {\bibinfo {author} {\bibfnamefont {Bin}\ \bibnamefont
  {Zhang}}, \bibinfo {author} {\bibfnamefont {Yan}\ \bibnamefont {Zhang}}, \
  and\ \bibinfo {author} {\bibfnamefont {Daoben}\ \bibnamefont {Zhu}},\
  }\bibfield  {title} {\enquote {\bibinfo {title} {{[(C2H5)3NH]2Cu2(C2O4)3: a
  three-dimensional metal-oxalato framework showing structurally related
  dielectric and magnetic transitions at around 165 K}},}\ }\href {\doibase
  10.1039/C2DT30818E} {\bibfield  {journal} {\bibinfo  {journal} {Dalton
  Trans.}\ }\textbf {\bibinfo {volume} {41}},\ \bibinfo {pages} {8509--8511}
  (\bibinfo {year} {2012})}\BibitemShut {NoStop}%
\bibitem [{\citenamefont {Zhang}\ \emph {et~al.}(2014)\citenamefont {Zhang},
  \citenamefont {Zhang}, \citenamefont {Wang}, \citenamefont {Wang},
  \citenamefont {Baker}, \citenamefont {Pratt},\ and\ \citenamefont
  {Zhu}}]{Zhang2014honeycomb}%
  \BibitemOpen
  \bibfield  {author} {\bibinfo {author} {\bibfnamefont {Bin}\ \bibnamefont
  {Zhang}}, \bibinfo {author} {\bibfnamefont {Yan}\ \bibnamefont {Zhang}},
  \bibinfo {author} {\bibfnamefont {Zheming}\ \bibnamefont {Wang}}, \bibinfo
  {author} {\bibfnamefont {Dongwei}\ \bibnamefont {Wang}}, \bibinfo {author}
  {\bibfnamefont {Peter~J.}\ \bibnamefont {Baker}}, \bibinfo {author}
  {\bibfnamefont {Francis~L.}\ \bibnamefont {Pratt}}, \ and\ \bibinfo {author}
  {\bibfnamefont {Daoben}\ \bibnamefont {Zhu}},\ }\bibfield  {title} {\enquote
  {\bibinfo {title} {{Candidate Quantum Spin Liquid due to Dimensional
  Reduction of a Two-Dimensional Honeycomb Lattice}},}\ }\href {\doibase
  10.1038/srep06451} {\bibfield  {journal} {\bibinfo  {journal} {Sci. Rep.}\
  }\textbf {\bibinfo {volume} {4}},\ \bibinfo {pages} {6451} (\bibinfo {year}
  {2014})}\BibitemShut {NoStop}%
\bibitem [{\citenamefont {Coronado}\ \emph {et~al.}(2001)\citenamefont
  {Coronado}, \citenamefont {Gal\'an-Mascar\'os}, \citenamefont
  {G\'omez-Garc\'ia},\ and\ \citenamefont
  {Mart\'inez-Agudo}}]{Coronado2001hyperoctagon}%
  \BibitemOpen
  \bibfield  {author} {\bibinfo {author} {\bibfnamefont {E.}~\bibnamefont
  {Coronado}}, \bibinfo {author} {\bibfnamefont {J.~R.}\ \bibnamefont
  {Gal\'an-Mascar\'os}}, \bibinfo {author} {\bibfnamefont {C.~J.}\ \bibnamefont
  {G\'omez-Garc\'ia}}, \ and\ \bibinfo {author} {\bibfnamefont {J.~M.}\
  \bibnamefont {Mart\'inez-Agudo}},\ }\bibfield  {title} {\enquote {\bibinfo
  {title} {{Molecule-Based Magnets Formed by Bimetallic Three-Dimensional
  Oxalate Networks and Chiral Tris(bipyridyl) Complex Cations. The Series
  [ZII(bpy)3][ClO4][MIICrIII(ox)3] (ZII = Ru, Fe, Co, and Ni; MII = Mn, Fe, Co,
  Ni, Cu, and Zn; ox = Oxalate Dianion)}},}\ }\href {\doibase
  10.1021/ic0008870} {\bibfield  {journal} {\bibinfo  {journal} {Inorg. Chem.}\
  }\textbf {\bibinfo {volume} {40}},\ \bibinfo {pages} {113--120} (\bibinfo
  {year} {2001})}\BibitemShut {NoStop}%
\bibitem [{\citenamefont {Lieb}(1994)}]{Lieb1994flux}%
  \BibitemOpen
  \bibfield  {author} {\bibinfo {author} {\bibfnamefont {Elliott~H.}\
  \bibnamefont {Lieb}},\ }\bibfield  {title} {\enquote {\bibinfo {title} {{Flux
  Phase of the Half-Filled Band}},}\ }\href {\doibase
  10.1103/PhysRevLett.73.2158} {\bibfield  {journal} {\bibinfo  {journal}
  {Phys. Rev. Lett.}\ }\textbf {\bibinfo {volume} {73}},\ \bibinfo {pages}
  {2158--2161} (\bibinfo {year} {1994})}\BibitemShut {NoStop}%
\bibitem [{\citenamefont {Wen}(2002)}]{Wen2002quantum}%
  \BibitemOpen
  \bibfield  {author} {\bibinfo {author} {\bibfnamefont {Xiao-Gang}\
  \bibnamefont {Wen}},\ }\bibfield  {title} {\enquote {\bibinfo {title}
  {{Quantum orders and symmetric spin liquids}},}\ }\href {\doibase
  10.1103/PhysRevB.65.165113} {\bibfield  {journal} {\bibinfo  {journal} {Phys.
  Rev. B}\ }\textbf {\bibinfo {volume} {65}},\ \bibinfo {pages} {165113}
  (\bibinfo {year} {2002})}\BibitemShut {NoStop}%
\bibitem [{\citenamefont {Kitaev}\ and\ \citenamefont
  {Laumann}(2009)}]{Kitaev2009topological}%
  \BibitemOpen
  \bibfield  {author} {\bibinfo {author} {\bibfnamefont {Alexei}\ \bibnamefont
  {Kitaev}}\ and\ \bibinfo {author} {\bibfnamefont {Chris}\ \bibnamefont
  {Laumann}},\ }\bibfield  {title} {\enquote {\bibinfo {title} {{Topological
  phases and quantum computation --- Lectures given by Alexei Kitaev at the
  2008 Les Houches Summer School "Exact methods in low-dimensional physics and
  quantum computing}},}\ }\href@noop {} {\bibfield  {journal} {\bibinfo
  {journal} {arXiv:0904.2771}\ } (\bibinfo {year} {2009})}\BibitemShut
  {NoStop}%
\bibitem [{\citenamefont {Eschmann}\ and\ \citenamefont
  {Trebst}(2017)}]{QMCunpublished}%
  \BibitemOpen
  \bibfield  {author} {\bibinfo {author} {\bibfnamefont {Tim}\ \bibnamefont
  {Eschmann}}\ and\ \bibinfo {author} {\bibfnamefont {Simon}\ \bibnamefont
  {Trebst}},\ }\href@noop {} {\bibfield  {journal} {\bibinfo  {journal}
  {private communication}\ } (\bibinfo {year} {2017})}\BibitemShut {NoStop}%
\bibitem [{\citenamefont {Nasu}\ \emph {et~al.}(2014)\citenamefont {Nasu},
  \citenamefont {Udagawa},\ and\ \citenamefont
  {Motome}}]{Nasu2014vaporization}%
  \BibitemOpen
  \bibfield  {author} {\bibinfo {author} {\bibfnamefont {Joji}\ \bibnamefont
  {Nasu}}, \bibinfo {author} {\bibfnamefont {Masafumi}\ \bibnamefont
  {Udagawa}}, \ and\ \bibinfo {author} {\bibfnamefont {Yukitoshi}\ \bibnamefont
  {Motome}},\ }\bibfield  {title} {\enquote {\bibinfo {title} {{Vaporization of
  Kitaev Spin Liquids}},}\ }\href {\doibase 10.1103/PhysRevLett.113.197205}
  {\bibfield  {journal} {\bibinfo  {journal} {Phys. Rev. Lett.}\ }\textbf
  {\bibinfo {volume} {113}},\ \bibinfo {pages} {197205} (\bibinfo {year}
  {2014})}\BibitemShut {NoStop}%
\bibitem [{Note4()}]{Note4}%
  \BibitemOpen
  \bibinfo {note} {Note that the value of the chiral invariant is shifted by 1
  compared to the full model.}\BibitemShut {Stop}%
\bibitem [{\citenamefont {Gruselle}\ \emph {et~al.}(2006)\citenamefont
  {Gruselle}, \citenamefont {Train}, \citenamefont {Boubekeur}, \citenamefont
  {Gredin},\ and\ \citenamefont {Ovanesyan}}]{Gruselle2006enantioselective}%
  \BibitemOpen
  \bibfield  {author} {\bibinfo {author} {\bibfnamefont {Michel}\ \bibnamefont
  {Gruselle}}, \bibinfo {author} {\bibfnamefont {Cyrille}\ \bibnamefont
  {Train}}, \bibinfo {author} {\bibfnamefont {Kamal}\ \bibnamefont
  {Boubekeur}}, \bibinfo {author} {\bibfnamefont {Patrick}\ \bibnamefont
  {Gredin}}, \ and\ \bibinfo {author} {\bibfnamefont {Nicholas}\ \bibnamefont
  {Ovanesyan}},\ }\bibfield  {title} {\enquote {\bibinfo {title}
  {{Enantioselective self-assembly of chiral bimetallic oxalate-based
  networks}},}\ }\href {\doibase https://doi.org/10.1016/j.ccr.2006.03.020}
  {\bibfield  {journal} {\bibinfo  {journal} {Coordination Chemistry Reviews}\
  }\textbf {\bibinfo {volume} {250}},\ \bibinfo {pages} {2491 -- 2500}
  (\bibinfo {year} {2006})}\BibitemShut {NoStop}%
\bibitem [{\citenamefont {Tranchemontagne}\ \emph {et~al.}(2008)\citenamefont
  {Tranchemontagne}, \citenamefont {Hunt},\ and\ \citenamefont
  {Yaghi}}]{Tranchemontagne2008room}%
  \BibitemOpen
  \bibfield  {author} {\bibinfo {author} {\bibfnamefont {David~J.}\
  \bibnamefont {Tranchemontagne}}, \bibinfo {author} {\bibfnamefont
  {Joseph~R.}\ \bibnamefont {Hunt}}, \ and\ \bibinfo {author} {\bibfnamefont
  {Omar~M.}\ \bibnamefont {Yaghi}},\ }\bibfield  {title} {\enquote {\bibinfo
  {title} {{Room temperature synthesis of metal-organic frameworks: MOF-5,
  MOF-74, MOF-177, MOF-199, and IRMOF-0}},}\ }\href {\doibase
  http://dx.doi.org/10.1016/j.tet.2008.06.036} {\bibfield  {journal} {\bibinfo
  {journal} {Tetrahedron}\ }\textbf {\bibinfo {volume} {64}},\ \bibinfo {pages}
  {8553 -- 8557} (\bibinfo {year} {2008})}\BibitemShut {NoStop}%
\bibitem [{\citenamefont {Rosi}\ \emph {et~al.}(2005)\citenamefont {Rosi},
  \citenamefont {Kim}, \citenamefont {Eddaoudi}, \citenamefont {Chen},
  \citenamefont {O'Keeffe},\ and\ \citenamefont {Yaghi}}]{Rosi2005rod}%
  \BibitemOpen
  \bibfield  {author} {\bibinfo {author} {\bibfnamefont {Nathaniel~L.}\
  \bibnamefont {Rosi}}, \bibinfo {author} {\bibfnamefont {Jaheon}\ \bibnamefont
  {Kim}}, \bibinfo {author} {\bibfnamefont {Mohamed}\ \bibnamefont {Eddaoudi}},
  \bibinfo {author} {\bibfnamefont {Banglin}\ \bibnamefont {Chen}}, \bibinfo
  {author} {\bibfnamefont {Michael}\ \bibnamefont {O'Keeffe}}, \ and\ \bibinfo
  {author} {\bibfnamefont {Omar~M.}\ \bibnamefont {Yaghi}},\ }\bibfield
  {title} {\enquote {\bibinfo {title} {{Rod Packings and Metal-Organic
  Frameworks Constructed from Rod-Shaped Secondary Building Units}},}\ }\href
  {\doibase 10.1021/ja045123o} {\bibfield  {journal} {\bibinfo  {journal}
  {Journal of the American Chemical Society}\ }\textbf {\bibinfo {volume}
  {127}},\ \bibinfo {pages} {1504--1518} (\bibinfo {year} {2005})},\ \bibinfo
  {note} {pMID: 15686384}\BibitemShut {NoStop}%
\bibitem [{\citenamefont {Chaloupka}\ \emph {et~al.}(2010)\citenamefont
  {Chaloupka}, \citenamefont {Jackeli},\ and\ \citenamefont
  {Khaliullin}}]{Chaloupka2010}%
  \BibitemOpen
  \bibfield  {author} {\bibinfo {author} {\bibfnamefont {Ji\v{r}\'{\i}}\
  \bibnamefont {Chaloupka}}, \bibinfo {author} {\bibfnamefont {George}\
  \bibnamefont {Jackeli}}, \ and\ \bibinfo {author} {\bibfnamefont {Giniyat}\
  \bibnamefont {Khaliullin}},\ }\bibfield  {title} {\enquote {\bibinfo {title}
  {{Kitaev-Heisenberg Model on a Honeycomb Lattice: Possible Exotic Phases in
  Iridium Oxides ${A}_{2}{\mathrm{IrO}}_{3}$}},}\ }\href {\doibase
  10.1103/PhysRevLett.105.027204} {\bibfield  {journal} {\bibinfo  {journal}
  {Phys. Rev. Lett.}\ }\textbf {\bibinfo {volume} {105}},\ \bibinfo {pages}
  {027204} (\bibinfo {year} {2010})}\BibitemShut {NoStop}%
\bibitem [{\citenamefont {Jiang}\ \emph {et~al.}(2011)\citenamefont {Jiang},
  \citenamefont {Gu}, \citenamefont {Qi},\ and\ \citenamefont
  {Trebst}}]{Jiang2011possible}%
  \BibitemOpen
  \bibfield  {author} {\bibinfo {author} {\bibfnamefont {Hong-Chen}\
  \bibnamefont {Jiang}}, \bibinfo {author} {\bibfnamefont {Zheng-Cheng}\
  \bibnamefont {Gu}}, \bibinfo {author} {\bibfnamefont {Xiao-Liang}\
  \bibnamefont {Qi}}, \ and\ \bibinfo {author} {\bibfnamefont {Simon}\
  \bibnamefont {Trebst}},\ }\bibfield  {title} {\enquote {\bibinfo {title}
  {{Possible proximity of the Mott insulating iridate Na${}_{2}$IrO${}_{3}$ to
  a topological phase: Phase diagram of the Heisenberg-Kitaev model in a
  magnetic field}},}\ }\href {\doibase 10.1103/PhysRevB.83.245104} {\bibfield
  {journal} {\bibinfo  {journal} {Phys. Rev. B}\ }\textbf {\bibinfo {volume}
  {83}},\ \bibinfo {pages} {245104} (\bibinfo {year} {2011})}\BibitemShut
  {NoStop}%
\bibitem [{\citenamefont {Rousochatzakis}\ \emph {et~al.}(2015)\citenamefont
  {Rousochatzakis}, \citenamefont {Reuther}, \citenamefont {Thomale},
  \citenamefont {Rachel},\ and\ \citenamefont
  {Perkins}}]{Rousochatzakis2015phase}%
  \BibitemOpen
  \bibfield  {author} {\bibinfo {author} {\bibfnamefont {Ioannis}\ \bibnamefont
  {Rousochatzakis}}, \bibinfo {author} {\bibfnamefont {Johannes}\ \bibnamefont
  {Reuther}}, \bibinfo {author} {\bibfnamefont {Ronny}\ \bibnamefont
  {Thomale}}, \bibinfo {author} {\bibfnamefont {Stephan}\ \bibnamefont
  {Rachel}}, \ and\ \bibinfo {author} {\bibfnamefont {N.~B.}\ \bibnamefont
  {Perkins}},\ }\bibfield  {title} {\enquote {\bibinfo {title} {{Phase Diagram
  and Quantum Order by Disorder in the Kitaev ${K}_{1}\ensuremath{-}{K}_{2}$
  Honeycomb Magnet}},}\ }\href {\doibase 10.1103/PhysRevX.5.041035} {\bibfield
  {journal} {\bibinfo  {journal} {Phys. Rev. X}\ }\textbf {\bibinfo {volume}
  {5}},\ \bibinfo {pages} {041035} (\bibinfo {year} {2015})}\BibitemShut
  {NoStop}%
\bibitem [{\citenamefont {Winter}\ \emph {et~al.}(2016)\citenamefont {Winter},
  \citenamefont {Li}, \citenamefont {Jeschke},\ and\ \citenamefont
  {Valent\'{\i}}}]{WinterValenti2016}%
  \BibitemOpen
  \bibfield  {author} {\bibinfo {author} {\bibfnamefont {Stephen~M.}\
  \bibnamefont {Winter}}, \bibinfo {author} {\bibfnamefont {Ying}\ \bibnamefont
  {Li}}, \bibinfo {author} {\bibfnamefont {Harald~O.}\ \bibnamefont {Jeschke}},
  \ and\ \bibinfo {author} {\bibfnamefont {Roser}\ \bibnamefont
  {Valent\'{\i}}},\ }\bibfield  {title} {\enquote {\bibinfo {title} {Challenges
  in design of kitaev materials: Magnetic interactions from competing energy
  scales},}\ }\href {\doibase 10.1103/PhysRevB.93.214431} {\bibfield  {journal}
  {\bibinfo  {journal} {Phys. Rev. B}\ }\textbf {\bibinfo {volume} {93}},\
  \bibinfo {pages} {214431} (\bibinfo {year} {2016})}\BibitemShut {NoStop}%
\bibitem [{\citenamefont {Dikhtiarenko}\ \emph {et~al.}(2016)\citenamefont
  {Dikhtiarenko}, \citenamefont {Villanueva-Delgado}, \citenamefont {Valiente},
  \citenamefont {Garc\'ia},\ and\ \citenamefont {Gimeno}}]{Dikhtiarenko2016}%
  \BibitemOpen
  \bibfield  {author} {\bibinfo {author} {\bibfnamefont {Alla}\ \bibnamefont
  {Dikhtiarenko}}, \bibinfo {author} {\bibfnamefont {Pedro}\ \bibnamefont
  {Villanueva-Delgado}}, \bibinfo {author} {\bibfnamefont {Rafael}\
  \bibnamefont {Valiente}}, \bibinfo {author} {\bibfnamefont {Jos\'e~R.}\
  \bibnamefont {Garc\'ia}}, \ and\ \bibinfo {author} {\bibfnamefont {Jos\'e}\
  \bibnamefont {Gimeno}},\ }\bibfield  {title} {\enquote {\bibinfo {title}
  {{Tris(bipyridine)Metal(II)-Templated Assemblies of 3D Alkali-Ruthenium
  Oxalate Coordination Frameworks: Crystal Structures, Characterization and
  Photocatalytic Activity in Water Reduction}},}\ }\href {\doibase
  10.3390/polym8020048} {\bibfield  {journal} {\bibinfo  {journal} {Polymers}\
  }\textbf {\bibinfo {volume} {8}},\ \bibinfo {pages} {48} (\bibinfo {year}
  {2016})}\BibitemShut {NoStop}%
\bibitem [{\citenamefont {Knolle}\ \emph {et~al.}(2014)\citenamefont {Knolle},
  \citenamefont {Kovrizhin}, \citenamefont {Chalker},\ and\ \citenamefont
  {Moessner}}]{Knolle2014dynamics}%
  \BibitemOpen
  \bibfield  {author} {\bibinfo {author} {\bibfnamefont {J.}~\bibnamefont
  {Knolle}}, \bibinfo {author} {\bibfnamefont {D.~L.}\ \bibnamefont
  {Kovrizhin}}, \bibinfo {author} {\bibfnamefont {J.~T.}\ \bibnamefont
  {Chalker}}, \ and\ \bibinfo {author} {\bibfnamefont {R.}~\bibnamefont
  {Moessner}},\ }\bibfield  {title} {\enquote {\bibinfo {title} {{Dynamics of a
  Two-Dimensional Quantum Spin Liquid: Signatures of Emergent Majorana Fermions
  and Fluxes}},}\ }\href {\doibase 10.1103/PhysRevLett.112.207203} {\bibfield
  {journal} {\bibinfo  {journal} {Phys. Rev. Lett.}\ }\textbf {\bibinfo
  {volume} {112}},\ \bibinfo {pages} {207203} (\bibinfo {year}
  {2014})}\BibitemShut {NoStop}%
\bibitem [{\citenamefont {Smith}\ \emph {et~al.}(2015)\citenamefont {Smith},
  \citenamefont {Knolle}, \citenamefont {Kovrizhin}, \citenamefont {Chalker},\
  and\ \citenamefont {Moessner}}]{Smith2015neutron}%
  \BibitemOpen
  \bibfield  {author} {\bibinfo {author} {\bibfnamefont {A.}~\bibnamefont
  {Smith}}, \bibinfo {author} {\bibfnamefont {J.}~\bibnamefont {Knolle}},
  \bibinfo {author} {\bibfnamefont {D.~L.}\ \bibnamefont {Kovrizhin}}, \bibinfo
  {author} {\bibfnamefont {J.~T.}\ \bibnamefont {Chalker}}, \ and\ \bibinfo
  {author} {\bibfnamefont {R.}~\bibnamefont {Moessner}},\ }\bibfield  {title}
  {\enquote {\bibinfo {title} {{Neutron scattering signatures of the 3D
  hyperhoneycomb Kitaev quantum spin liquid}},}\ }\href {\doibase
  10.1103/PhysRevB.92.180408} {\bibfield  {journal} {\bibinfo  {journal} {Phys.
  Rev. B}\ }\textbf {\bibinfo {volume} {92}},\ \bibinfo {pages} {180408}
  (\bibinfo {year} {2015})}\BibitemShut {NoStop}%
\bibitem [{\citenamefont {Banerjee}\ \emph {et~al.}(2017)\citenamefont
  {Banerjee}, \citenamefont {Yan}, \citenamefont {Knolle}, \citenamefont
  {Bridges}, \citenamefont {Stone}, \citenamefont {Lumsden}, \citenamefont
  {Mandrus}, \citenamefont {Tennant},\ and\ \citenamefont
  {Nagler}}]{Banerjee2016neutron}%
  \BibitemOpen
  \bibfield  {author} {\bibinfo {author} {\bibfnamefont {Arnab}\ \bibnamefont
  {Banerjee}}, \bibinfo {author} {\bibfnamefont {Jiaqiang}\ \bibnamefont
  {Yan}}, \bibinfo {author} {\bibfnamefont {Johannes}\ \bibnamefont {Knolle}},
  \bibinfo {author} {\bibfnamefont {Craig~A.}\ \bibnamefont {Bridges}},
  \bibinfo {author} {\bibfnamefont {Matthew~B.}\ \bibnamefont {Stone}},
  \bibinfo {author} {\bibfnamefont {Mark~D.}\ \bibnamefont {Lumsden}}, \bibinfo
  {author} {\bibfnamefont {David~G.}\ \bibnamefont {Mandrus}}, \bibinfo
  {author} {\bibfnamefont {David A. Moessner~Roderich}\ \bibnamefont
  {Tennant}}, \ and\ \bibinfo {author} {\bibfnamefont {Stephen~E.}\
  \bibnamefont {Nagler}},\ }\bibfield  {title} {\enquote {\bibinfo {title}
  {{{Neutron scattering in proximate quantum spin liquid
  $\alpha$-RuCl$_3$}}},}\ }\href@noop {} {\bibfield  {journal} {\bibinfo
  {journal} {Science}\ }\textbf {\bibinfo {volume} {356}},\ \bibinfo {pages}
  {1055} (\bibinfo {year} {2017})}\BibitemShut {NoStop}%
\bibitem [{\citenamefont {Perreault}\ \emph {et~al.}(2015)\citenamefont
  {Perreault}, \citenamefont {Knolle}, \citenamefont {Perkins},\ and\
  \citenamefont {Burnell}}]{Perreault2015theory}%
  \BibitemOpen
  \bibfield  {author} {\bibinfo {author} {\bibfnamefont {Brent}\ \bibnamefont
  {Perreault}}, \bibinfo {author} {\bibfnamefont {Johannes}\ \bibnamefont
  {Knolle}}, \bibinfo {author} {\bibfnamefont {Natalia~B.}\ \bibnamefont
  {Perkins}}, \ and\ \bibinfo {author} {\bibfnamefont {F.~J.}\ \bibnamefont
  {Burnell}},\ }\bibfield  {title} {\enquote {\bibinfo {title} {{Theory of
  Raman response in three-dimensional Kitaev spin liquids: Application to
  $\ensuremath{\beta}$- and
  $\ensuremath{\gamma}\ensuremath{-}{\mathrm{Li}}_{2}{\mathrm{IrO}}_{3}$
  compounds}},}\ }\href {\doibase 10.1103/PhysRevB.92.094439} {\bibfield
  {journal} {\bibinfo  {journal} {Phys. Rev. B}\ }\textbf {\bibinfo {volume}
  {92}},\ \bibinfo {pages} {094439} (\bibinfo {year} {2015})}\BibitemShut
  {NoStop}%
\bibitem [{\citenamefont {Sandilands}\ \emph {et~al.}(2015)\citenamefont
  {Sandilands}, \citenamefont {Tian}, \citenamefont {Plumb}, \citenamefont
  {Kim},\ and\ \citenamefont {Burch}}]{Sandilands2015scattering}%
  \BibitemOpen
  \bibfield  {author} {\bibinfo {author} {\bibfnamefont {Luke~J.}\ \bibnamefont
  {Sandilands}}, \bibinfo {author} {\bibfnamefont {Yao}\ \bibnamefont {Tian}},
  \bibinfo {author} {\bibfnamefont {Kemp~W.}\ \bibnamefont {Plumb}}, \bibinfo
  {author} {\bibfnamefont {Young-June}\ \bibnamefont {Kim}}, \ and\ \bibinfo
  {author} {\bibfnamefont {Kenneth~S.}\ \bibnamefont {Burch}},\ }\bibfield
  {title} {\enquote {\bibinfo {title} {{Scattering Continuum and Possible
  Fractionalized Excitations in
  $\ensuremath{\alpha}\text{\ensuremath{-}}{\mathrm{RuCl}}_{3}$}},}\ }\href
  {\doibase 10.1103/PhysRevLett.114.147201} {\bibfield  {journal} {\bibinfo
  {journal} {Phys. Rev. Lett.}\ }\textbf {\bibinfo {volume} {114}},\ \bibinfo
  {pages} {147201} (\bibinfo {year} {2015})}\BibitemShut {NoStop}%
\bibitem [{\citenamefont {Nasu}\ \emph {et~al.}(2016)\citenamefont {Nasu},
  \citenamefont {Knolle}, \citenamefont {Kovrizhin}, \citenamefont {Motome},\
  and\ \citenamefont {Moessner}}]{Nasu2016fermionic}%
  \BibitemOpen
  \bibfield  {author} {\bibinfo {author} {\bibfnamefont {Joji}\ \bibnamefont
  {Nasu}}, \bibinfo {author} {\bibfnamefont {Johannes}\ \bibnamefont {Knolle}},
  \bibinfo {author} {\bibfnamefont {Dima~L.}\ \bibnamefont {Kovrizhin}},
  \bibinfo {author} {\bibfnamefont {Yukitoshi}\ \bibnamefont {Motome}}, \ and\
  \bibinfo {author} {\bibfnamefont {Roderich}\ \bibnamefont {Moessner}},\
  }\bibfield  {title} {\enquote {\bibinfo {title} {{Fermionic response from
  fractionalization in an insulating two-dimensional magnet}},}\ }\href
  {\doibase 10.1038/nphys3809} {\bibfield  {journal} {\bibinfo  {journal} {Nat.
  Phys.}\ }\textbf {\bibinfo {volume} {12}},\ \bibinfo {pages} {912--915}
  (\bibinfo {year} {2016})}\BibitemShut {NoStop}%
\bibitem [{\citenamefont {Hal\'{a}sz}\ and\ \citenamefont
  {Perkins}(2017)}]{RIXS3D}%
  \BibitemOpen
  \bibfield  {author} {\bibinfo {author} {\bibfnamefont {Brent}\ \bibnamefont
  {Hal\'{a}sz}, \bibfnamefont {G\'{a}bor B.~Perreault}}\ and\ \bibinfo {author}
  {\bibfnamefont {Natalia~B.}\ \bibnamefont {Perkins}},\ }\bibfield  {title}
  {\enquote {\bibinfo {title} {{Probing spinon nodal structures in
  three-dimensional Kitaev spin liquids}},}\ }\href@noop {} {\bibfield
  {journal} {\bibinfo  {journal} {arXiv:1705.05894}\ } (\bibinfo {year}
  {2017})}\BibitemShut {NoStop}%
\bibitem [{\citenamefont {Tanaka}\ \emph {et~al.}(2012)\citenamefont {Tanaka},
  \citenamefont {Ren}, \citenamefont {Sato}, \citenamefont {Nakayama},
  \citenamefont {Souma}, \citenamefont {Takahashi}, \citenamefont {Segawa},\
  and\ \citenamefont {Ando}}]{Tanaka2012SnTe}%
  \BibitemOpen
  \bibfield  {author} {\bibinfo {author} {\bibfnamefont {Y}~\bibnamefont
  {Tanaka}}, \bibinfo {author} {\bibfnamefont {Zhi}\ \bibnamefont {Ren}},
  \bibinfo {author} {\bibfnamefont {T}~\bibnamefont {Sato}}, \bibinfo {author}
  {\bibfnamefont {K}~\bibnamefont {Nakayama}}, \bibinfo {author} {\bibfnamefont
  {S}~\bibnamefont {Souma}}, \bibinfo {author} {\bibfnamefont {T}~\bibnamefont
  {Takahashi}}, \bibinfo {author} {\bibfnamefont {Kouji}\ \bibnamefont
  {Segawa}}, \ and\ \bibinfo {author} {\bibfnamefont {Yoichi}\ \bibnamefont
  {Ando}},\ }\bibfield  {title} {\enquote {\bibinfo {title} {{Experimental
  realization of a topological crystalline insulator in SnTe}},}\ }\href@noop
  {} {\bibfield  {journal} {\bibinfo  {journal} {Nature Physics}\ }\textbf
  {\bibinfo {volume} {8}},\ \bibinfo {pages} {800--803} (\bibinfo {year}
  {2012})}\BibitemShut {NoStop}%
\bibitem [{\citenamefont {Xu}\ \emph {et~al.}(2012)\citenamefont {Xu},
  \citenamefont {Liu}, \citenamefont {Alidoust}, \citenamefont {Neupane},
  \citenamefont {Qian}, \citenamefont {Belopolski}, \citenamefont {Denlinger},
  \citenamefont {Wang}, \citenamefont {Lin}, \citenamefont {Wray} \emph
  {et~al.}}]{Yang2012SnTe}%
  \BibitemOpen
  \bibfield  {author} {\bibinfo {author} {\bibfnamefont {Su-Yang}\ \bibnamefont
  {Xu}}, \bibinfo {author} {\bibfnamefont {Chang}\ \bibnamefont {Liu}},
  \bibinfo {author} {\bibfnamefont {N}~\bibnamefont {Alidoust}}, \bibinfo
  {author} {\bibfnamefont {M}~\bibnamefont {Neupane}}, \bibinfo {author}
  {\bibfnamefont {D}~\bibnamefont {Qian}}, \bibinfo {author} {\bibfnamefont
  {I}~\bibnamefont {Belopolski}}, \bibinfo {author} {\bibfnamefont
  {JD}~\bibnamefont {Denlinger}}, \bibinfo {author} {\bibfnamefont
  {YJ}~\bibnamefont {Wang}}, \bibinfo {author} {\bibfnamefont {H}~\bibnamefont
  {Lin}}, \bibinfo {author} {\bibfnamefont {LA}~\bibnamefont {Wray}},  \emph
  {et~al.},\ }\bibfield  {title} {\enquote {\bibinfo {title} {{Observation of a
  topological crystalline insulator phase and topological phase transition in
  Pb$_{1-x}$Sn$_x$Te}},}\ }\href@noop {} {\bibfield  {journal} {\bibinfo
  {journal} {Nature communications}\ }\textbf {\bibinfo {volume} {3}},\
  \bibinfo {pages} {1192} (\bibinfo {year} {2012})}\BibitemShut {NoStop}%
\bibitem [{\citenamefont {Dziawa}\ \emph {et~al.}(2012)\citenamefont {Dziawa},
  \citenamefont {Kowalski}, \citenamefont {Dybko}, \citenamefont {Buczko},
  \citenamefont {Szczerbakow}, \citenamefont {Szot}, \citenamefont
  {{\L}usakowska}, \citenamefont {Balasubramanian}, \citenamefont {Wojek},
  \citenamefont {Berntsen} \emph {et~al.}}]{Sziawa2012SnTe}%
  \BibitemOpen
  \bibfield  {author} {\bibinfo {author} {\bibfnamefont {P}~\bibnamefont
  {Dziawa}}, \bibinfo {author} {\bibfnamefont {BJ}~\bibnamefont {Kowalski}},
  \bibinfo {author} {\bibfnamefont {K}~\bibnamefont {Dybko}}, \bibinfo {author}
  {\bibfnamefont {R}~\bibnamefont {Buczko}}, \bibinfo {author} {\bibfnamefont
  {A}~\bibnamefont {Szczerbakow}}, \bibinfo {author} {\bibfnamefont
  {M}~\bibnamefont {Szot}}, \bibinfo {author} {\bibfnamefont {E}~\bibnamefont
  {{\L}usakowska}}, \bibinfo {author} {\bibfnamefont {T}~\bibnamefont
  {Balasubramanian}}, \bibinfo {author} {\bibfnamefont {Bastian~M}\
  \bibnamefont {Wojek}}, \bibinfo {author} {\bibfnamefont {MH}~\bibnamefont
  {Berntsen}},  \emph {et~al.},\ }\bibfield  {title} {\enquote {\bibinfo
  {title} {{Topological crystalline insulator states in Pb$_{1-x}$Sn$_x$Se}},}\
  }\href@noop {} {\bibfield  {journal} {\bibinfo  {journal} {Nature materials}\
  }\textbf {\bibinfo {volume} {11}},\ \bibinfo {pages} {1023--1027} (\bibinfo
  {year} {2012})}\BibitemShut {NoStop}%
\bibitem [{\citenamefont {Delgado~Friedrichs}\ \emph
  {et~al.}(2003{\natexlab{a}})\citenamefont {Delgado~Friedrichs}, \citenamefont
  {O'Keeffe},\ and\ \citenamefont {Yaghi}}]{OKeeffe2003regular}%
  \BibitemOpen
  \bibfield  {author} {\bibinfo {author} {\bibfnamefont {Olaf}\ \bibnamefont
  {Delgado~Friedrichs}}, \bibinfo {author} {\bibfnamefont {Michael}\
  \bibnamefont {O'Keeffe}}, \ and\ \bibinfo {author} {\bibfnamefont {Omar~M.}\
  \bibnamefont {Yaghi}},\ }\bibfield  {title} {\enquote {\bibinfo {title}
  {{Three-periodic nets and tilings: regular and quasiregular nets}},}\ }\href
  {\doibase 10.1107/S0108767302018494} {\bibfield  {journal} {\bibinfo
  {journal} {Acta Crystallogr. Sect. A}\ }\textbf {\bibinfo {volume} {59}},\
  \bibinfo {pages} {22--27} (\bibinfo {year} {2003}{\natexlab{a}})}\BibitemShut
  {NoStop}%
\bibitem [{\citenamefont {Delgado~Friedrichs}\ \emph
  {et~al.}(2003{\natexlab{b}})\citenamefont {Delgado~Friedrichs}, \citenamefont
  {O'Keeffe},\ and\ \citenamefont {Yaghi}}]{OKeeffe2003semiregular}%
  \BibitemOpen
  \bibfield  {author} {\bibinfo {author} {\bibfnamefont {Olaf}\ \bibnamefont
  {Delgado~Friedrichs}}, \bibinfo {author} {\bibfnamefont {Michael}\
  \bibnamefont {O'Keeffe}}, \ and\ \bibinfo {author} {\bibfnamefont {Omar~M.}\
  \bibnamefont {Yaghi}},\ }\bibfield  {title} {\enquote {\bibinfo {title}
  {Three-periodic nets and tilings: semiregular nets},}\ }\href {\doibase
  10.1107/S0108767303017100} {\bibfield  {journal} {\bibinfo  {journal} {Acta
  Crystallogr. Sect. A}\ }\textbf {\bibinfo {volume} {59}},\ \bibinfo {pages}
  {515--525} (\bibinfo {year} {2003}{\natexlab{b}})}\BibitemShut {NoStop}%
\bibitem [{\citenamefont {Heesch}\ and\ \citenamefont {Laves}(1933)}]{laves}%
  \BibitemOpen
  \bibfield  {author} {\bibinfo {author} {\bibfnamefont {H.}~\bibnamefont
  {Heesch}}\ and\ \bibinfo {author} {\bibfnamefont {F.}~\bibnamefont {Laves}},\
  }\bibfield  {title} {\enquote {\bibinfo {title} {{\"Uber d\"unne
  Kugelpackungen}},}\ }\href {http://dx.doi.org/10.1524/zkri.1933.85.1.443}
  {\bibfield  {journal} {\bibinfo  {journal} {Z. Kristallogr.}\ }\textbf
  {\bibinfo {volume} {85}},\ \bibinfo {pages} {443} (\bibinfo {year}
  {1933})}\BibitemShut {NoStop}%
\bibitem [{\citenamefont {Sunada}(2008)}]{K4}%
  \BibitemOpen
  \bibfield  {author} {\bibinfo {author} {\bibfnamefont {T.}~\bibnamefont
  {Sunada}},\ }\bibfield  {title} {\enquote {\bibinfo {title} {{Crystals That
  Nature Might Miss Creating}},}\ }\href@noop {} {\bibfield  {journal}
  {\bibinfo  {journal} {Notices Amer. Math. Soc.}\ }\textbf {\bibinfo {volume}
  {55}},\ \bibinfo {pages} {208} (\bibinfo {year} {2008})}\BibitemShut
  {NoStop}%
\end{thebibliography}%

\end{document}